\documentclass[preprint,3p,12pt]{elsarticle}

\makeatletter
\def\ps@pprintTitle{%
	\let\@oddhead\@empty
	\let\@evenhead\@empty
	\def\@oddfoot{\footnotesize\itshape
		\hfill\today}%
	\let\@evenfoot\@oddfoot}
\makeatother

\usepackage{amsmath}
\usepackage{amsfonts}
\usepackage{amssymb}
\usepackage{graphicx}
\usepackage{subfigure}
\usepackage{float}

\biboptions{sort&compress}

\newcommand{\degree}{^\circ}
\renewcommand{\vec}[1]{\boldsymbol{#1}}

\begin{document}

\begin{frontmatter}
\title{Ray Effect in Rarefied Flow Simulation}

\author[ad1]{Yajun Zhu}
\ead{zhuyajun@mail.nwpu.edu.cn}
\author[ad1]{Chengwen Zhong}
\ead{zhongcw@nwpu.edu.cn}
\author[ad2,ad3]{Kun Xu\corref{cor1}}
\ead{makxu@ust.hk}
\address[ad1]{National Key Laboratory of Science and Technology on Aerodynamic Design and Research, Northwestern Polytechnical University, Xi'an, Shaanxi 710072, China}
\address[ad2]{Department of Mathematics, Hong Kong University of Science and Technology, Hong Kong, China}
\address[ad3]{HKUST Shenzhen Research Institute, Shenzhen 518057, China}
\cortext[cor1]{Corresponding author}

\begin{abstract}
Ray effect usually appears in the radiative transfer when using discrete ordinates method (DOM) in the simulations.
The cause and remedy for the ray effect have been intensively investigated in the radiation community.
For rarefied gas flow, the ray effect is also associated with the discrete velocity method (DVM).
However, few studies have been carried out in the rarefied community.
In this paper, we take a detailed investigation of the ray effect in the rarefied flow simulations.
Starting from a few commonly used benchmark tests, the root of the ray effect has been analyzed theoretically and validated numerically.
At the same time, the influence of the ray effect on the quality of the numerical results of rarefied flow is estimated quantitatively.
After understanding the nature of the ray effect, the strategy to minimize the ray effect through the discretization of the particle velocity space is presented and applied in the numerical simulations.
An optimal velocity discretization for DVM is problem dependent and can be hardly obtained in the complex flow simulations.
Due to the intrinsic self-adaptation of particle velocity, the stochastic particle methods are free from the ray effect.
In rarefied regimes, the particle method seems more appropriate in the capturing of highly non-equilibrium flow behavior.
\end{abstract}

\end{frontmatter}

\section{Introduction}
Ray effect is one of the most commonly unsolved problem in numerical simulations of radiative transfer problems when using the discrete ordinates method (DOM).
In 1960s, unphysical spatial distortion, i.e., the so-called ray effect, was found in the numerical solutions obtained from the discrete ordinates approximations to the Boltzmann transport equation.
It was gradually realized that the cause of ray effect is due to the discrete ordinates formulation itself instead of numerical treatments.
Specifically, a limited number of discrete angular directions could  only provide a finite capability to resolve the local distribution function and the information propagate only along some specific directions.
After Lathrop gave a detailed analysis about the ray effect \cite{lathrop1968ray}, many remedies have been proposed to eliminate or mitigate it in the numerical simulations.

In radiative transfer under the discrete ordinate method (DOM) framework, all efforts to reduce the ray effect are about the angular discretization in the velocity space.
Theoretically, the technique without angular discretization, such as the fictitious source method and the modified discrete ordinates methods, can fully eliminate the ray effect; while the optimal angular discretization method can only mitigate the ray effect to certain extent.
The fictitious source method presented by Lathrop \cite{lathrop1971remedies} was about to convert the discrete ordinates approximation into a spherical harmonics-like equation by introducing a fictitious source term.
From the original idea of modified differential approximation \cite{olfe1967modification,modest1989modified}, Liou et al. \cite{liou1997ray}, Ramankutty et al. \cite{ramankutty1997modified}, and Sakami et al. \cite{sakami2000application}  employed a modified DOM to reduce the ray effect, where the radiation intensity is divided into two components, i.e., the emission and reflection from the walls and the emission and scattering from the medium, and the contribution from the wall is computed by analytic method, while that from the medium is treated by discrete ordinates.
A modified DOM was also given by Coelho \cite{coelho2002role} for the ray effect caused by the sharp gradients of medium temperature.
Instead of adopting analytic computation, Baek et al. \cite{baek2000combined} used the Monte Carlo method to solve the wall related intensity in the modified DOM.
From the analysis of the ray effect, it is commonly agreed that increasing the number of discrete angular directions would effectively mitigate the problem.
However, for the standard quadrature sets of DOM, undesirable negative weights will occur for $N>12$, which gives an upper limit on the number of angular directions and constrains the $S_N$ method to fully remove the ray effect.
Therefore, many strategies have been proposed to adjust the angular discretization and quadrature sets in order to mitigate the ray effect, such as $T_N$ quadrature set \cite{thurgood1995tn}, staggered and adaptive quadrature sets \cite{cumber2000jet}, and quadrature rotation \cite{tencer2016ray,camminady2019ray}.
Detailed analysis and reviews can be found in \cite{chai1993ray,li2003mitigation,coelho2014advances}.

In rarefied gas simulations, in order to capture the non-equilibrium physics, many numerical schemes have been developed based on the discrete velocity distribution function, such as the discrete velocity method (DVM) \cite{yang1995rarefied,mieussens2000discrete}, gas-kinetic unified algorithm (GKUA) \cite{li2004gkua}, and unified gas-kinetic scheme (UGKS) \cite{xu2010unified,xu2015direct} and discrete UGKS (DUGKS) \cite{guo2013discrete,guo2015discrete}.
 The discontinuities induced in the velocity space by solid boundary or initial state \cite{sone1992discontinuity,aoki2001note} are associated with  the ray effect in rarefied flow simulation with discrete velocity space.
For example, several plateaus were found by Brull et al. \cite{brull2014local} in the numerical solutions of Sod test case, and it is pointed out that the size of the plateaus is related to the discrete velocity interval and evolution time \cite{brull2014local}.
In the computations of lid-driven cavity flow, diagonal wiggles were observed in the temperature distribution obtained by the DVM \cite{sekaran2018analysis,su2019dg} and implicit UGKS \cite{zhu2016implicit}.
Ray effect was also mentioned in the DUGKS computations \cite{zhu2016unstructure,guo2016dphonon}.
Similar to the mitigation techniques in the radiation community, semi-analytic method \cite{naris2005driven,aoki2001note} has been introduced to remove the influence of ray effect by identifying the propagation of the discontinuities.
Without discretization in the velocity space, the integro-moment method (IMM) \cite{varoutis2008application} does not suffer from the ray effect either.
However, for highly non-equilibrium flows with complex geometry, the numerical schemes mostly use discrete particle velocities.
In comparison with discrete angular directions in radiative transfer, the discretization of the particle velocity space in rarefied flow
is more flexible.
Therefore, in the current paper we will focus mainly on the numerical discretization of the particle velocity space.
Since stochastic particle methods, such as the direct simulation Monte Carlo (DSMC) \cite{bird1994book}, unified stochastic particle (USP) method \cite{fei2018particle}, unified gas-kinetic particle (UGKP) and wave-particle (UGKWP) methods \cite{li2018ugkp,liu2018ugkwp,zhu2019ugkwp,li2019ugkwp}, are ray effect free and can be regarded as an adaptive method in the velocity space, the particle method will be analyzed as well in the present paper.

The rest of the paper is organized as follows.
In Section \ref{sec:analysis}, theoretical analysis is carried out on the cause and main features of the ray effect.
In Section \ref{sec:validation}, the theoretical analysis will be validated by several numerical test cases.
Different mitigation techniques are compared in Section \ref{sec:mitigation}.
A conclusion will be drawn in the last section.

\section{Theoretical analysis of ray effect} \label{sec:analysis}
In this section, we will give a detailed analysis of the ray effect in rarefied flow.
For simplicity, we will introduce the problem based on a one dimensional test with an initial discontinuity in the physical space, i.e.,
\begin{equation}\label{eq:problem_setup}
(\rho, U, T) =
\begin{cases}
\left(\rho_l, U_l, T_l \right), &\quad x < 0;\\
\left(\rho_r, U_r, T_r \right), &\quad x > 0;
\end{cases}
\end{equation}
where $\rho$, $U$, $T$ denote macroscopic density, velocity and temperature, respectively.

\begin{figure}[H]
	\subfigure[Density]{\includegraphics[width=0.32\textwidth]{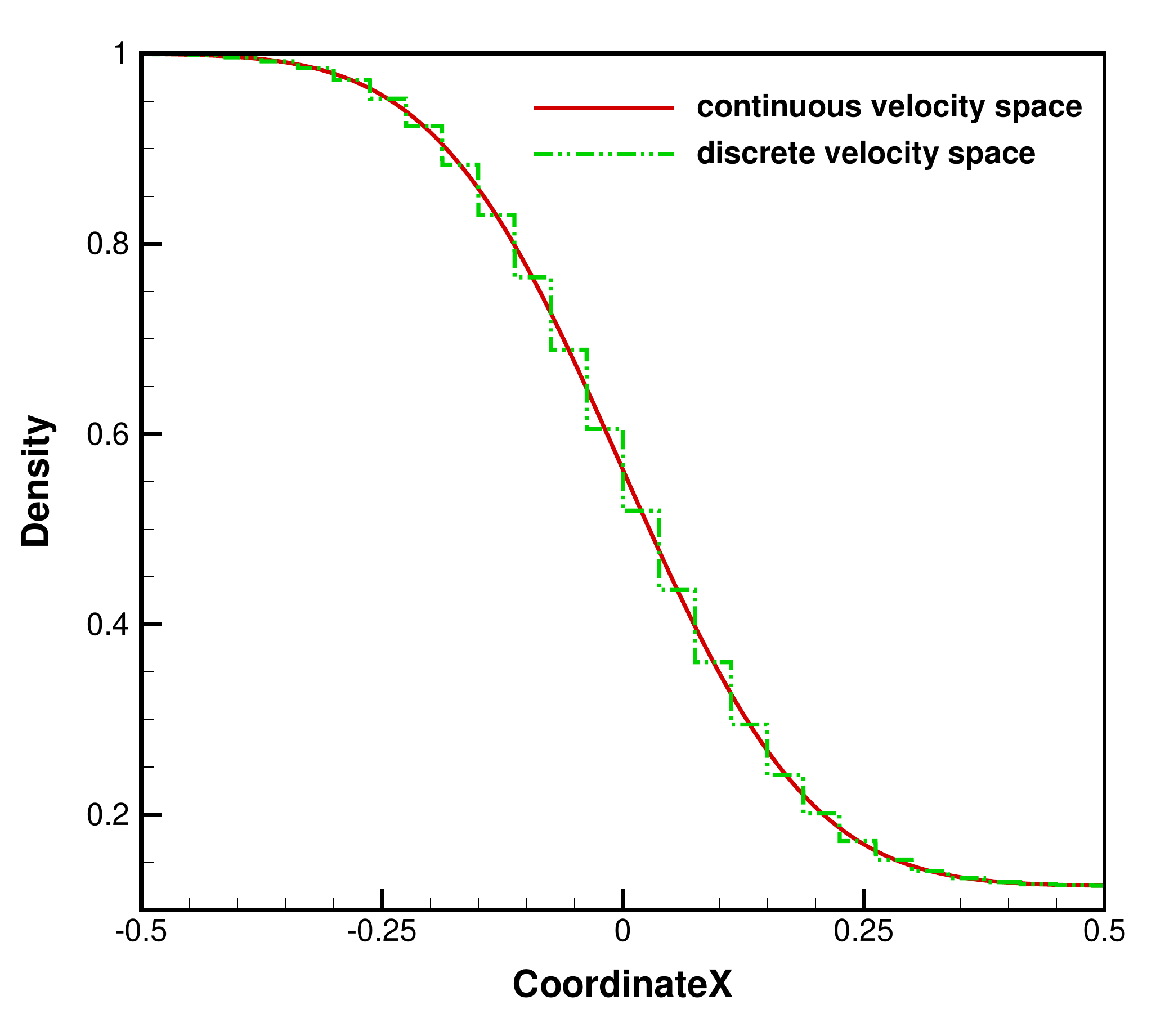}}
	\subfigure[Velocity]{\includegraphics[width=0.32\textwidth]{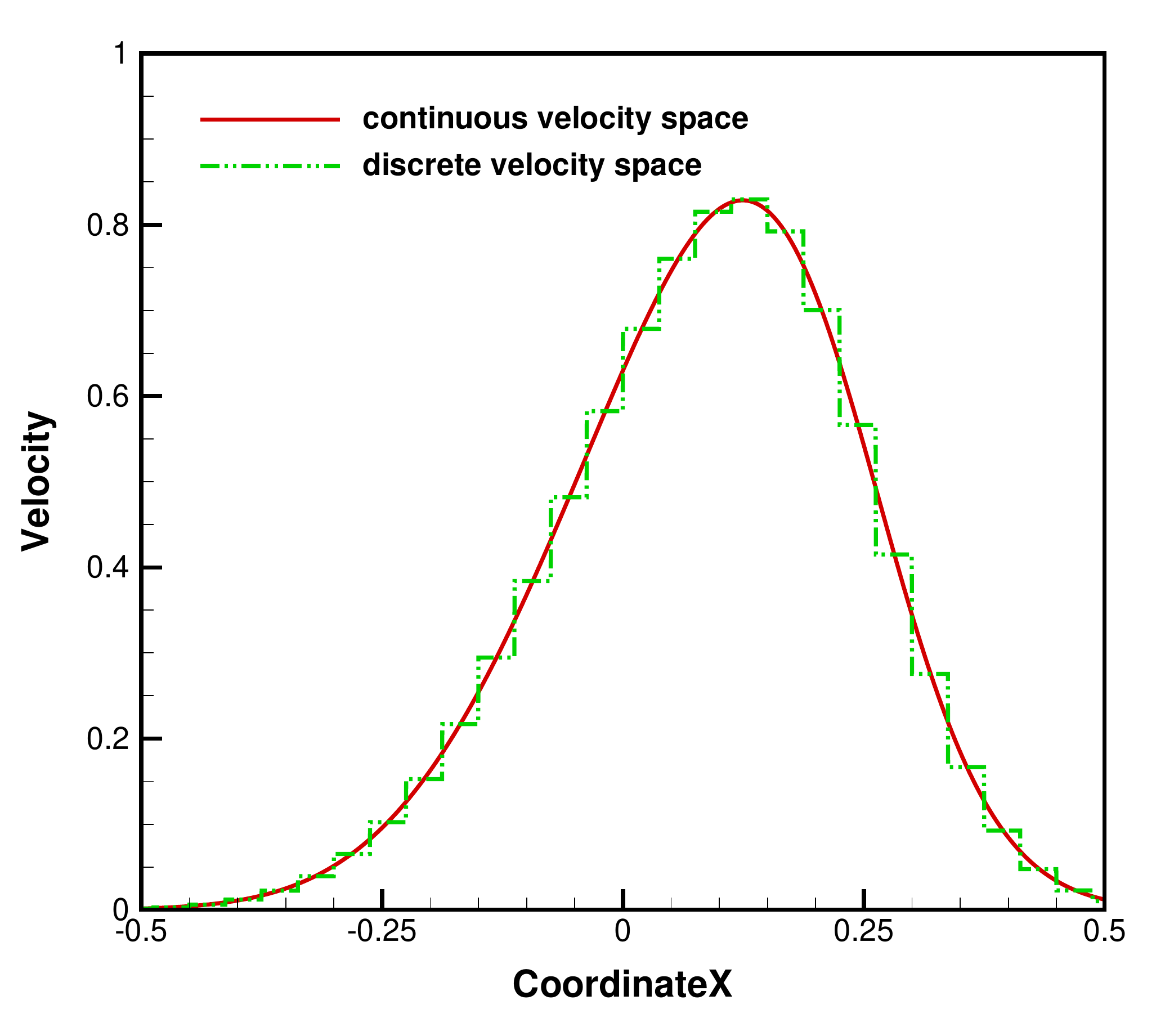}}
	\subfigure[Temperature]{\includegraphics[width=0.32\textwidth]{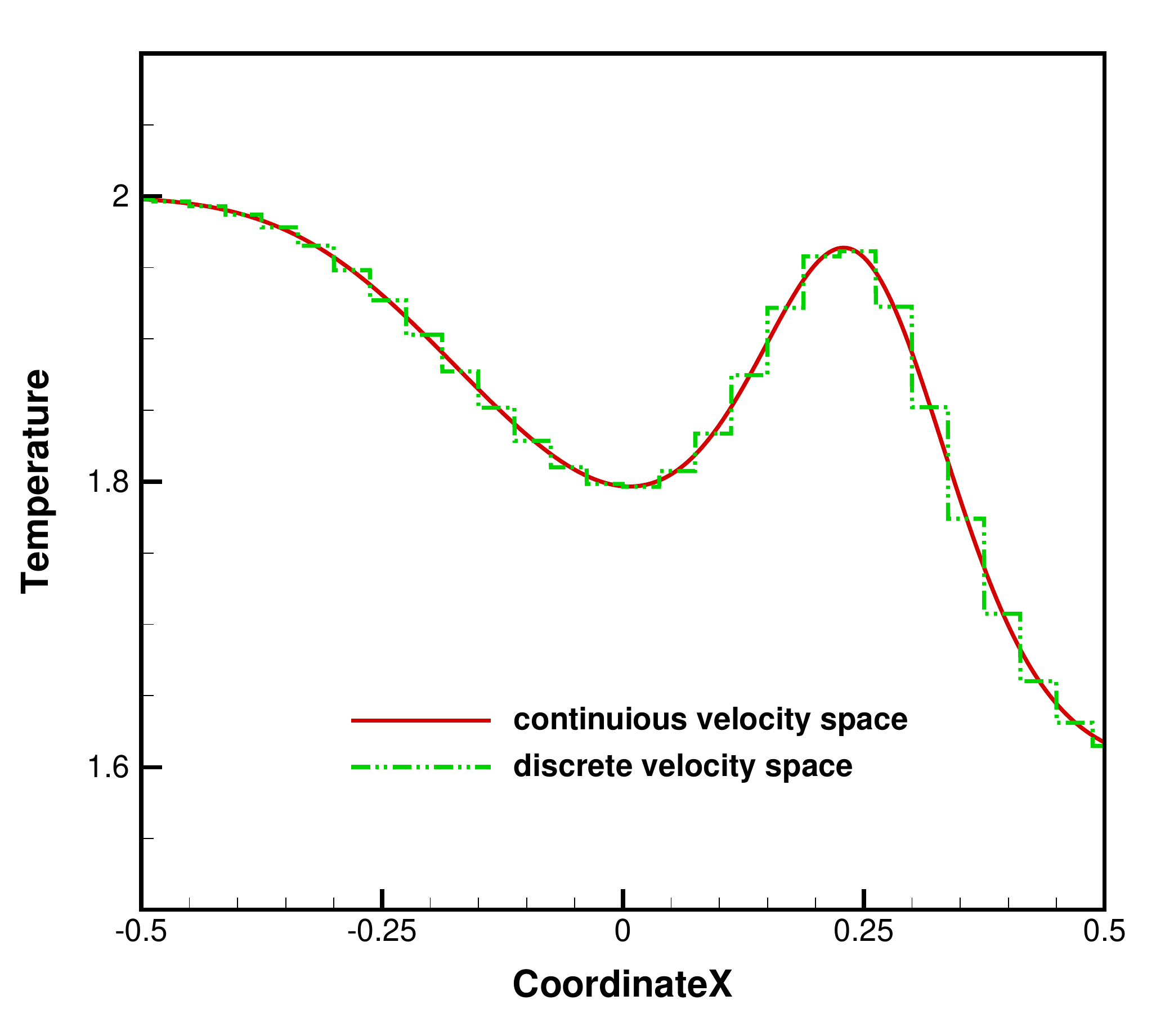}}
	\caption{\label{fig:sod_analytic_solutions} Analytic solutions to Eq.(\ref{eq:continuous_macro_solution}) for continuous velocity space and to Eq.(\ref{eq:discrete_macro_solution}) for discrete velocity space. The so-called ray effect appears in the cases with discrete velocity space.}
\end{figure}

In Fig.~\ref{fig:sod_analytic_solutions}, we first plot the analytical collisionless solutions at time $t=0.15$ for the case with initial data  $(\rho_l, U_l, T_l) = (1,0,2)$ and $(\rho_r, U_r, T_r)=(0.125,0,1.6)$ from the modelings of a continuous velocity space and a discretized velocity space with a uniform interval $\Delta u = 0.25$.
Clearly, the stepped flow structure, which is similar to the ray effect,  has been observed in the exact solution  with a discretized velocity space.

\subsection{Collisionless limit}
In the collisionless limit, the governing equation of gas distribution function $f$ for the free transport process is
\begin{equation}\label{eq:advection}
\dfrac{\partial f}{\partial t} + \vec{u} \cdot \dfrac{\partial f}{\partial \vec{x}} = \vec{0},
\end{equation}
where $\vec{u}$ is the microscopic velocity of gas particles.
The macroscopic variables $\vec{w}$ can be obtained by taking moments of the gas distribution function,
\begin{equation}\label{eq:micro_to_macro}
\vec{w} = \int{f \vec{\psi} d\vec{u}},
\end{equation}
where $\vec{\psi} = (1, \vec{u}, |\vec{u}|^2 / 2)^T$. The vector of conservative variables $\vec{w}$ denotes the densities of mass $\rho$, momentum $\rho \vec{U}$ and energy $\rho E$.
Corresponding to the conservative flow variables, the local equilibrium state is a Maxwellian distribution
\begin{equation}\label{eq:equilibrium}
g(\vec{u}) = \rho \left(\dfrac{\lambda}{\pi}\right)^{d/2} {\rm Exp}\left[-\lambda (\vec{u} - \vec{U})^2\right],
\end{equation}
where $d$ is the degree of freedoms. $\lambda$ is related to the temperature $T$ by $\lambda = {m_0}/{2 k_B T}$, where $m_0$ is the molecular mass and $k_B$ is the Boltzmann constant.

\begin{figure}[H]
\centering
\subfigure[\label{fig:initial_phase}]
{\includegraphics[width=0.48\textwidth]{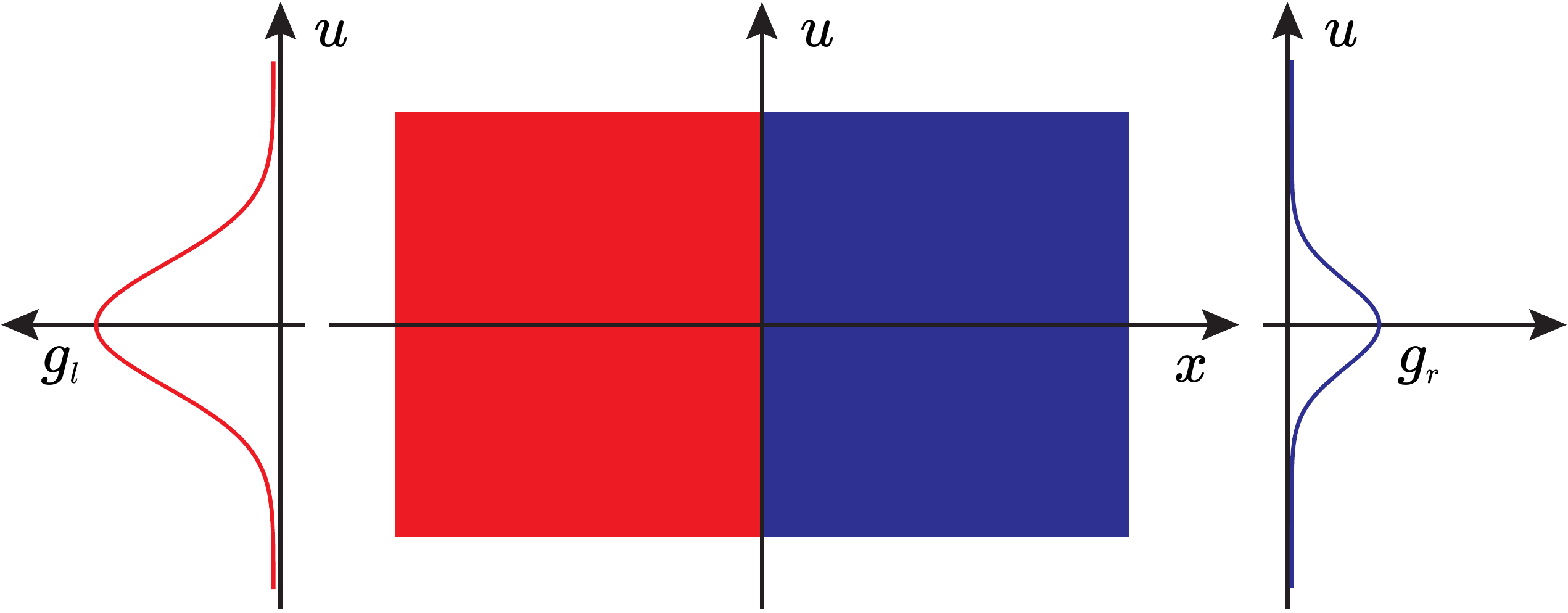}}
\subfigure[\label{fig:continuous_phase}]
{\includegraphics[width=0.48\textwidth]{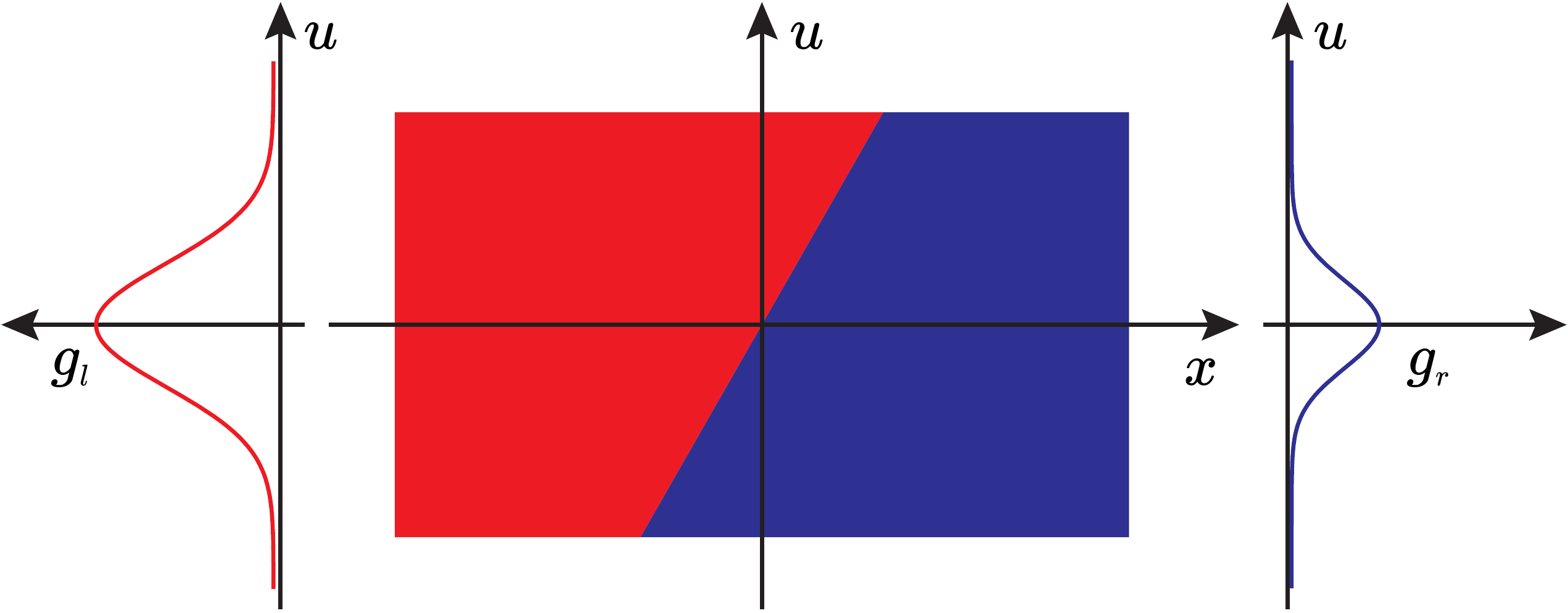}}
\caption{\label{fig:continuous_solution} Distribution in the phase space. (a) Initial state at time $t=0$; (b) distribution at time $t$.}
\end{figure}

\begin{figure}[H]
	\centering
	\subfigure[\label{fig:vdf_a}]{\includegraphics[width=0.32\textwidth]{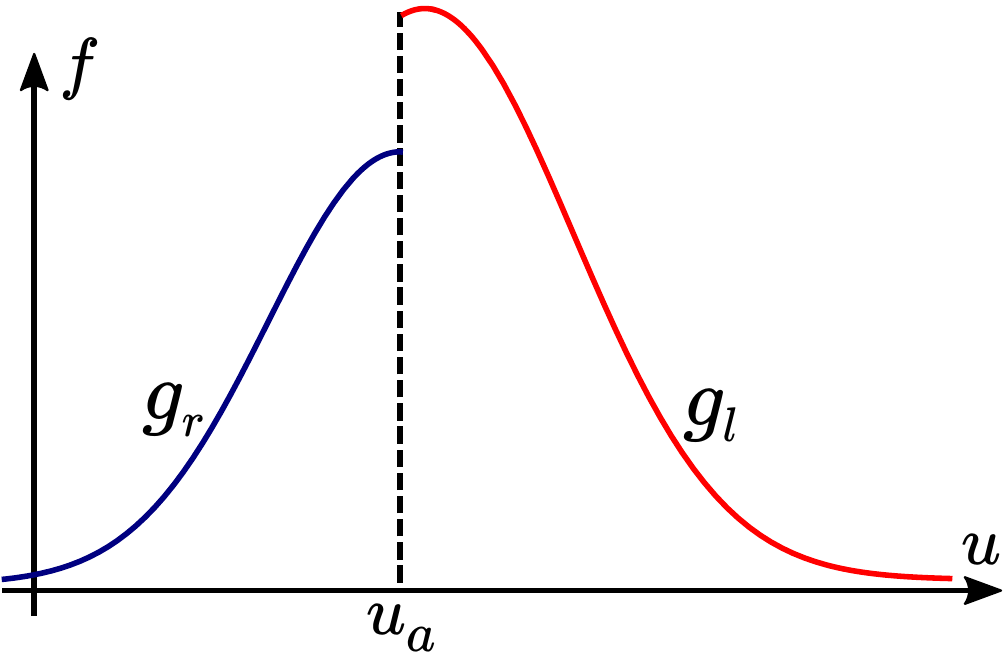}}
	\subfigure[\label{fig:vdf_b}]{\includegraphics[width=0.32\textwidth]{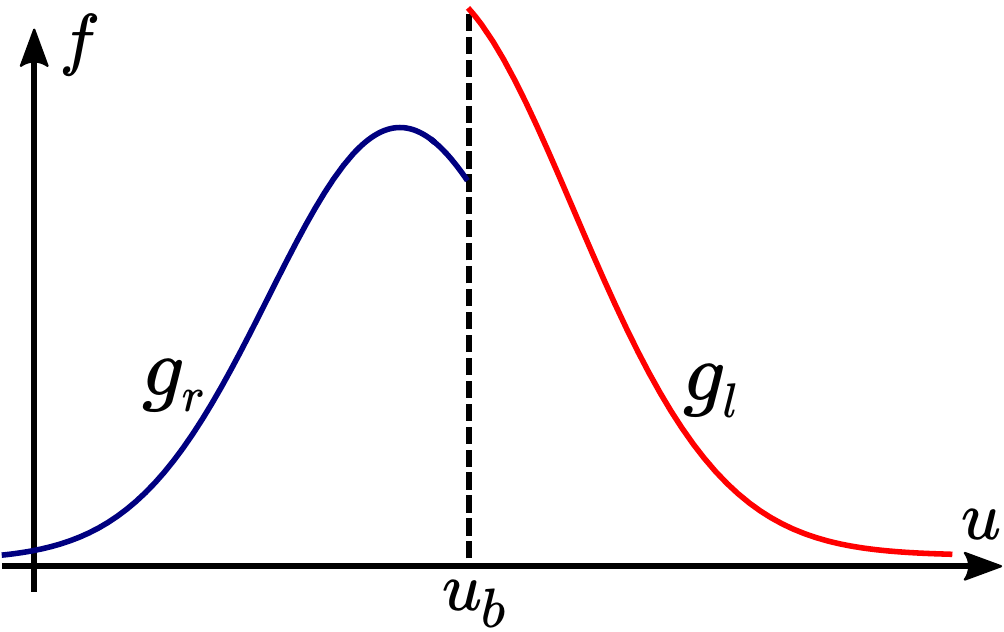}}
	\subfigure[\label{fig:vdf_ab}]{\includegraphics[width=0.32\textwidth]{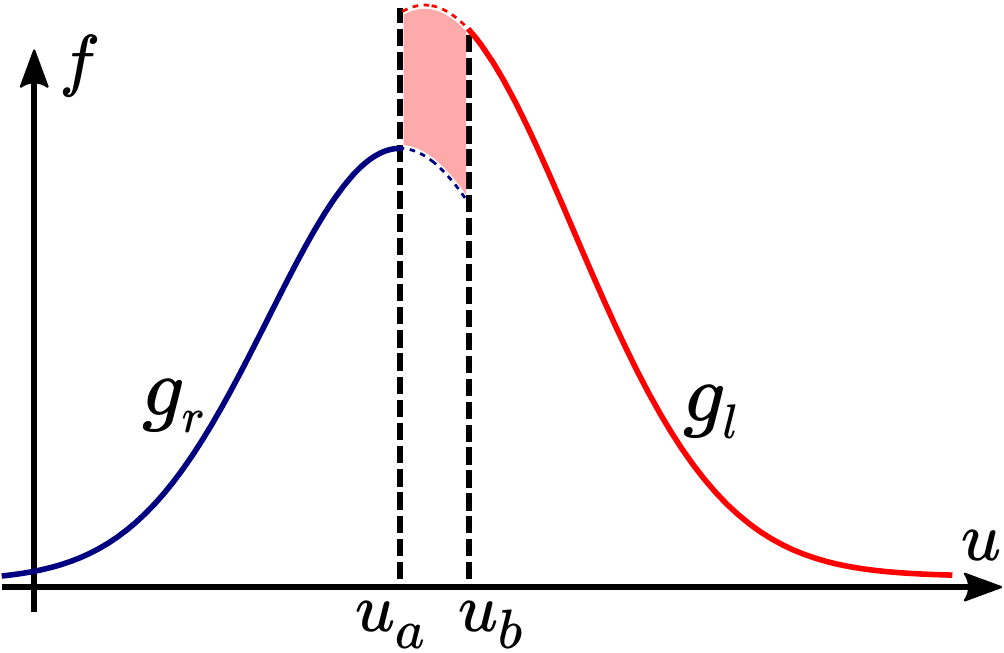}}
	\caption{\label{fig:continuous_velocity_space}Distribution function in the velocity space at time $t$. (a) Distribution function at location $x_a$; (b) distribution function at location $x_b$; (c) distribution function difference between $x_a$ and $x_b$.}
\end{figure}

For the problem with an initial discontinuity described in Eq.~(\ref{eq:problem_setup}), the initial distribution function is given by the Maxwellian distribution $g_l$ and $g_r$ for the left and right states.
The initial state in the phase space is illustrated in the Fig.~\ref{fig:initial_phase}, where there is a spatial discontinuity at $x=0$ for the distribution function $f$.
In the collisionless limit, the gas particles keep free streaming with their initial microscopic velocity in the physical space, and the distribution function will evolve into a highly non-equilibrium state which deviates far from the local Maxwellian equilibrium state.
As shown in Fig.~\ref{fig:continuous_phase}, the initial discontinuity line $x=0$ evolves into a discontinuity line $x = u t$ in the phase space at time $t$.
For each physical location point $x$, the distribution function in the velocity space is a combination of two half Maxwellian distributions, which is split at the velocity point $u = x / t$.
Therefore, the analytic solution of the distribution function at time $t$ is
\begin{equation}\label{eq:continuous_micro_solution}
f(x, t, u) =
\begin{cases}
g_l (u), \quad &u > x / t, \\
g_r (u), \quad &u < x / t.
\end{cases}
\end{equation}
At time $t$, the discontinuity at point $u$ in the velocity space has one-to-one correspondence with the physical location $x$, and when the discontinuity point moves in the velocity space, it gives the distribution function in different physical locations.
Figure \ref{fig:vdf_a} and \ref{fig:vdf_b} show an example of the distribution function in the velocity space for the location points $x_a$ and $x_b$.
For physical location $x$, the macroscopic variables $\vec{w}$ are
\begin{equation}\label{eq:continuous_macro_solution}
\vec{w}(x,t) = \int_{-\infty}^{u=x/t} {g_r(u^\prime) \psi du^\prime}
	+\int_{u=x/t}^{\infty} {g_l(u^\prime) \psi du^\prime},
\end{equation}
and the variation of the macroscopic flow variables in the physical space between two arbitrary points $x_a$ and $x_b$ would be
\begin{equation}\label{eq:continuous_variation}
\vec{w}(x_b) - \vec{w}(x_a) = -\int_{u_a}^{u_b} {\left[g_l(u^\prime)-g_r(u^\prime)\right] \psi du^\prime},
\end{equation}
which corresponds to the integration of the distribution function difference in the interval $(u_a, u_b)$, see the illustration in Fig.~\ref{fig:vdf_ab}.
As the discontinuity point $u$ gradually moves from $u_a$ to $u_b$ in the velocity space, it gives a smooth transition of macroscopic flow variables from  $x_a$ to $x_b$ in the physical space.

\begin{figure}[H]
	\centering	
	\subfigure[\label{fig:discrete_phase_space}]
	{\includegraphics[width=0.32\textwidth]{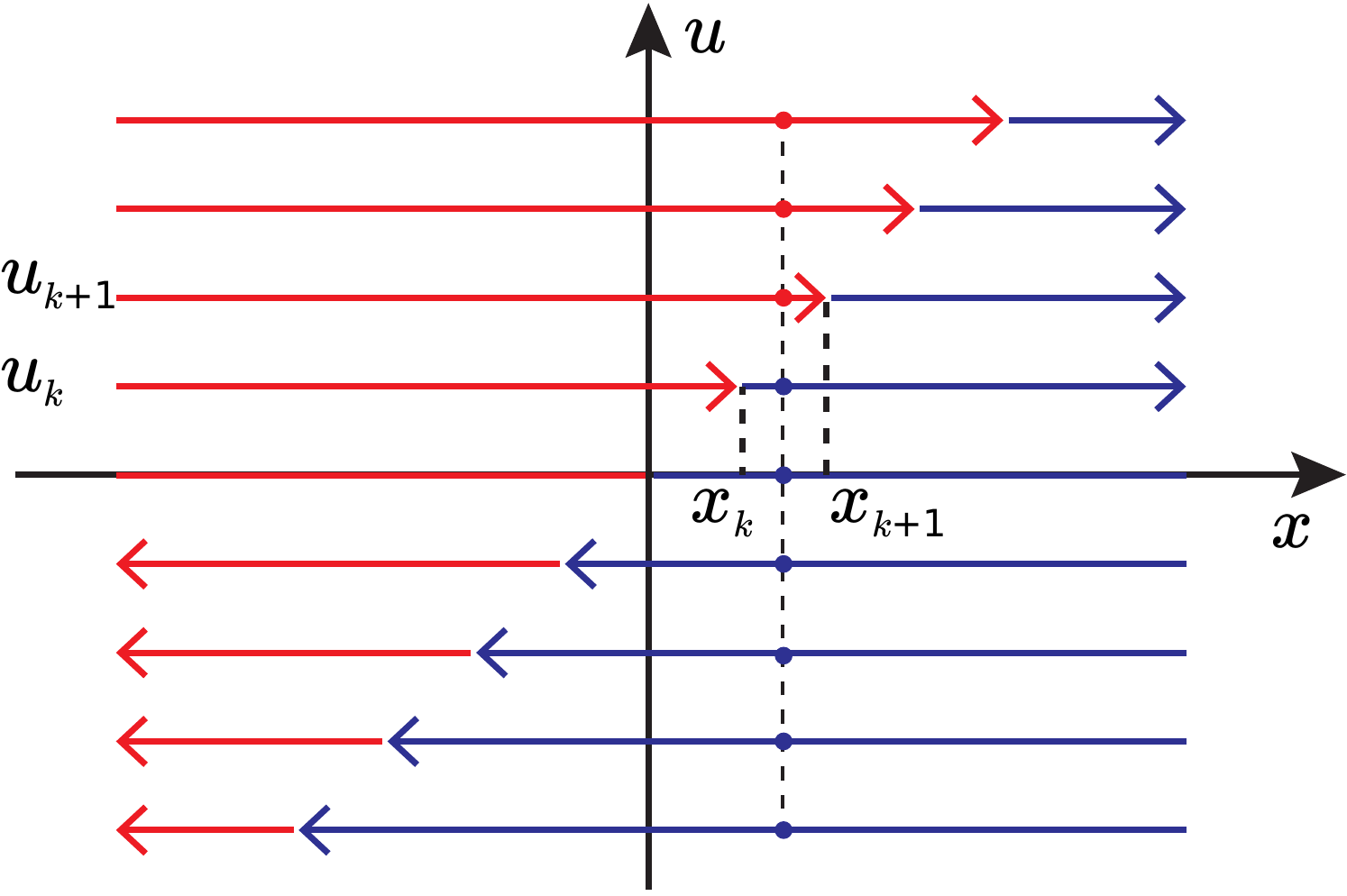}}
	\subfigure[\label{fig:discrete_velocity_space1}]
	{\includegraphics[width=0.32\textwidth]{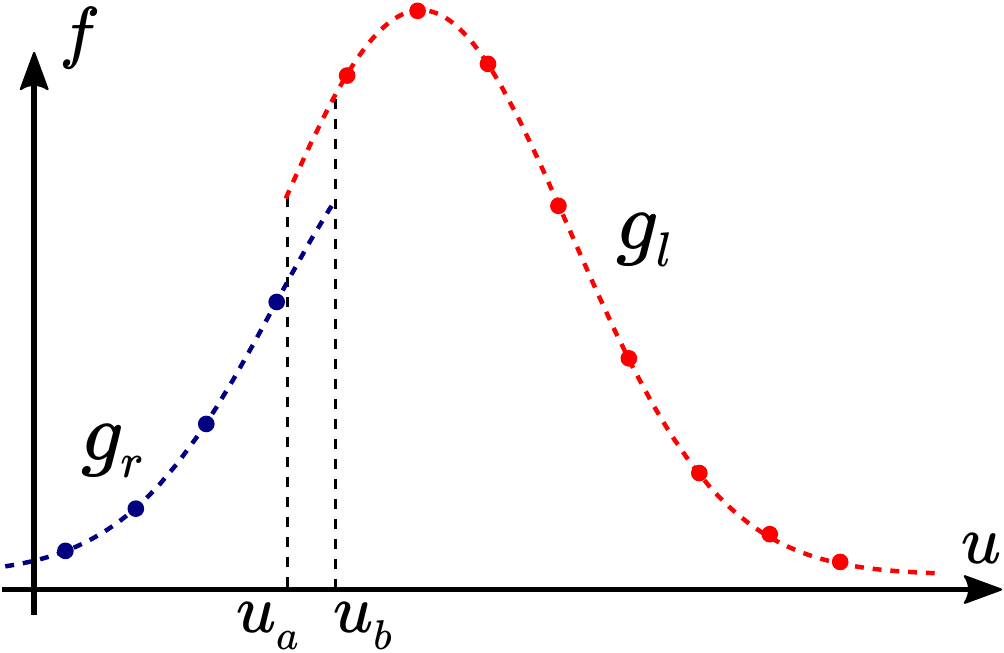}}
	\subfigure[\label{fig:discrete_velocity_space2}]
	{\includegraphics[width=0.32\textwidth]{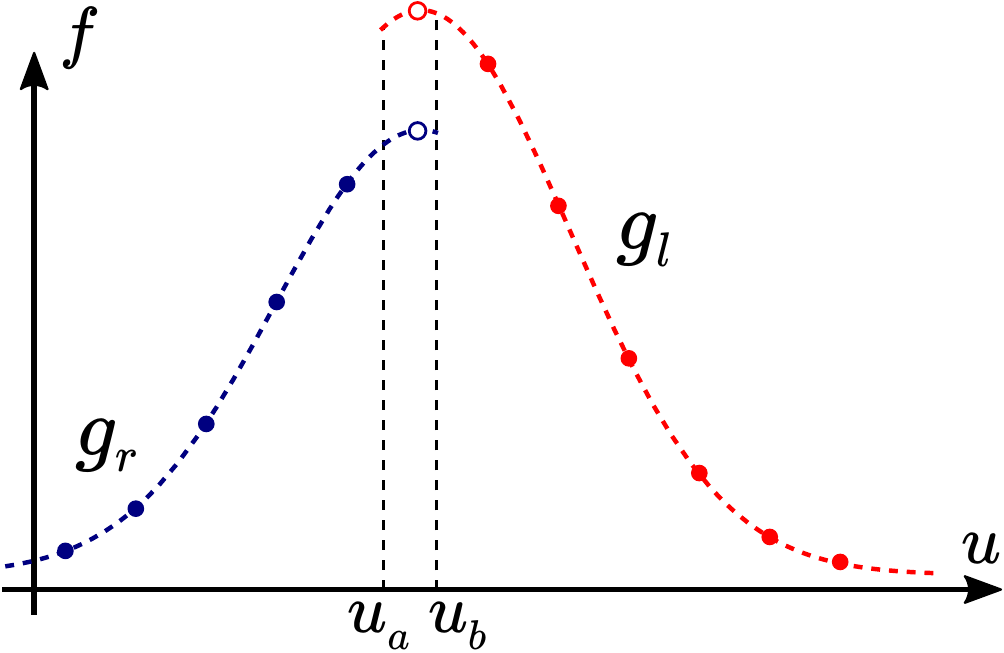}}
	\caption{\label{fig:discrete_solution} Discrete distribution at time $t$ in the phase space and velocity space. (a) Distribution in discrete phase space; (b) discrete distribution function at $x_a$ and $x_b$ with discontinuity points $u_a$ and $u_b$ locating within one discrete velocity interval; (c) discrete distribution function at $x_a$ and $x_b$ with discontinuity points $u_a$ and $u_b$ locating in two discrete velocity intervals. }
\end{figure}

In rarefied gas simulation, in order to capture the highly non-equilibrium state, a finite number of discrete velocities are usually employed to describe the distribution function in the velocity space.
From the discrete phase space illustrated in Fig.~\ref{fig:discrete_phase_space},
the analytic solution for the discrete distribution function at physical location $x$ can be obtained
\begin{equation}\label{eq:discrete_micro_solution}
f(x, t, u_k) = \left\{
\begin{aligned}
g_l (u_k), &\quad u_k > x / t, \\
g_r (u_k), &\quad u_k \le x / t,
\end{aligned}
\right.
\end{equation}
and the analytic solution of the conservative variables gives
\begin{equation}\label{eq:discrete_macro_solution}
\vec{w}(x, t) = \sum_{u_k \le x / t} {\vec{\psi}_k g_r (u_k) \omega_k} + \sum_{u_k > x / t}{\vec{\psi}_k g_l (u_k) \omega_k}
\end{equation}
where $\omega_k$ is the weight at the velocity point $u_k$ for numerical integration.
It can be found that even though the discontinuity line $u = x / t$ are accurately captured at the discrete velocity points, such as at $u_k$ and $u_{k+1}$, since there is no discrete velocity point to distinguish the variation between $u_k$ and $u_{k+1}$, the solution between the physical location points $x_k$ and $x_{k+1}$ can not be resolved.

In analogy to that in the continuous velocity space, the variation of the macroscopic flow variables between $x_a$ and $x_b$ would be the numerical integration of the distribution function difference over the velocity interval $(u_a, u_b)$, i.e,
\begin{equation}\label{eq:discrete_variation}
\vec{w}(x_b) - \vec{w}(x_a) = -\sum_{u_a \le u_k < u_b} {\vec{\psi}_k \left[g_l(u_k) - g_r(u_k) \right]\omega_k}.
\end{equation}
If $x_a$ locates near $x_b$ and the corresponding discontinuity points $u_a$ and $u_b$ are in one discrete velocity interval as shown in Fig.~\ref{fig:discrete_velocity_space1}, the discrete distribution functions given by Eq.~(\ref{eq:discrete_micro_solution}) and the macroscopic variables by Eq.~(\ref{eq:discrete_macro_solution}) have no difference at $x_a$ and $x_b$, and there is no discrete velocity point satisfying $u_a < u_k < u_b$ to contribute the spatial variations.
If $x_a$ is a little far away from $x_b$ and the corresponding discontinuity points $u_a$ and $u_b$ locate in different velocity intervals as illustrated in Fig.~\ref{fig:discrete_velocity_space2}, the discrete distribution functions for $x_a$ and $x_b$ would be different at some specific discrete velocities (e.g., the hollow circles illustrated in Fig.~\ref{fig:discrete_velocity_space2}), and the spatial variations between $x_a$ and $x_b$ will be the moments of distribution function difference at these discrete velocities.
As to the spatial distribution, it can be seen in Fig.~\ref{fig:discrete_phase_space} that the discrete velocity distribution function and macroscopic flow variables take sudden changes at the discrete-velocity-determined location points $x_k$, and keep unchanged within the spatial interval $(x_k, x_{k+1})$.
As a result, the numerical solution will have stepped structures in the physical space, which is the so called ray effect as shown in Fig.~\ref{fig:sod_analytic_solutions} where the analytic solutions to Eq.(\ref{eq:continuous_macro_solution}) for continuous velocity space and to Eq.(\ref{eq:discrete_macro_solution}) for a discrete velocity space with uniform interval $\Delta u = 0.25$ are plotted.

As mentioned, the discontinuity at $u_k$ in the discrete velocity space will cause a sudden change at $x_k = u_k t$ in the physical space, so the macroscopic variable jump at $x_k$ will be the moment of the distribution function change at $u_k$
\begin{equation}\label{eq:jump}
\Delta \vec{w}(x_k) = \vec{\psi}_k \left|g_r(u_k) - g_l(u_k)\right| \omega_k,
\end{equation}
and the length of the stepped flow structure is
\begin{equation}\label{eq:step}
	\Delta x_k = x_{k+1} - x_{k} = u_{i+k} t - u_k t = t \Delta u_k.
\end{equation}
In the following of this paper, we will refer to the macroscopic variable jump and the length of the stepped flow structure as the jump and step of ray effect, respectively.

In addition, it should be pointed out that the previous analysis is based on continuous physical space and discrete velocity space, and the distribution function at the discrete velocity point is given as the exact value.
Therefore, the result with the ray effect is a kind of analytic solution for the cases with discrete velocity space in the collisionless limit.
From this point of view, if any discrete velocity method doesn't present numerical solutions with ray effect, the method would be doubtful about its accuracy in capturing the correct discrete distribution function and its spatial resolution in resolving the stepped flow structure.
Once the resolution of velocity space is given, it sets a highest resolution limit for the discrete velocity method, the spatial variations in physical space induced by the distribution function due to the velocity discretization cannot be avoided, even with a much finer spatial mesh.
For the collisionless cases, with time evolution, the initial information spreads along the given discrete velocity directions without decaying, so the unresolved information with uncertainty would be extended in physical space and thus may deviate its results from the analytic ones with a continuous particle velocity space.
The similar phenomenon happens for moment method when solving rarefied flows.
Although each basis is continuous in the velocity space, a finite number of bases has limited capability (similar to the so-called resolution) to describe the non-equilibrium distribution function.
For collisionless limit, the moment equation will give the same number of characteristics as the employed bases, which restricts the number of directions and speeds for information spreading.
Thus, flow field evolution along a finite number of characteristics will result in distortion of the numerical solution in physical space.
For further comprehension, we mention the Riemann problem for one dimensional Euler equation where three characteristics give the solution with multiple stages.
However, the difference is that the three characteristics of Riemann problem are determined by the Euler equations, while the finite number of characteristics in DVM and moment method are due to the limited number of discrete velocity points or bases.

In the current study, we will not give further discussion on the moment method and mainly focus on the discrete velocity scheme.
Besides the velocity discretization, different numerical treatments, such as spatial discretization and evolution process, would influence the spatial distribution of the ray effect in practical numerical computations.
Thus, besides showing the appearance of ray effect in the rarefied flow, one main task in the current paper is to investigate the influence of different numerical treatments on the ray effect.

\subsection{Collision influence}\label{sec:collision}
In previous section, the cause of the ray effect has been analyzed based on the one dimensional case in the collisionless limit.
In this section, the influence of particle collision will be considered.

The Boltzmann equation for rarefied gas flow can be written as
\begin{equation}\label{eq:Boltzmann}
	\dfrac{\partial f}{\partial t} + \vec{u} \cdot \dfrac{\partial f}{\partial \vec{x}} = \Omega(f,f),
\end{equation}
where $\Omega(f,f)$ is the full Boltzmann collision term.
Since the evaluation of the full Boltzmann collision term is complicated,  we will adopt the BGK-type kinetic model in the current paper to approximate the collision term, where the mechanism of the ray effect is the same.
The BGK-type kinetic model equation can be written as
\begin{equation}
	\dfrac{\partial f}{\partial t} + \vec{u} \cdot \dfrac{\partial f}{\partial \vec{x}}  = \dfrac{g-f}{\tau},
\end{equation}
where $g$ is the local equilibrium state and $\tau$ is the mean collision time or the relaxation time.
For a local constant relaxation time $\tau$, the integral solution along the characteristic line can be obtained by
\begin{equation}
	f(\vec{x}_0, t) = \dfrac{1}{\tau}\int_{0}^{t} {g(\vec{x}^\prime, t^\prime) e^{-(t-t^\prime) / \tau} dt^\prime} + e^{-t/\tau} f_0(\vec{x}_0 - \vec{u} t),
\end{equation}
where $f_0(\vec{x})$ is the initial distribution function around $\vec{x}_0$ at time $t=0$, and $g(\vec{x}, t)$ is the equilibrium state in space and time.
The initial distribution function $f_0$ decays exponentially with the increment of time $t$.
Thus, the initial discontinuity of the distribution function in the velocity space will decay exponentially as well due to the particles' interaction.
Combined with Eq.~(\ref{eq:discrete_variation}), the jump of the ray effect will decay with the decreasing of the discontinuity in the distribution function.
Quantitatively, the jump of the ray effect would be reduced by $86.5\%$ and $90\%$ after $2$ and $3$ mean collision times.

\section{Numerical validations}\label{sec:validation}
In this section, numerical test cases including Sod problem, Rayleigh flow, and lid-driven cavity flow  are computed to show the ray effect in the rarefied gas flow.
The theoretical analysis in the previous section will be validated by the numerical solutions.
Without special statement, the reference values to normalize the flow variables in the current paper are chosen as $T_0 = 273{\rm K}$ and $U_0 = \sqrt{2 k_B T_0 / m_0}$.
The unified gas-kinetic scheme (UGKS) \cite{xu2010unified,xu2015direct} is employed to provide the numerical solutions of discrete velocity method.
For the collisionless cases, UGKS will give the same results as other discrete velocity method; while for collisional cases at smaller Knudsen numbers, UGKS considers the coupling of gas particles' transport and collision.
For better presentation, very fine meshes in the physical space will be used in the following cases, which ensures that the results can be recovered by all other DVMs.

\subsection{Sod test case}\label{sec:sod}

\begin{figure}[H]
	\centering
	\subfigure[Density]{\includegraphics[width=0.32\textwidth]{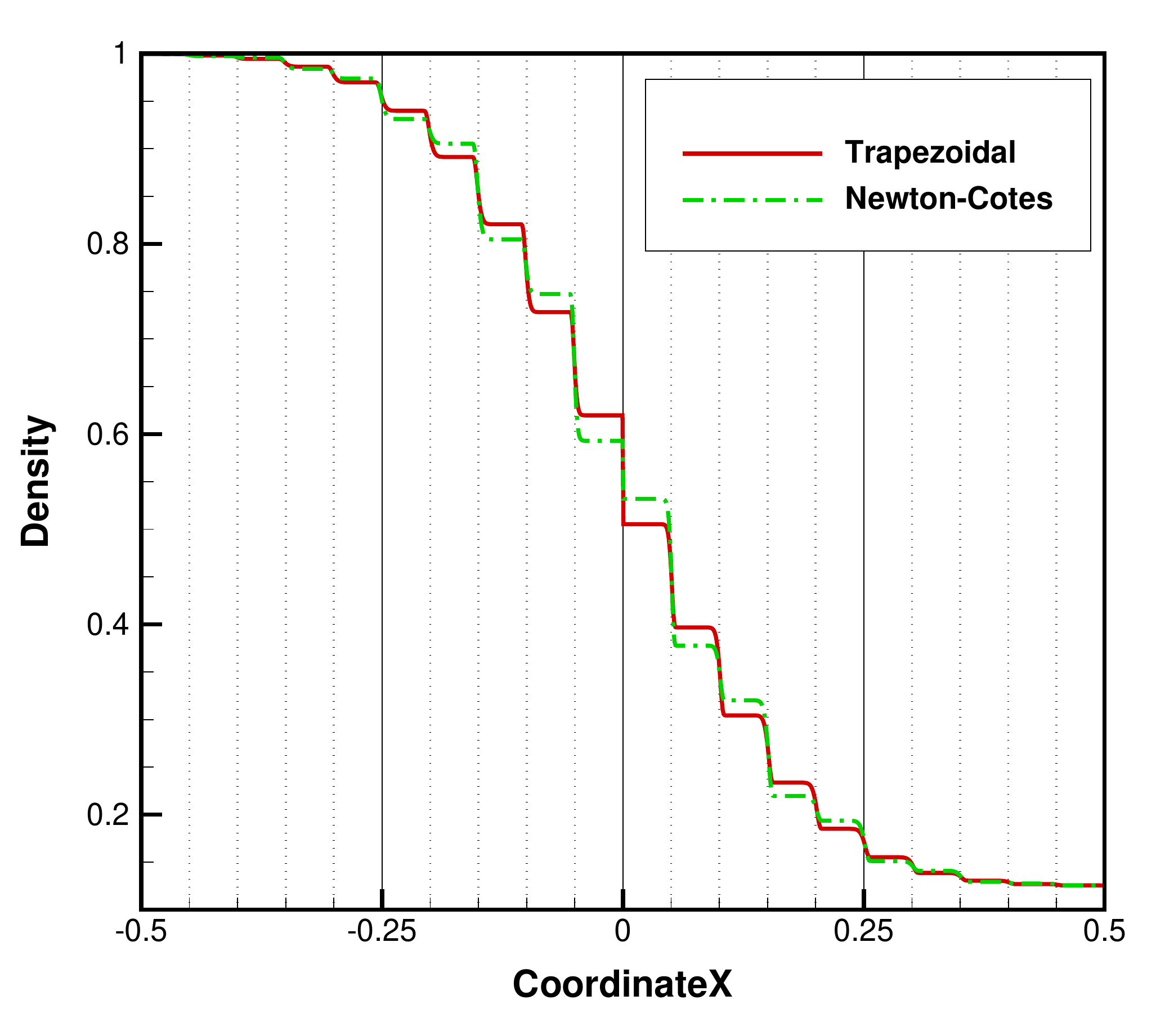}}
	\subfigure[Velocity]{\includegraphics[width=0.32\textwidth]{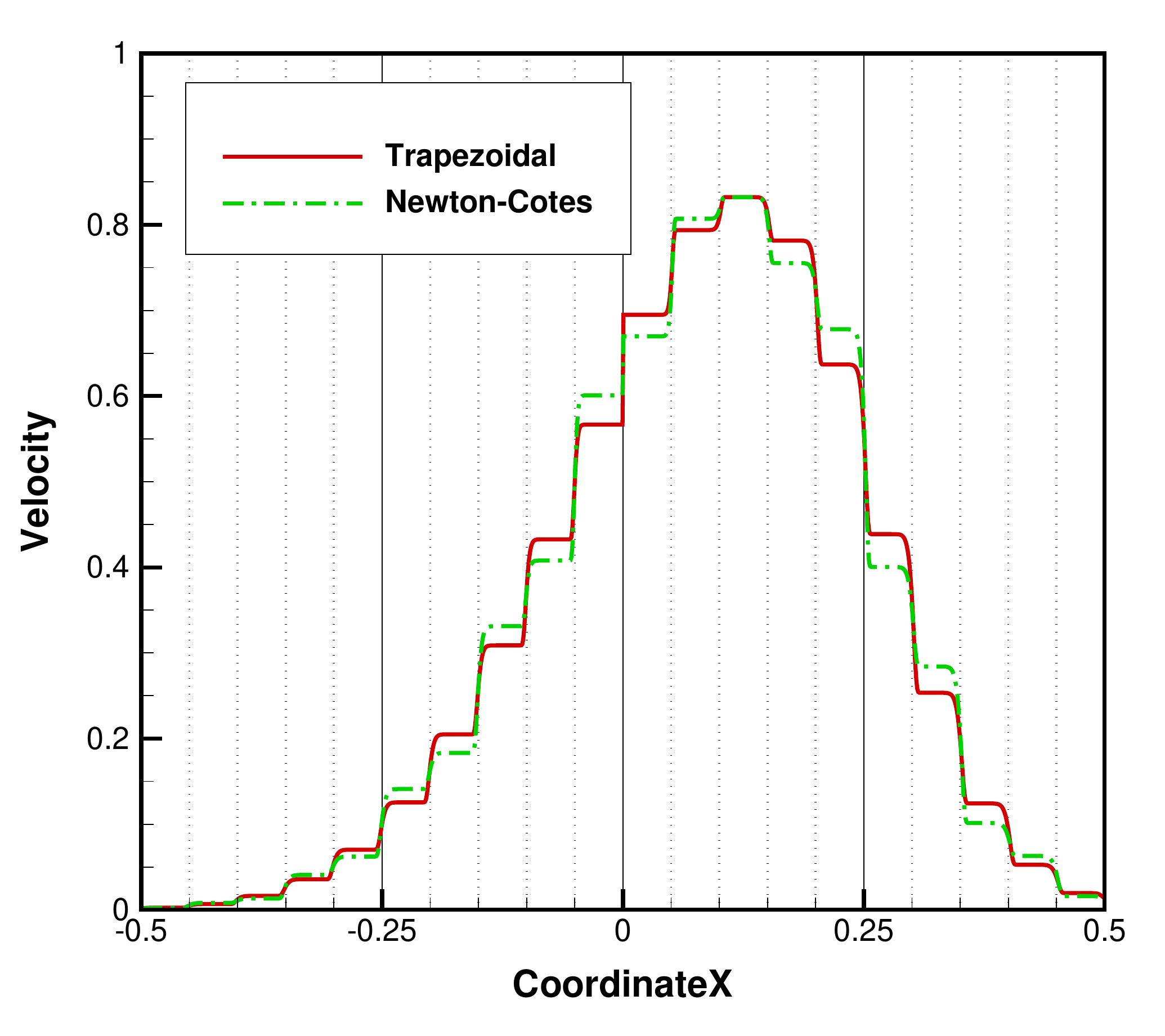}}
	\subfigure[Temperature]{\includegraphics[width=0.32\textwidth]{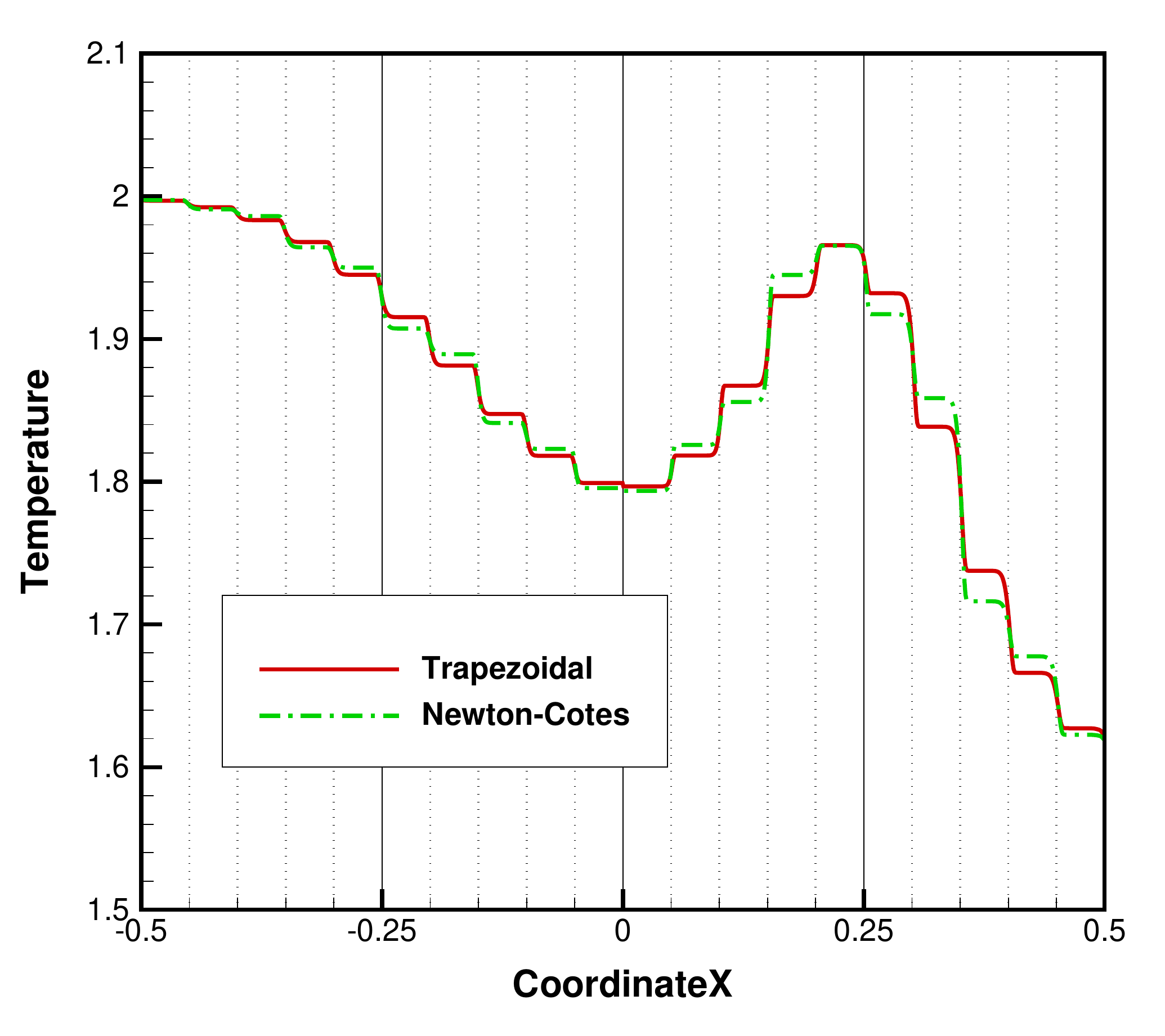}}
	\caption{\label{fig:sod_step}Numerical solutions for Sod test case at ${\rm Kn}\to \infty$ with different methods of numerical integration in the velocity space.}
\end{figure}

\begin{figure}[H]
	\centering
	\subfigure[]{\includegraphics[width=0.4\textwidth]{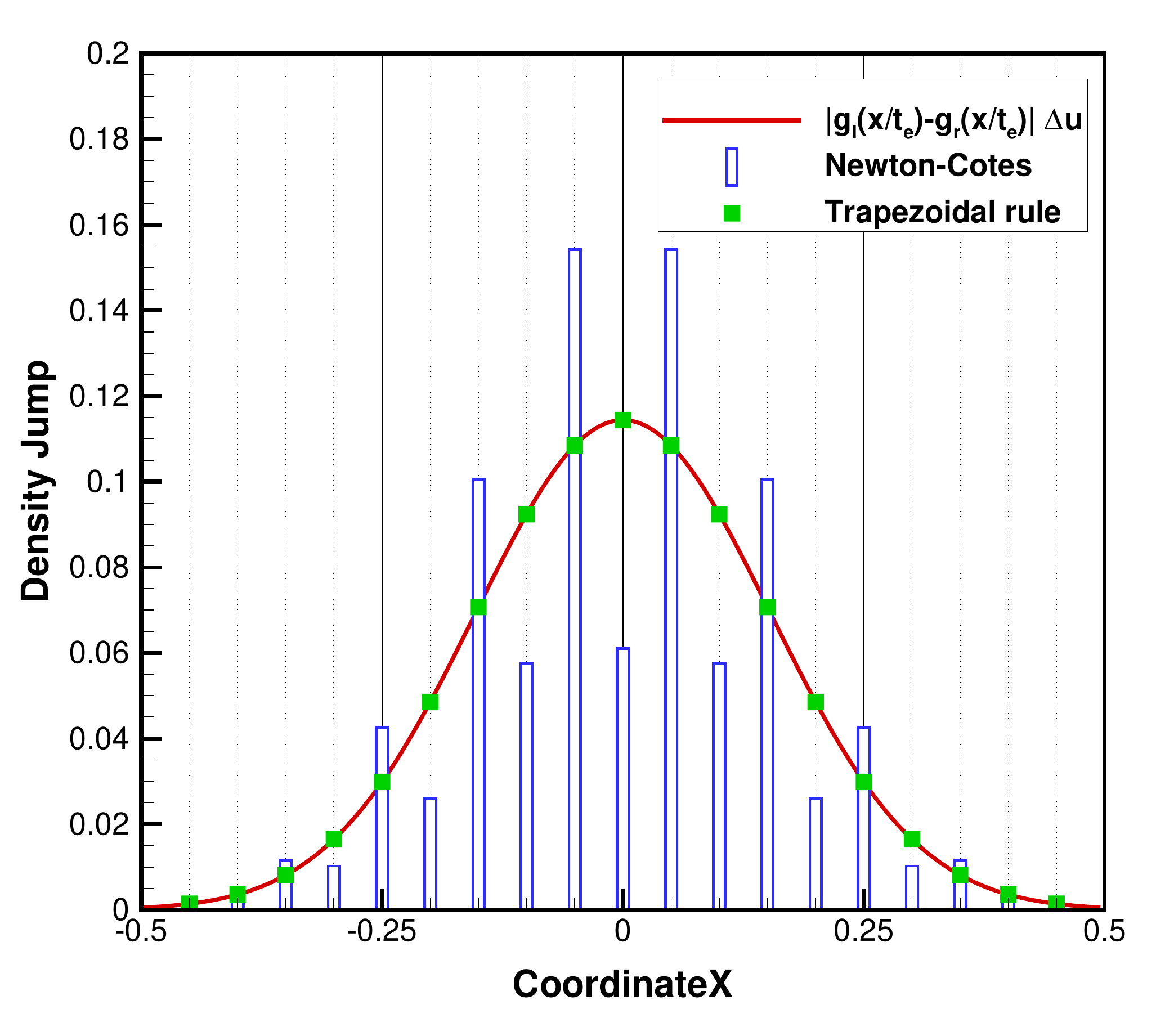}}\hspace{0.1\textwidth}
	\subfigure[]{\includegraphics[width=0.4\textwidth]{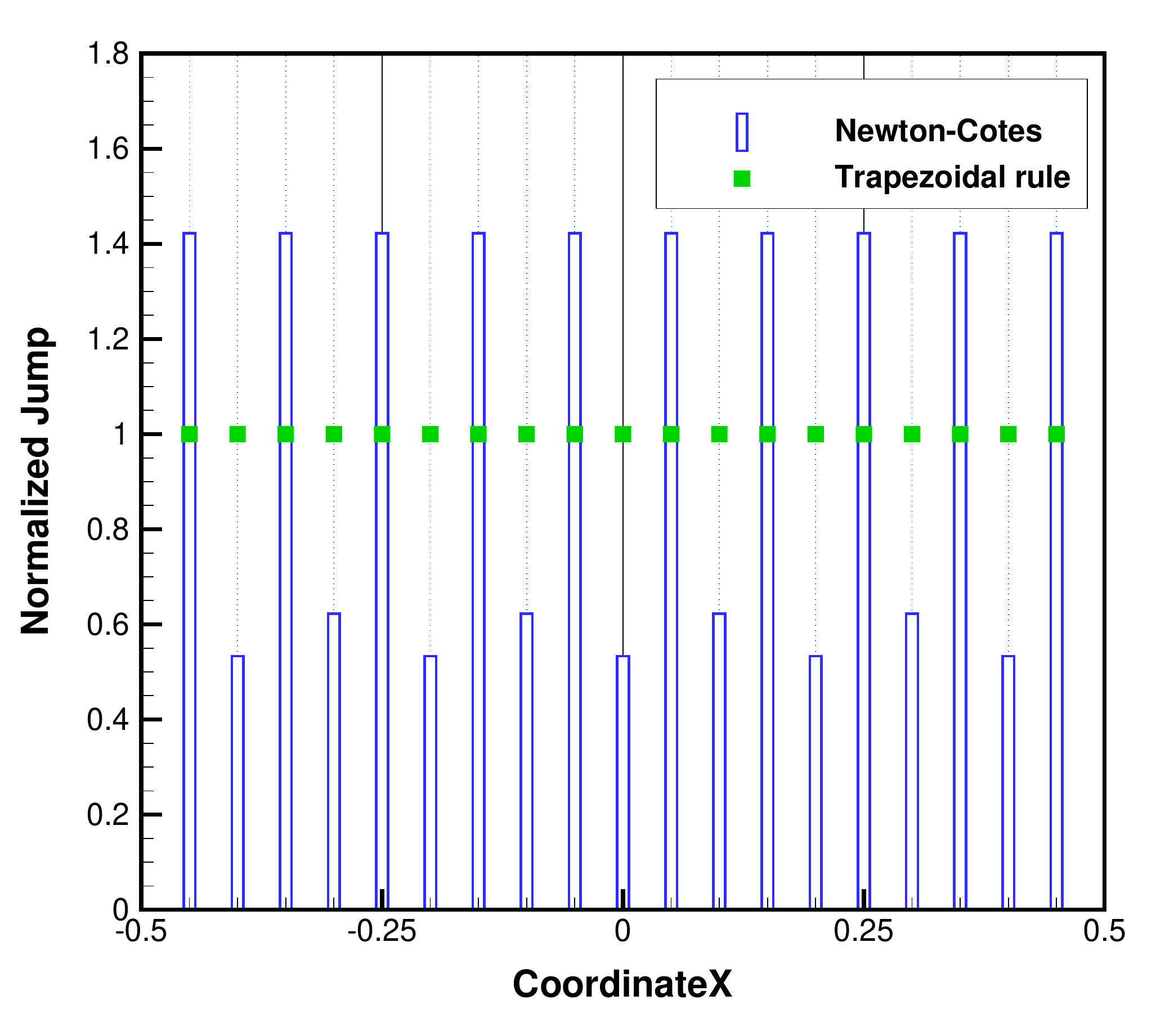}}
	\caption{\label{fig:sod_jump}The density jumps for Sod test cases. (a) Density jumps compared with the discontinuity of the distribution function; (b) density jumps normalized by $|g_l(x/t_e) - g_r(x/t_e)| \Delta u$.}
\end{figure}

The Sod shock tube problem is computed to validate the quantitative analysis of the ray effect in the previous section.
The initial condition is
\begin{equation}\label{eq:sod_setup}
(\rho, U, p) =
\begin{cases}
\left(1, 0, 1 \right), &\quad x < 0,\\
\left(0.125, 0, 0.1 \right), &\quad x > 0,
\end{cases}
\end{equation}
where $p$ is pressure.
In the collisionless limit at ${\rm Kn} \to \infty$, the velocity space is discretized by $61$ velocity points in the range of $\left[-10, 10\right]$.
In order to clearly show the ray effect, a uniform mesh with $1000$ cells is used in the computational physical space.
The numerical solutions at time $t_e = 0.15$ are presented in Fig.~\ref{fig:sod_step}.
It can be easily found that the step $\Delta x_s$ of the ray effect is about $0.05$, which is consistent with the analytic analysis of $\Delta x_s = t_e\Delta u$ in Eq.~(\ref{eq:step}).
Besides, the numerical solutions obtained by trapezoidal rule and Newton--Cotes method for velocity space integration are compared.
It can be seen that the results obtained by these two methods have the same step but different jumps of the ray effect,
because at the same discontinuity point $u_k$, the integration weights are different for these two methods.
As predicted in Eq.~(\ref{eq:jump}), the density jumps are supposed to be
\begin{equation}
	\Delta \rho(x_k) = |g_l(x_k / t_e) - g_r(x_k / t_e)| \omega_k.
\end{equation}
In order to give a quantitative demonstration, we pick up the density jumps and compare with the discontinuity of the distribution function between the left and right states.
As shown in Fig.~\ref{fig:sod_jump}, the density jumps of the trapezoidal solutions with $\omega_k = \Delta u$ are equal to the discontinuity in distribution function. 
The normalized jumps in the Newton--Cotes solutions from $|g_l(x_k / t_e) - g_r(x_k / t_e)|$ gives $\frac{28}{45}$, $\frac{64}{45}$, $\frac{24}{45}$, $\frac{64}{45}$, $\frac{28}{45}$ corresponding to the coefficients in the composite integration.
Hence, the analytic analysis in Eq.~(\ref{eq:jump}) and Eq.~(\ref{eq:step}) are fully validated in a quantitative way.

\subsection{Rayleigh flow}\label{sec:numerical_rayleigh}

The Rayleigh flow is computed at different Knudsen numbers to explore the ray effect in the rarefied flow simulations.
The Rayleigh flow is an unsteady gas flow around a vertical plate.
The initial argon gas around the plate is stationary with an initial temperature $273{\rm K}$, and the plate moves with a constant vertical velocity $30{\rm m/s}$ and a higher temperature $373{\rm K}$.
The computational domain is $1$ meter long, which is used to define the Knudsen number by the variable hard sphere model.
The dynamic viscosity is computed by the power law $\mu = \mu_0 (T/T_0)^\omega$ with $\omega = 0.81$.

\begin{figure}[H]
	\subfigure[Density\label{fig:rayleigh_density}]
	{\includegraphics[width=0.32\textwidth]{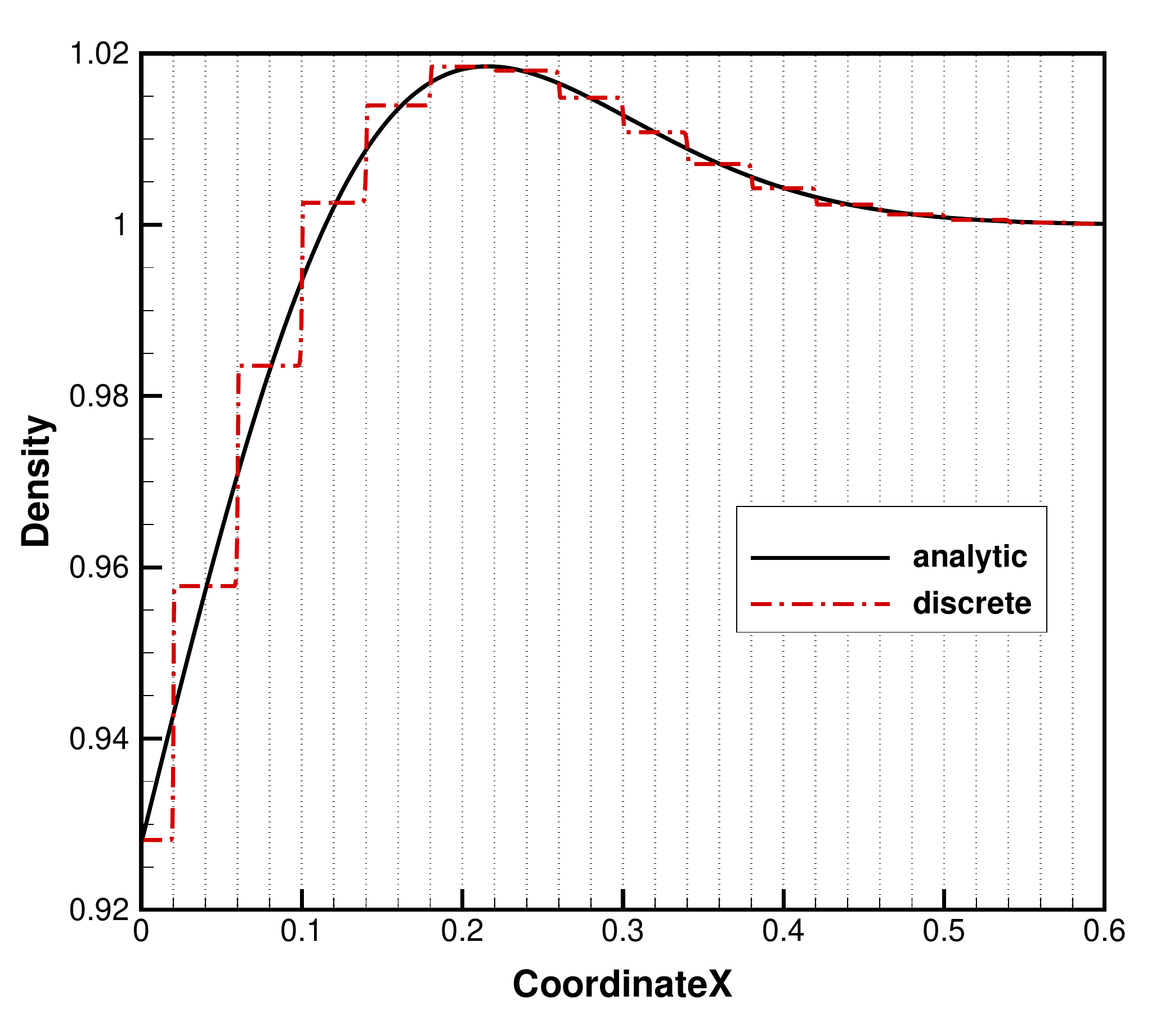}}
	\subfigure[Velocity\label{fig:rayleigh_velocity}]
	{\includegraphics[width=0.32\textwidth]{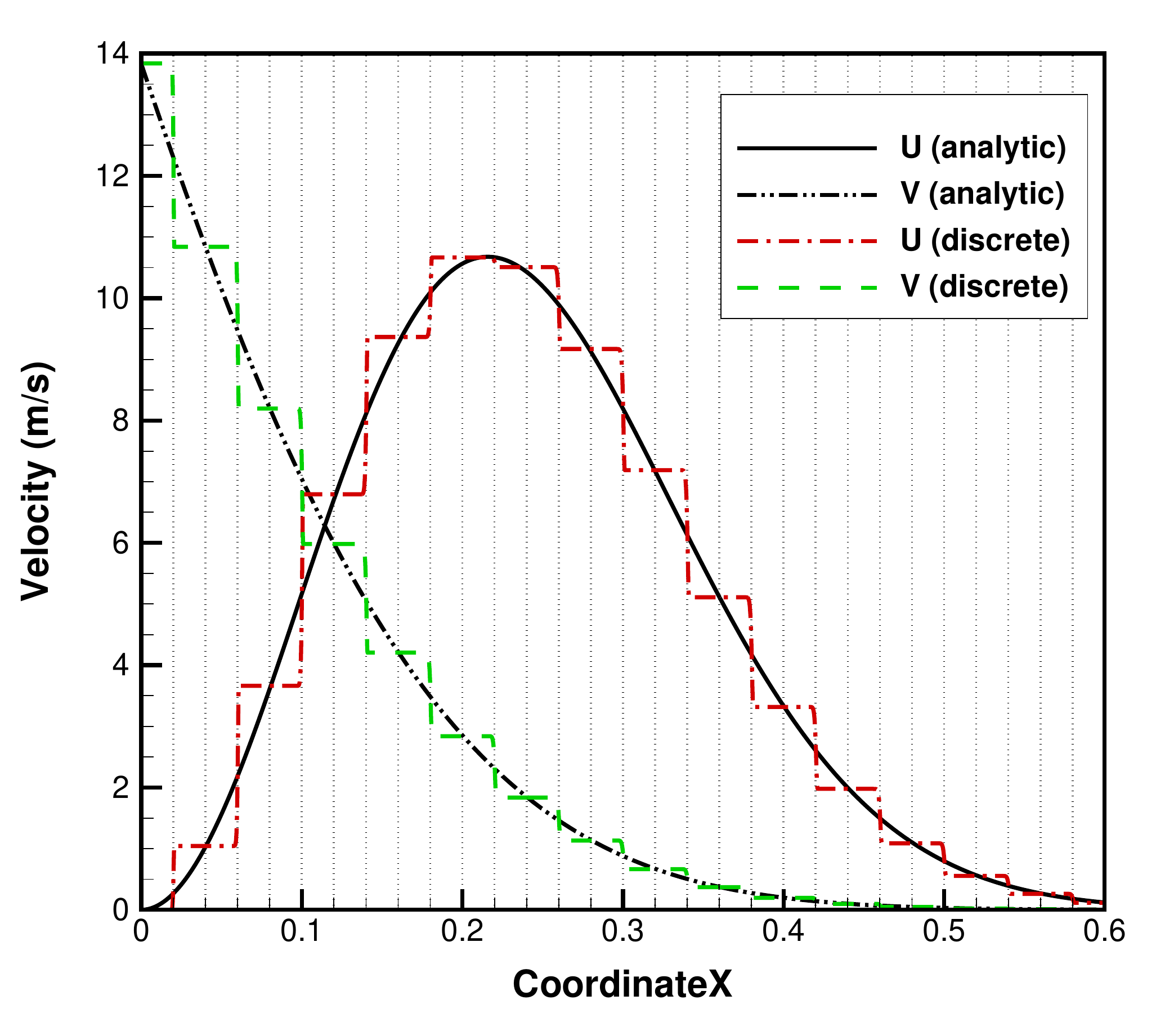}}
	\subfigure[Temperature\label{fig:rayleigh_temperature}]
	{\includegraphics[width=0.32\textwidth]{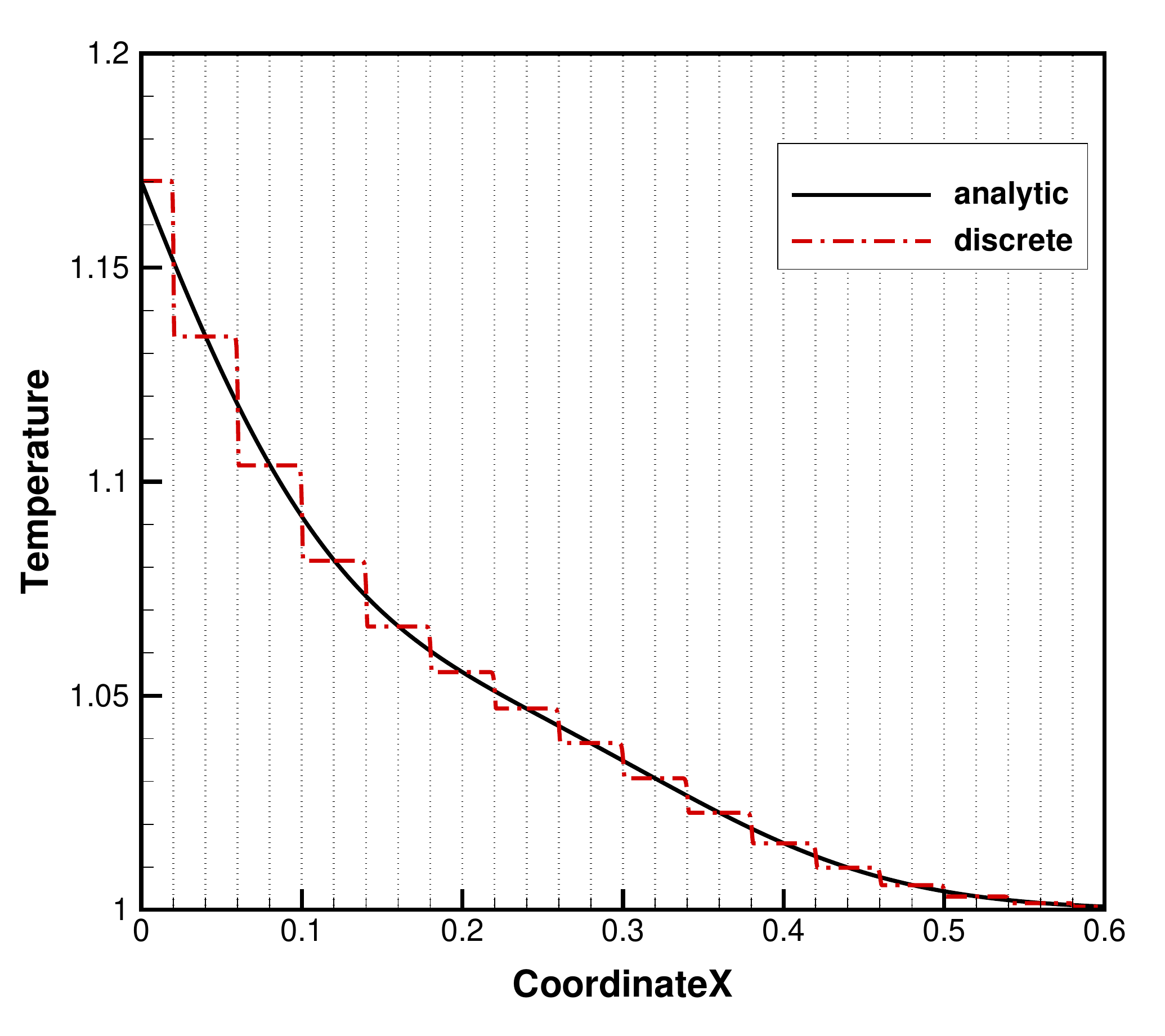}}
	\caption{\label{fig:rayleigh_collisionless}Numerical solutions of Rayleigh flow in the collisionless limit obtained by the discrete velocity method, and the results are compared with the analytic solutions for continuous velocity space.}
\end{figure}

\begin{figure}[H]
	\centering
	\subfigure[$g_0(u,v)$\label{fig:rayleigh_g0}]
	{\includegraphics[width=0.32\textwidth]{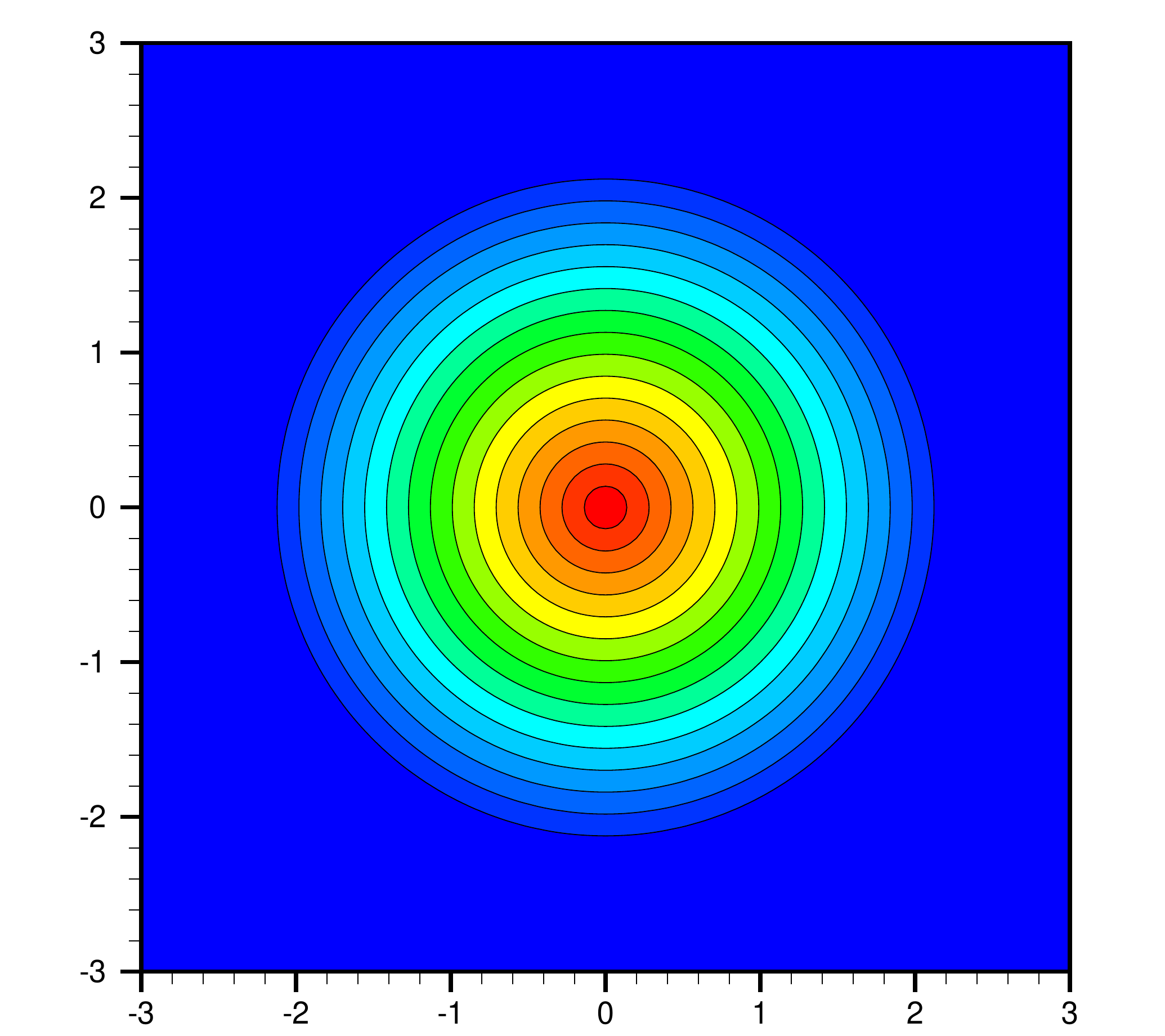}}
	\subfigure[$g_w(u,v)$\label{fig:rayleigh_gw}]
	{\includegraphics[width=0.32\textwidth]{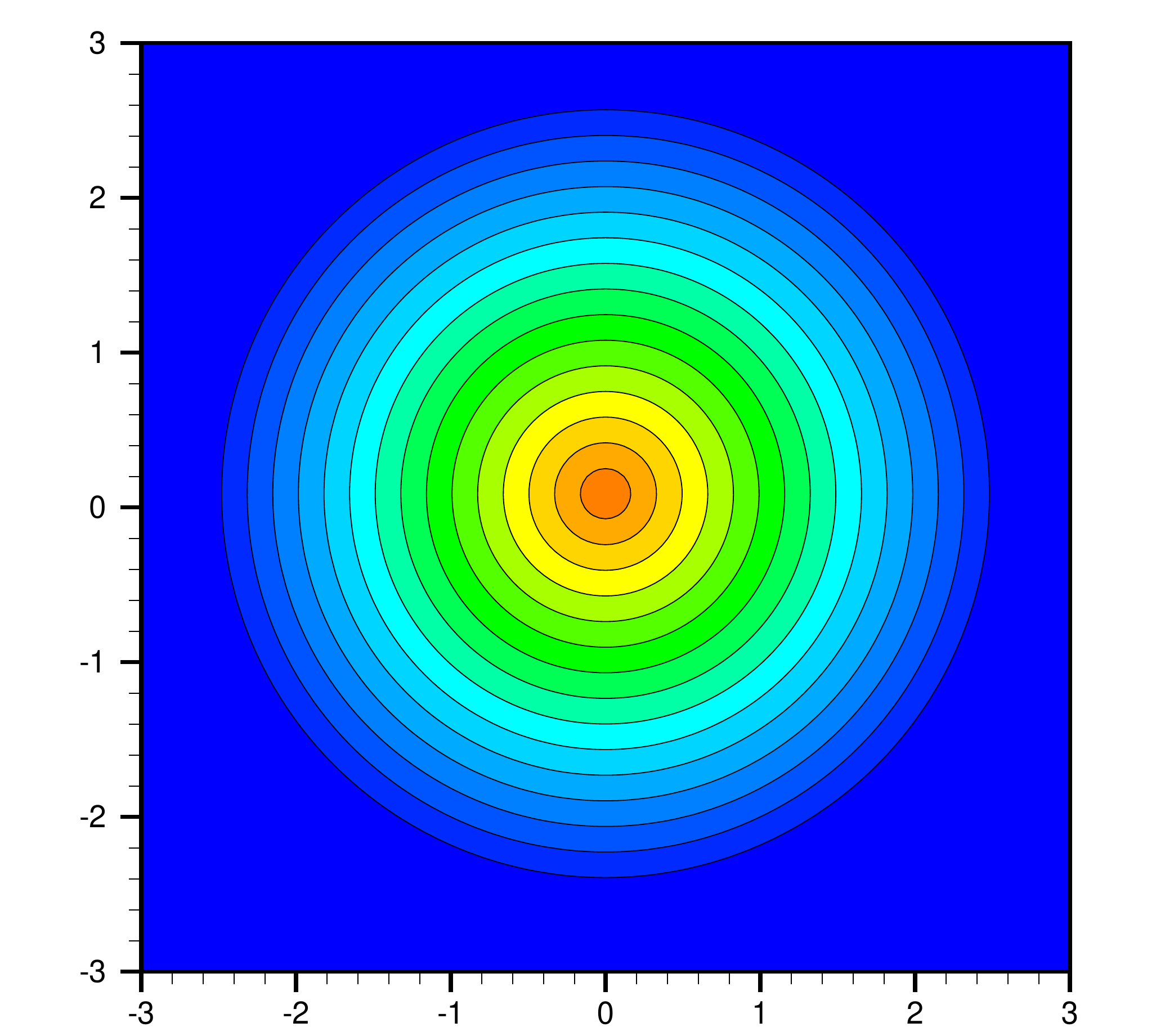}}
	\subfigure[$f(u,v)$\label{fig:rayleigh_vs}]
	{\includegraphics[width=0.32\textwidth]{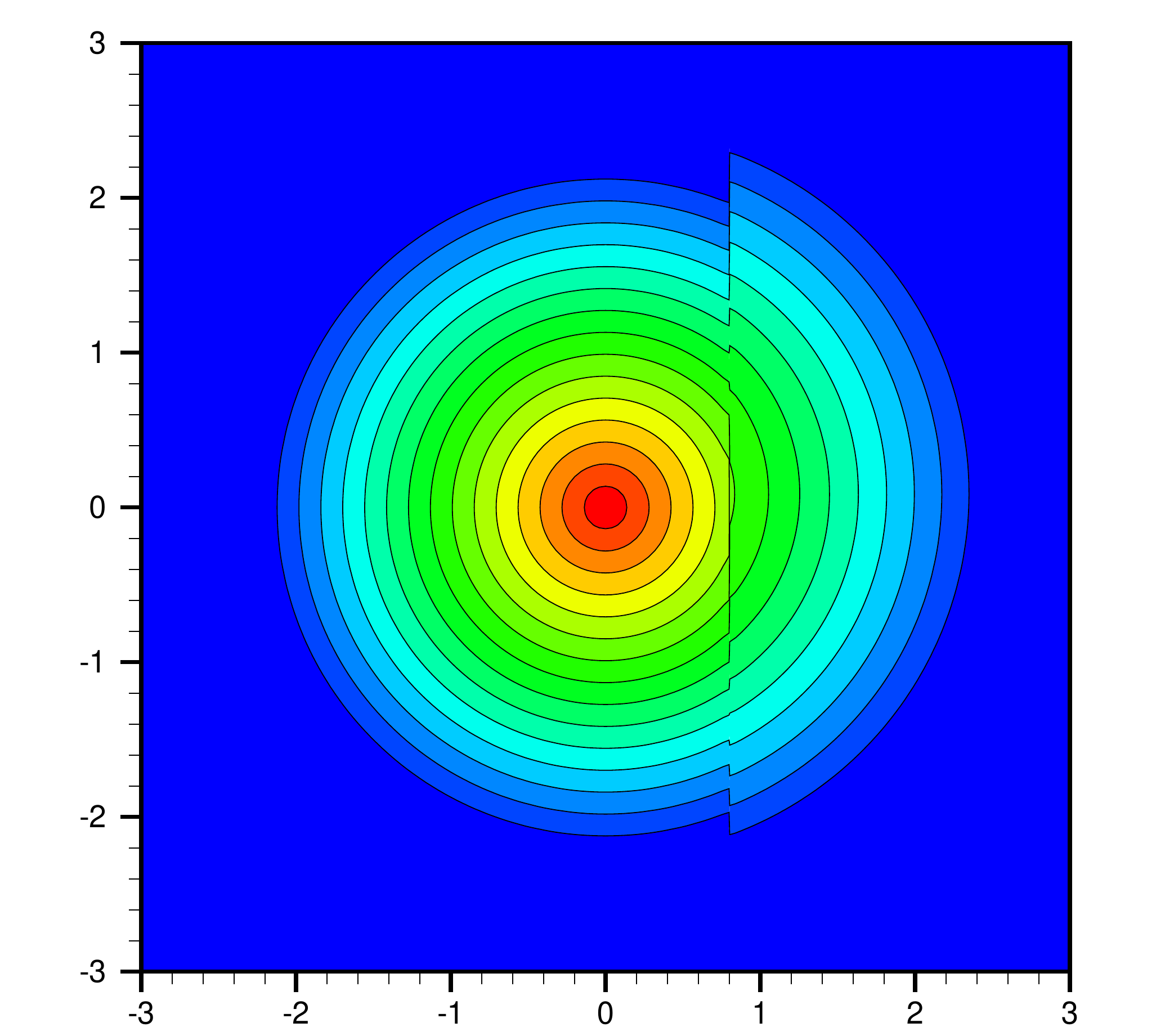}}\\
	\subfigure[$|g_0(u,v) - g_w(u,v)|$\label{fig:rayleigh_df}]
	{\includegraphics[width=0.32\textwidth]{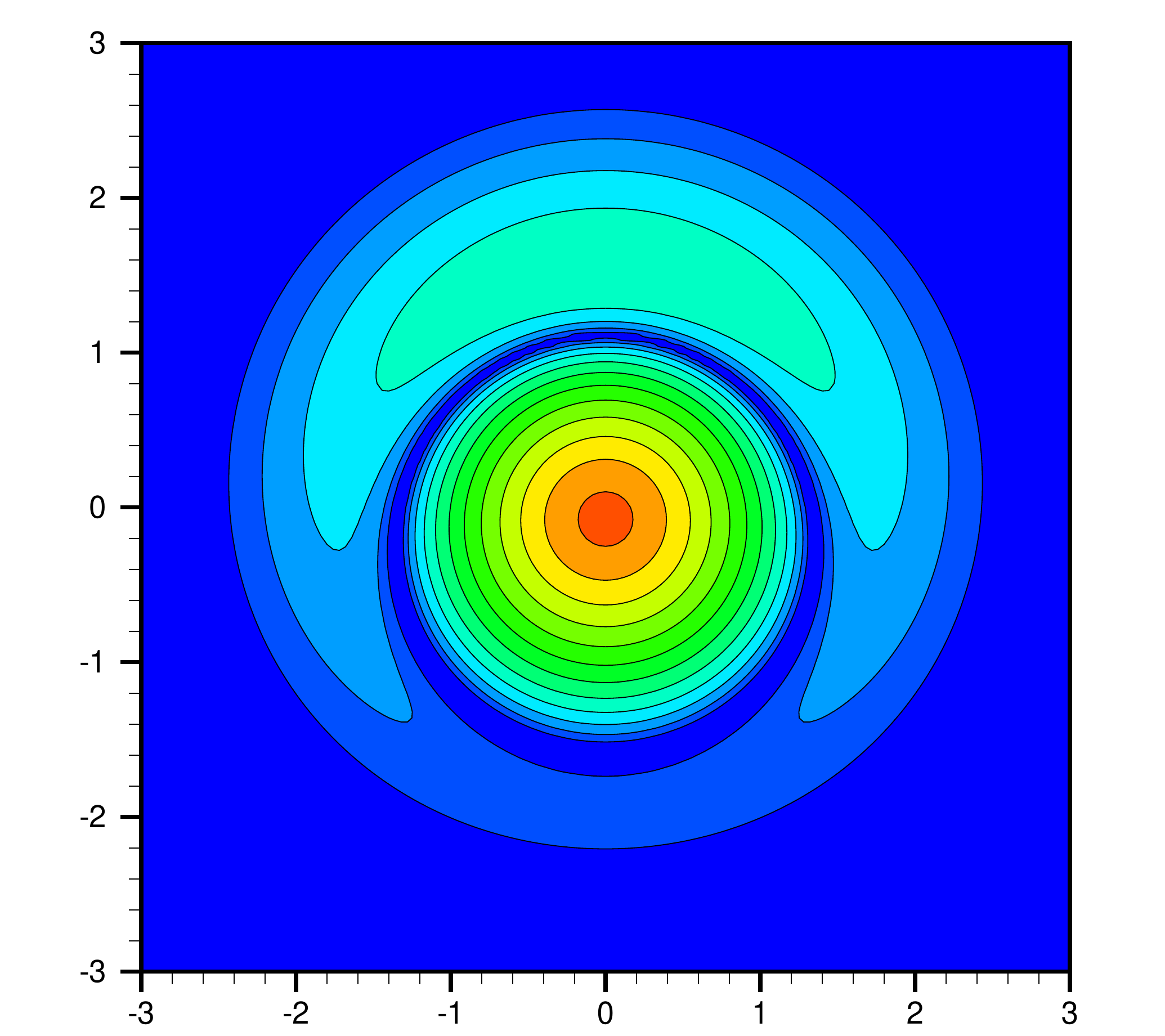}}
	\subfigure[$|u\left(g_0(u,v) - g_w(u,v)\right)|$\label{fig:rayleigh_dfu}]
	{\includegraphics[width=0.32\textwidth]{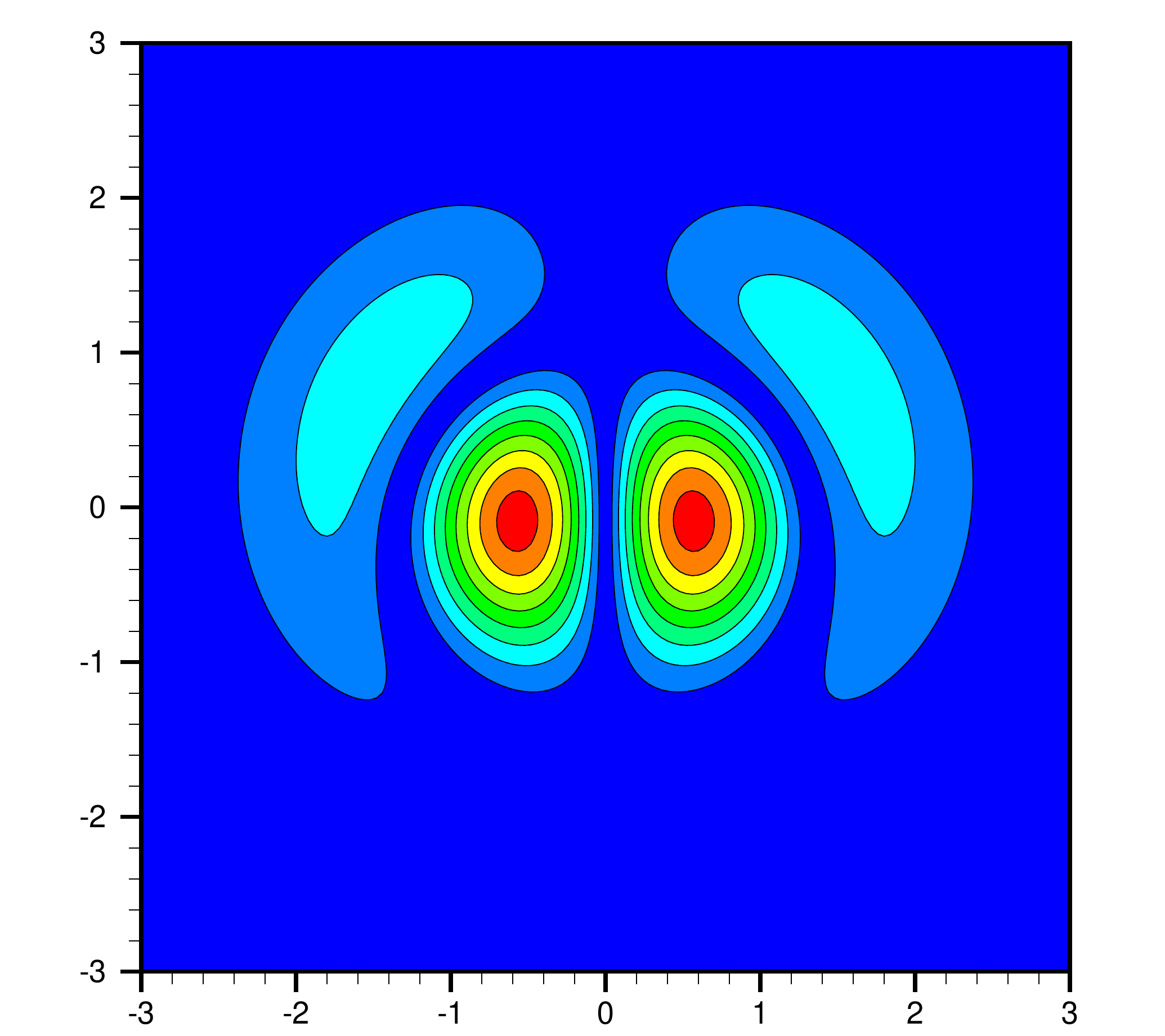}}
	\subfigure[$|v\left(g_0(u,v) - g_w(u,v)\right)|$\label{fig:rayleigh_dfv}]
	{\includegraphics[width=0.32\textwidth]{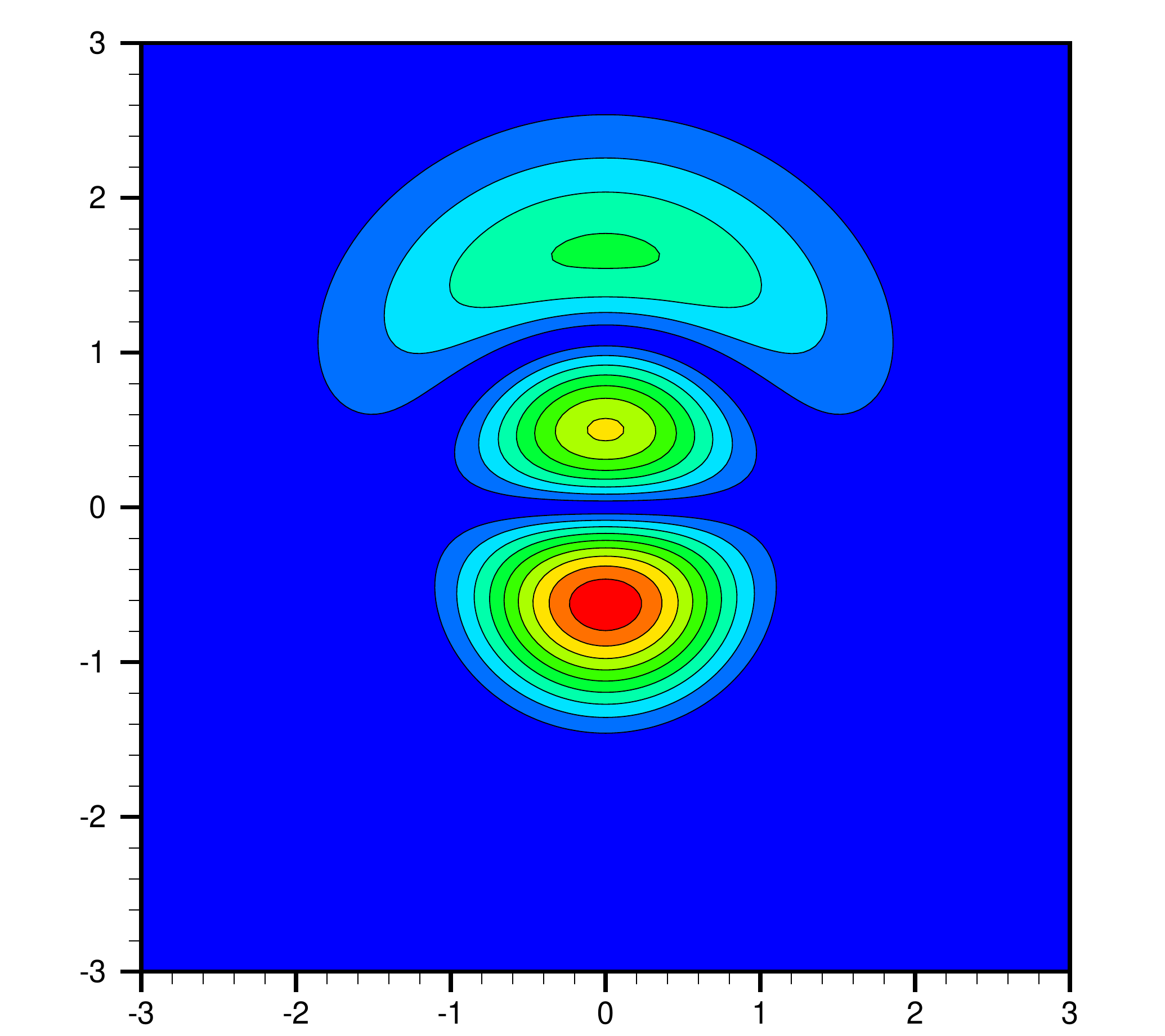}}
	\caption{Analytic distribution function in velocity space for Rayleigh flow at ${\rm Kn}\to \infty$. (a) Initial Maxwellian distribution $g_0(u,v)$; (b) reflected Maxwellian distribution  $g_w(u,v)$ from the wall; (c) distribution function $f(u,v)$ at physical location $x$ which is a combination of $g_0$ and $g_w$ at $u=x/t$; (d) (e) and (f) : distribution function difference between $g_0$ and $g_w$ for evaluation of ray effect jumps.}
\end{figure}

The computational domain is discretized by $10^4$ uniform cells to approximate the case of a continuous physical space.
The trapezoidal rule with $40\times40$ discrete velocity points covering a range of $\left[-4,4\right]\times\left[-4,4\right]$ is employed for the velocity space integration.
In the collisionless limit at ${\rm Kn}\to \infty$, the numerical solutions at time $t_e = 0.2$ are given in Fig.~\ref{fig:rayleigh_collisionless}.
It can be found that the step of the ray effect is $\Delta x_s = t_e \Delta u = 0.04$ as predicted by the theoretical analysis.
Theoretically, the distribution function at the physical location point $x$ at time $t_e$ is a combination of the initial half Maxwellian distribution $g_0$ and the wall reflected half Maxwellian distribution $g_w$ as shown in Fig.~\ref{fig:rayleigh_vs}.
Therefore, the jump of the ray effect at $x_k = u_k t_e$ would be the numerical integration of the distribution function difference along the discontinuity line $u=u_k$.
The analytic distribution function differences are plotted in Fig.~\ref{fig:rayleigh_df}, ~\ref{fig:rayleigh_dfu} and ~\ref{fig:rayleigh_dfv} for evaluation of the ray effect jumps.
In Fig.~\ref{fig:rayleigh_df}, with increment of $u$, the distribution function difference $|g_0 - g_w|$ decreases to zero first, and then slightly increases before decaying to zero again, which agree with the variations of the density jumps shown in Fig.~\ref{fig:rayleigh_density}.
Similarly in Fig.~\ref{fig:rayleigh_dfu} and \ref{fig:rayleigh_dfv}, the distribution function differences $|u(g_0 - g_w)|$ and $|v(g_0-g_w)|$ exactly indicate the velocity jumps of ray effect shown in Fig.~\ref{fig:rayleigh_velocity}.

\begin{figure}[H]
	\subfigure[Density]{\includegraphics[width=0.32\textwidth]{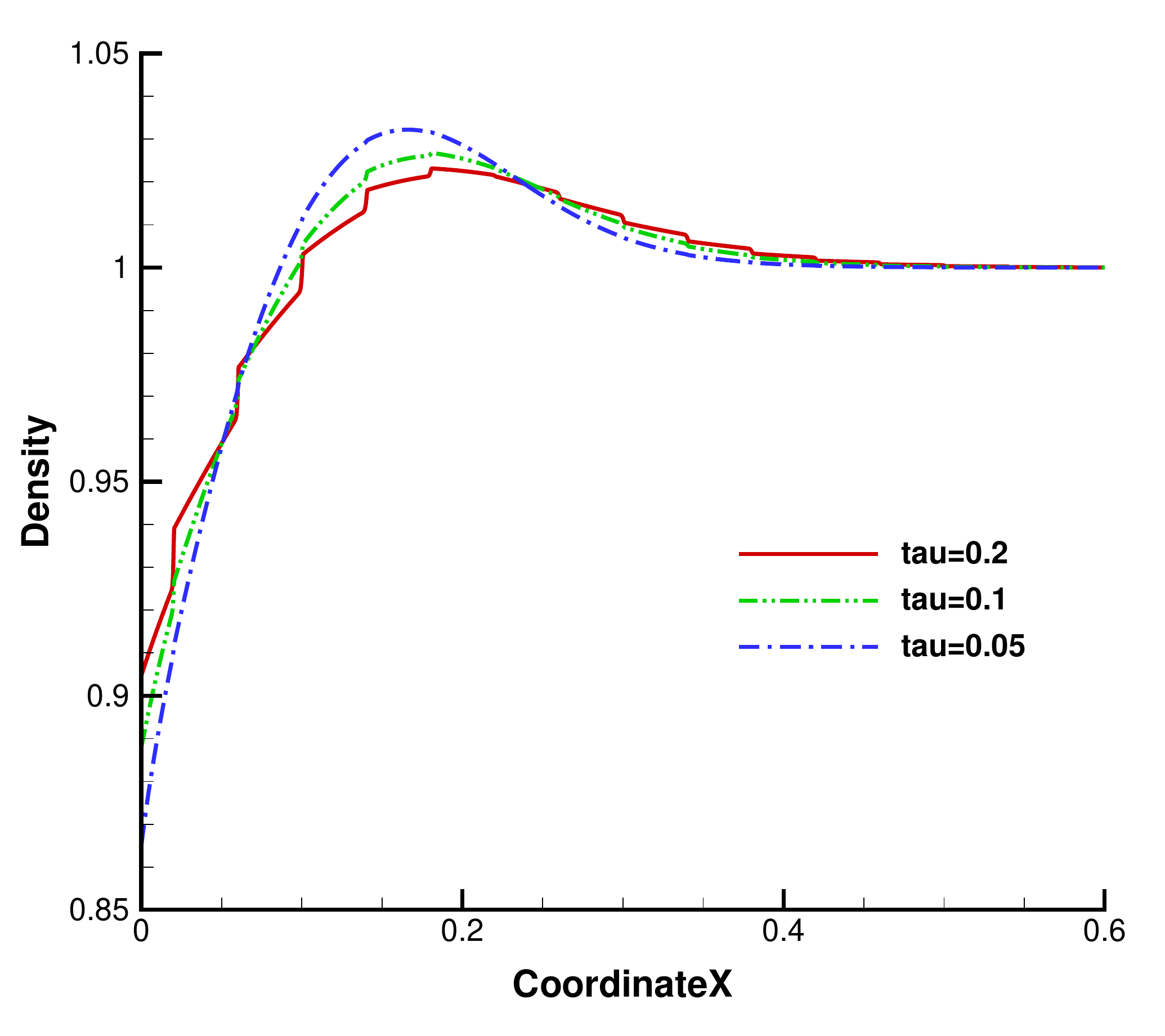}}
	\subfigure[Velocity]{\includegraphics[width=0.32\textwidth]{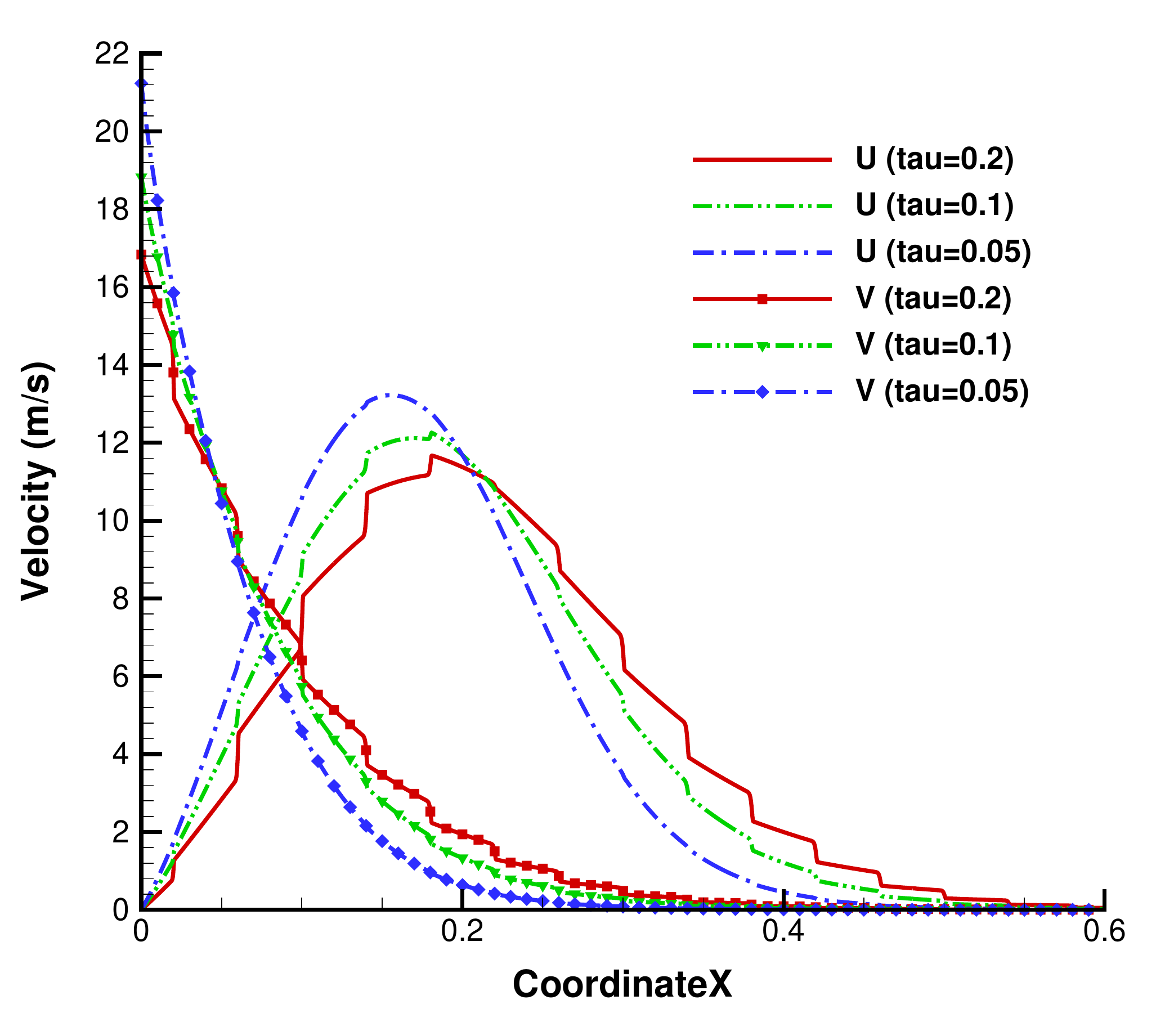}}
	\subfigure[Temperature]{\includegraphics[width=0.32\textwidth]{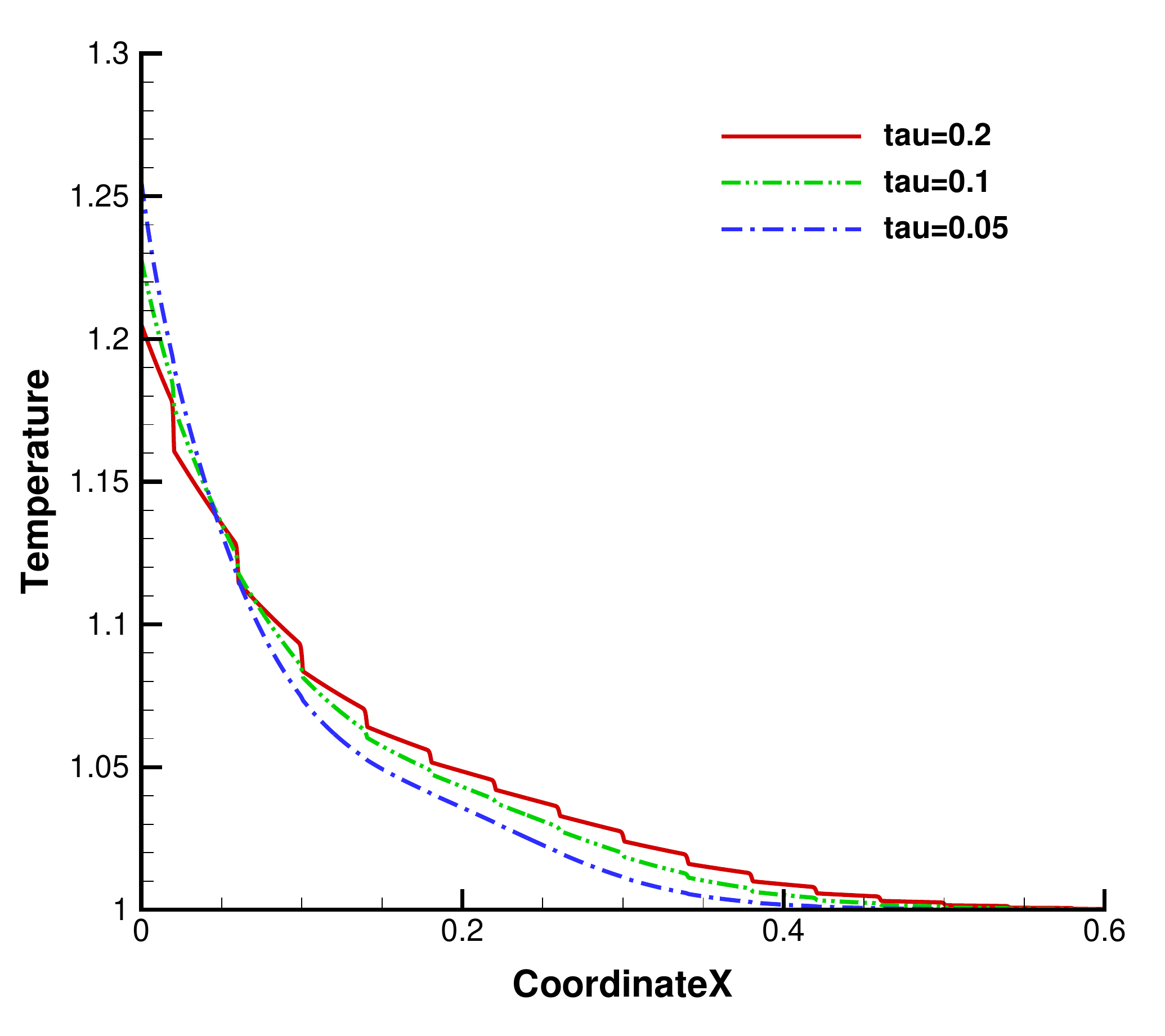}}
	\caption{\label{fig:rayleigh_collisional}Numerical solutions at time $t_e=0.2$ for Rayleigh flow at different Knudsen numbers.}
\end{figure}

\begin{figure}[H]
	\centering
	\subfigure[]{\includegraphics[width=0.48\textwidth]{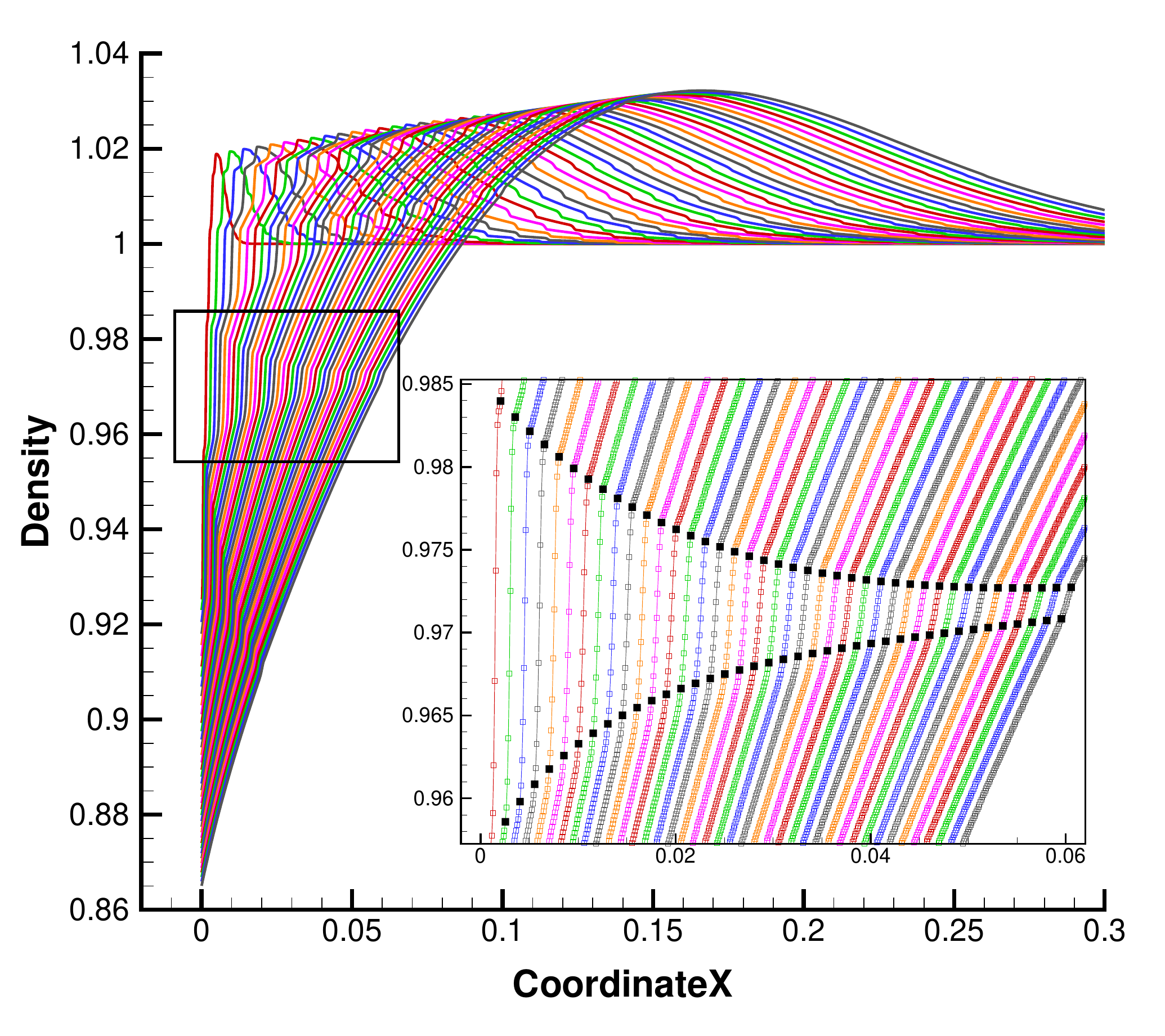}}
	\subfigure[]{\includegraphics[width=0.48\textwidth]{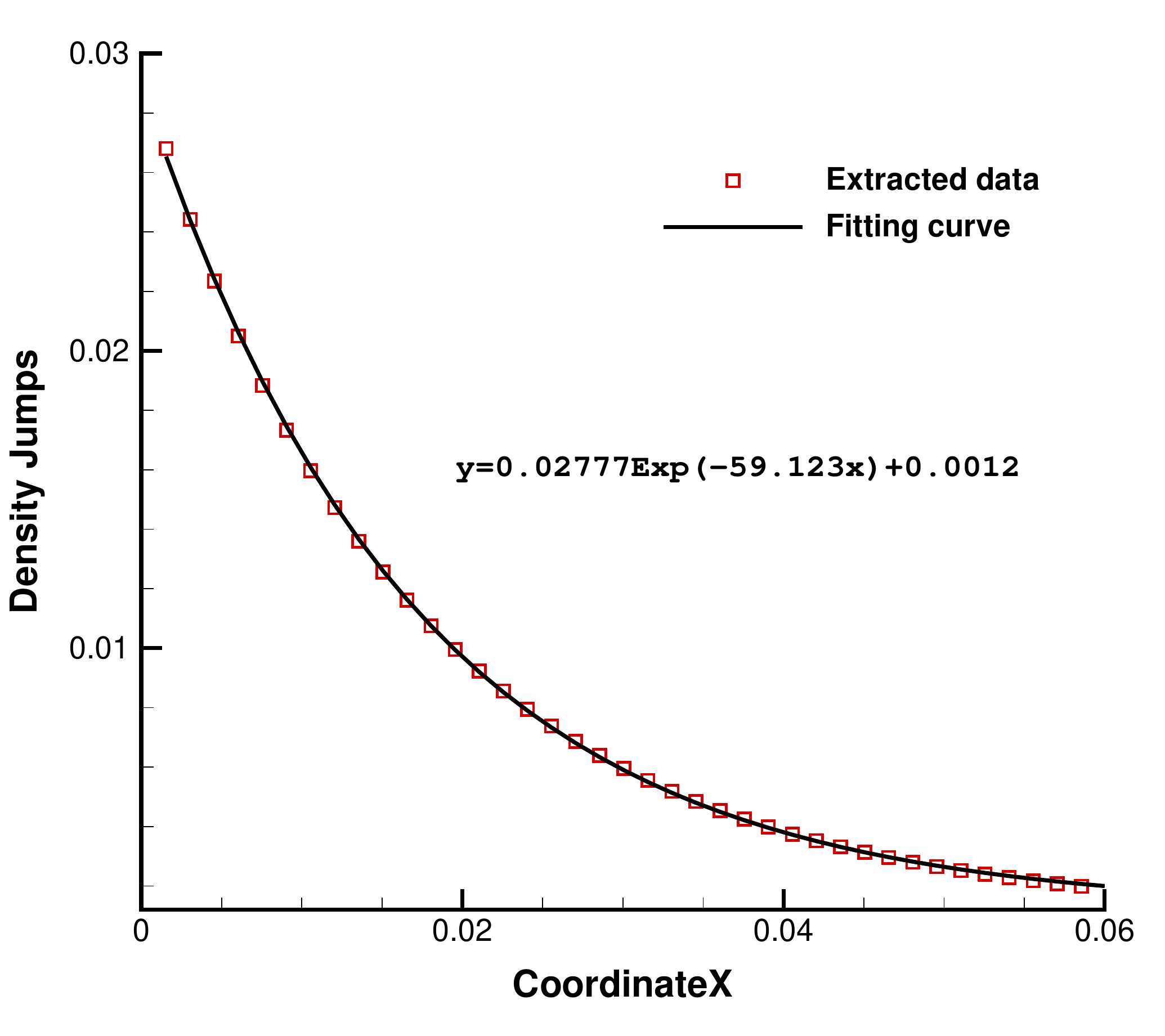}}
	\caption{\label{fig:rayleigh_decay}Time evolution of the Rayleigh flow for $\tau = 0.05$. (a) Time accurate density distribution and local enlargement of density jumps; (b) decay of the density jumps.}
\end{figure}

We also compute the Rayleigh flow at different Knudsen numbers to explore the influence of particles' collision on the ray effect.
As shown in Fig.~\ref{fig:rayleigh_collisional}, at different Knudsen numbers the jumps reduce when the increasing of particle collisions.
After $4$ collision time, i.e., $t>4\tau$, the ray effect is hardly observed.
For quantitative study, we plot the time evolution solutions for the case of $\tau=0.05$ in Fig.~\ref{fig:rayleigh_decay}.
The local density jumps due to the discontinuity of distribution function at $u_k=0.3$ are shown in details.
By fitting the extracted data of density jumps, we find that the jump decays exponentially.
Quantitatively, the decay rate is $e^{-59.123x} = e^{-59.123 u_k t} \approx e^{-t/0.056}$.
Since the local relaxation time is about $\tau \approx 0.053$, the theoretic analysis of the decay rate $e^{-t/\tau}$ in Section \ref{sec:collision} has been well validated.

\subsection{Cavity flow}

The lid-driven cavity flow is studied in rarefied regime to show the ray effect in the two dimensional case.
The stationary gas in the cavity is driven by the moving lid with a constant horizontal velocity $U_w = 100{\rm m/s}$.
The initial temperature $T_0$ is $273{\rm K}$ for the monatomic argon gas.
Isothermal boundary condition with a fixed temperature of $T_w = 273{\rm K}$ is applied to all side walls.
The Knudsen number is defined with respect to the length of the side wall.

\begin{figure}[H]
	\centering
	\subfigure[Density]
	{\includegraphics[width=0.32\textwidth]{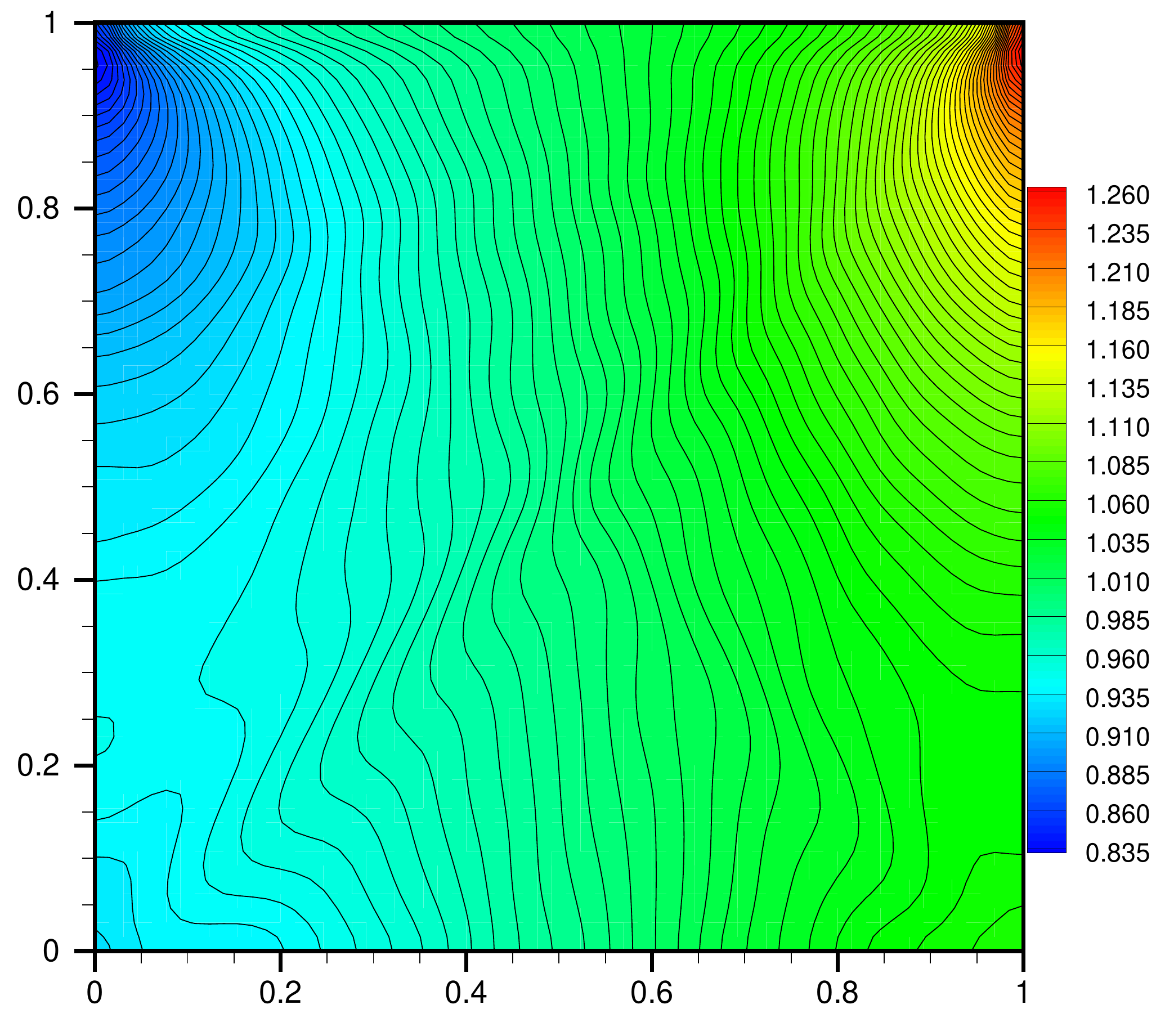}}
	\subfigure[Temperature\label{fig:cavity_collisionless_temperature}]
	{\includegraphics[width=0.32\textwidth]{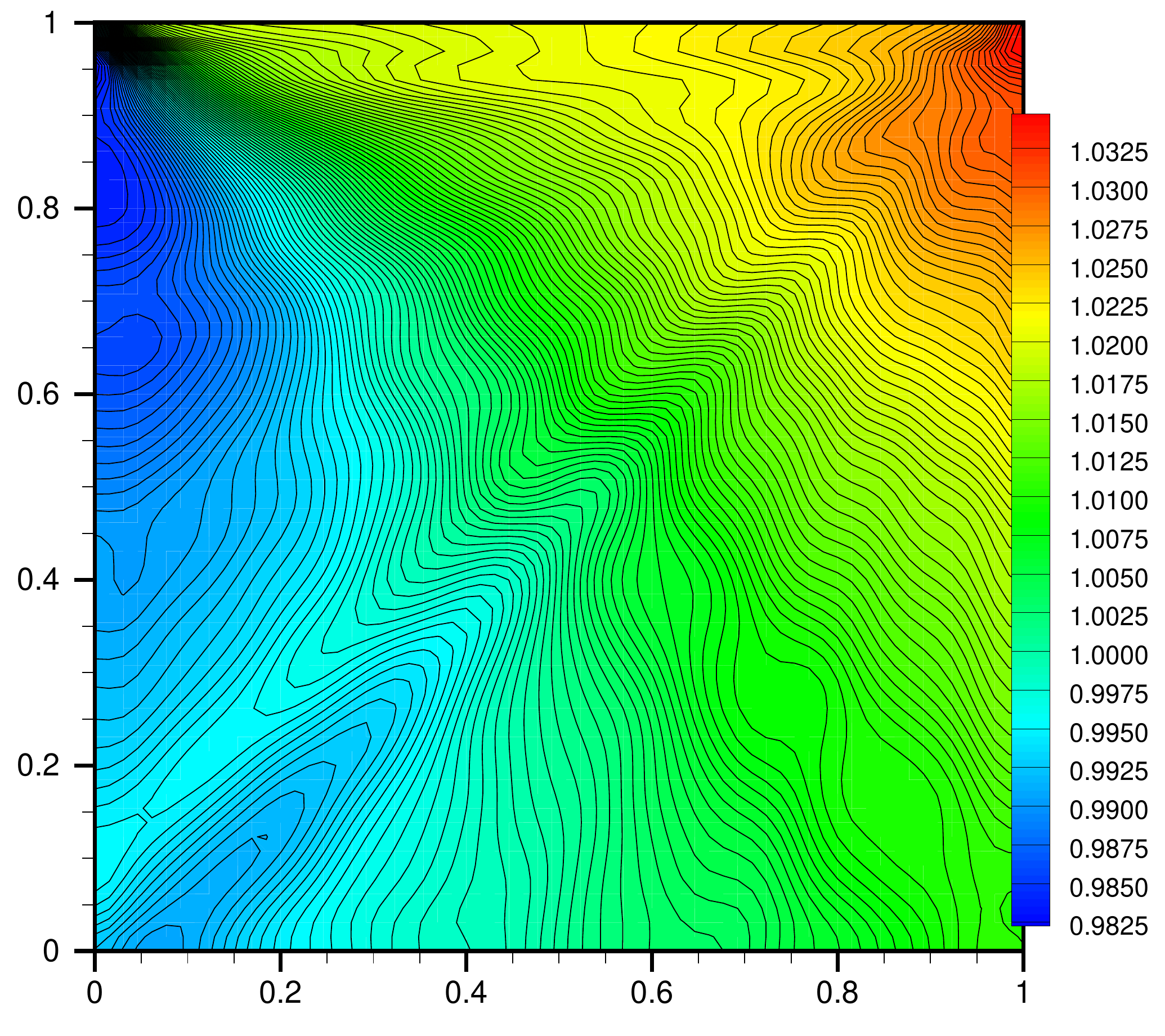}}
	\subfigure[Velocity]{\includegraphics[width=0.32\textwidth]{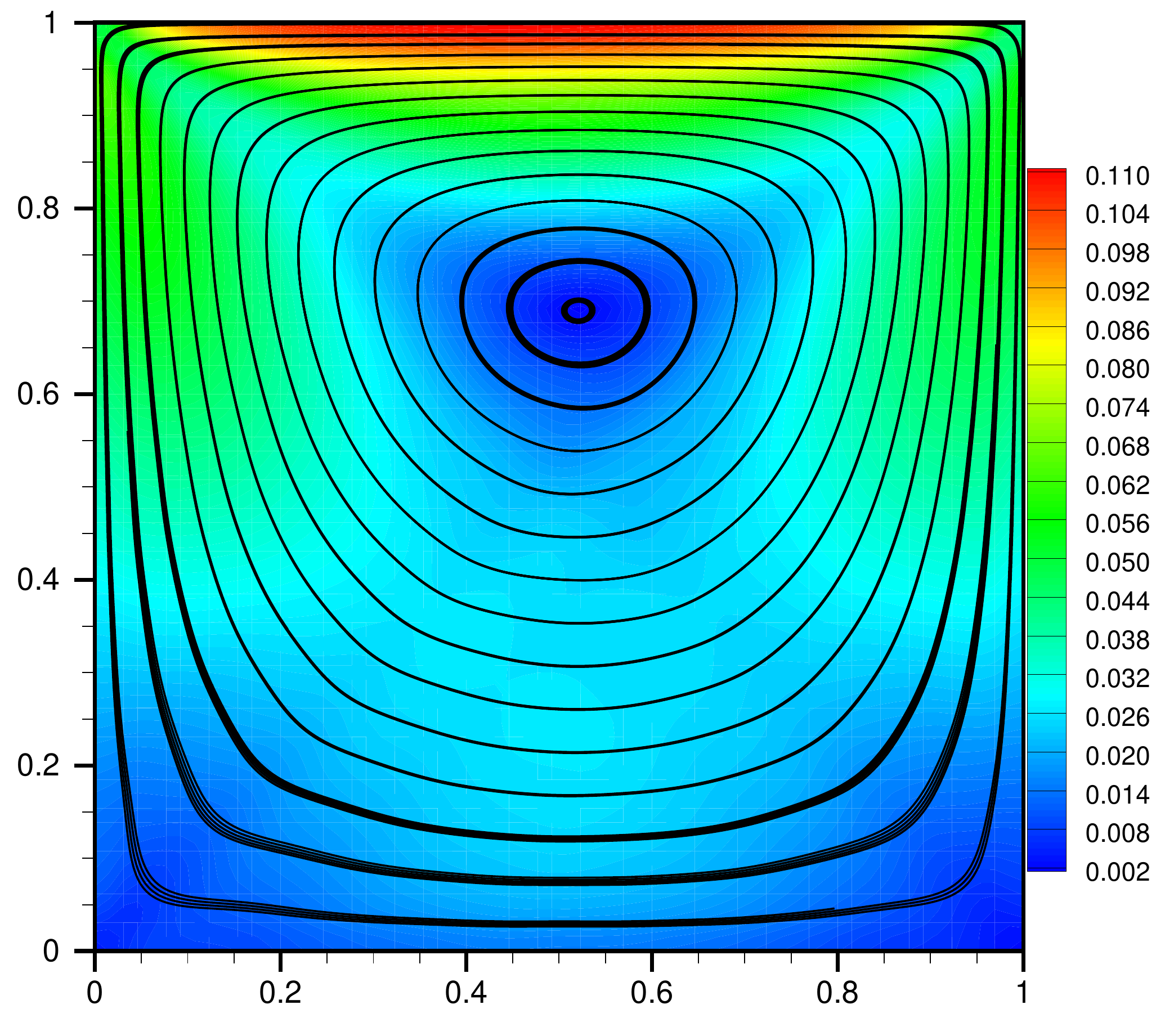}}
	\caption{\label{fig:cavity_collisionless}Cavity flow in the collisionless limit. (a) Density; (b) temperature; (c) magnitude of velocity and streamlines.}
\end{figure}

In the computation, the physical space and velocity space are discretized by the uniform meshes with $65\times65$ cells and $60\times60$ discrete velocity points, respectively.
The integration weight for each velocity point is $0.18\times0.18$.
In the collisionless limit, the flow field is shown in Fig.~\ref{fig:cavity_collisionless}, where obvious ray effect can be observed.
In details, two main phenomena are found:
(a) the ray effect shows radial pattern originated from the top corners;
(b) the ray effect looks more severe along the diagonal directions.
In the following, we will give the theoretical analysis and explanations.

\begin{figure}[H]
	\centering
	\subfigure[]{\includegraphics[width=0.42\textwidth]{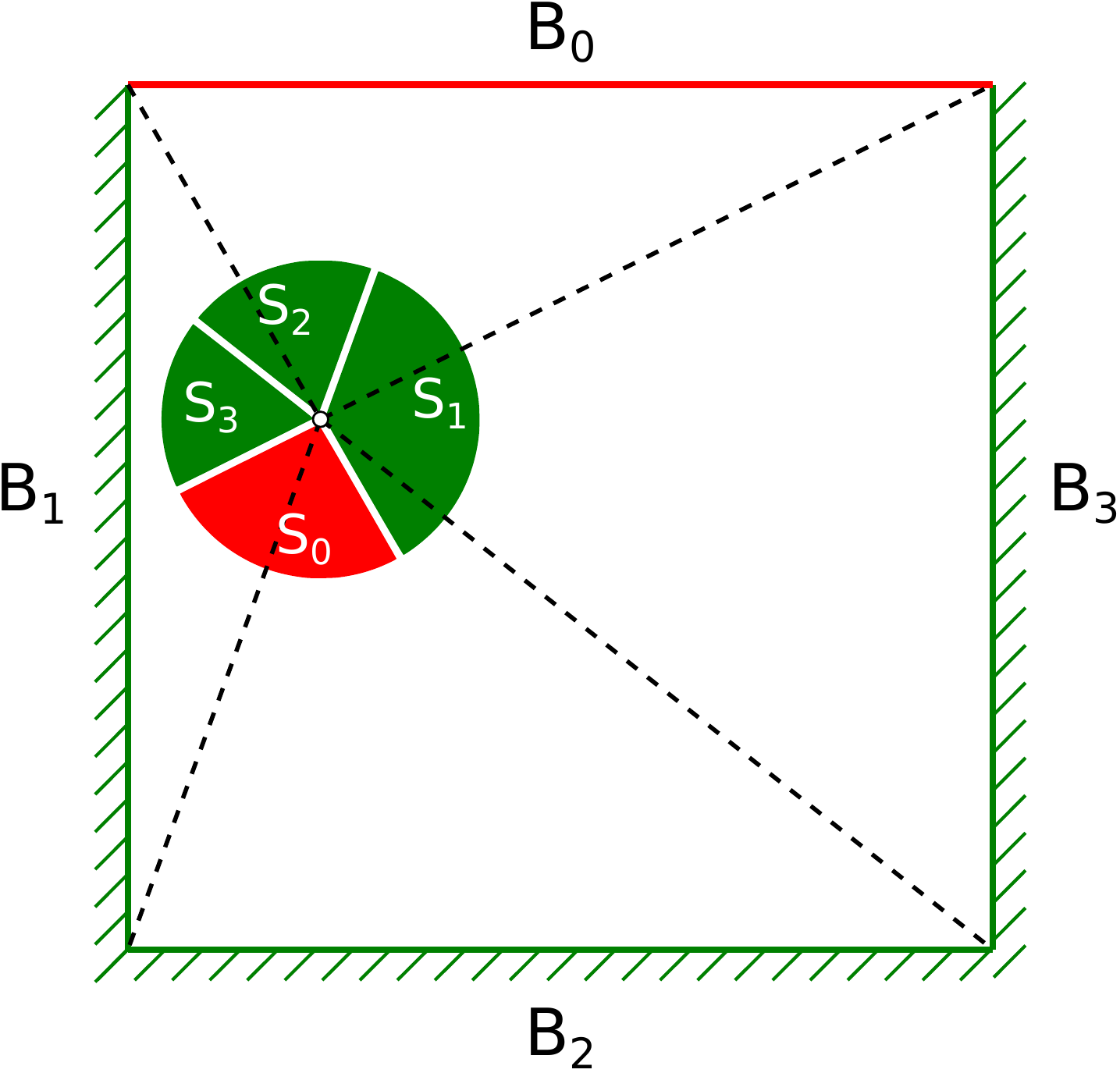}} \hspace{0.1\textwidth}
	\subfigure[]{\includegraphics[width=0.42\textwidth]{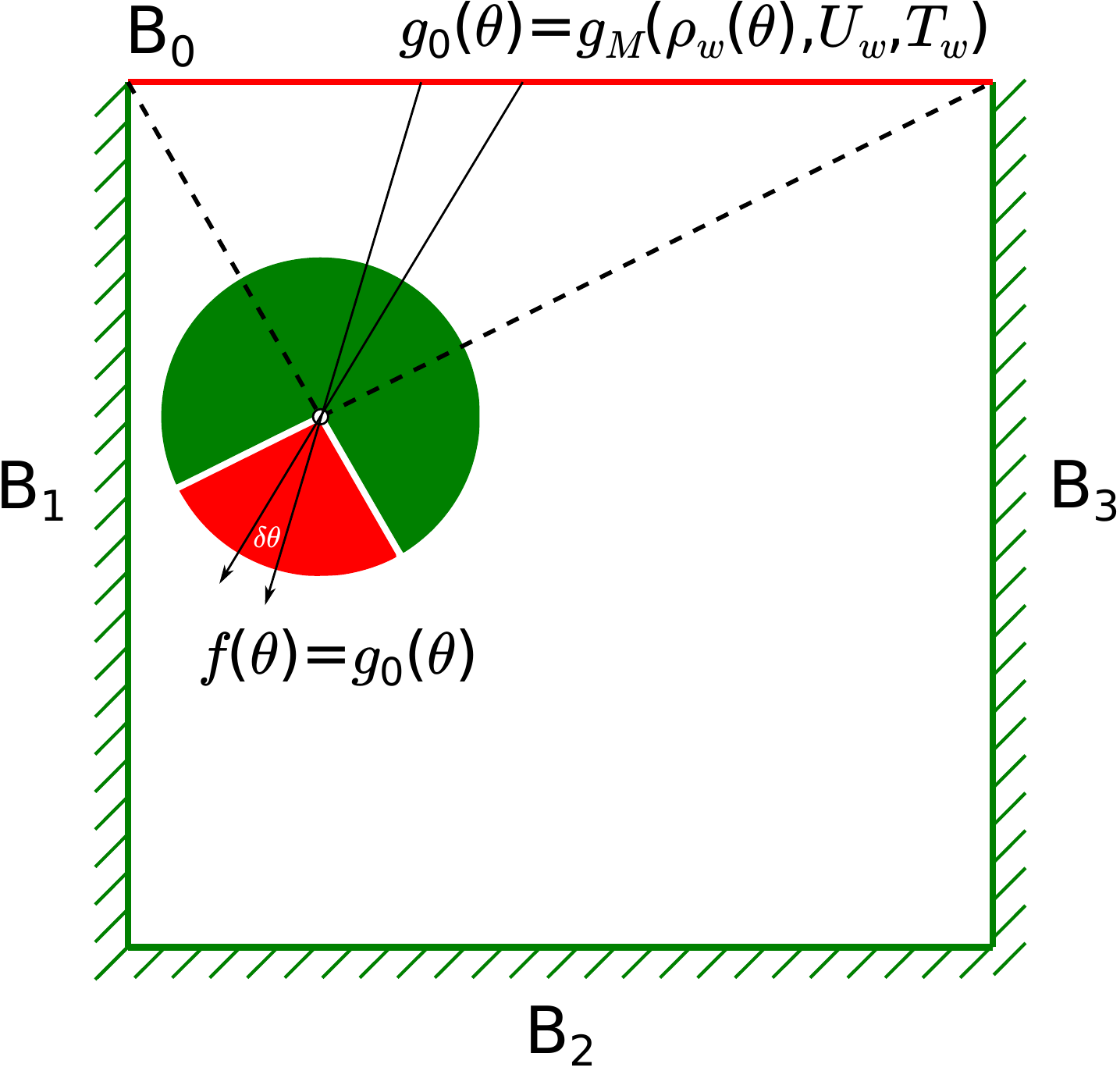}}
	\caption{\label{fig:cavity_phase}Composition of the distribution function for arbitrary location point in the cavity at ${\rm Kn}\to \infty$. $g_M$ denotes the reflected Maxwellian distribution function.}
\end{figure}

\begin{figure}[H]
	\centering
	\subfigure[Locations]{\includegraphics[width=0.45\textwidth]{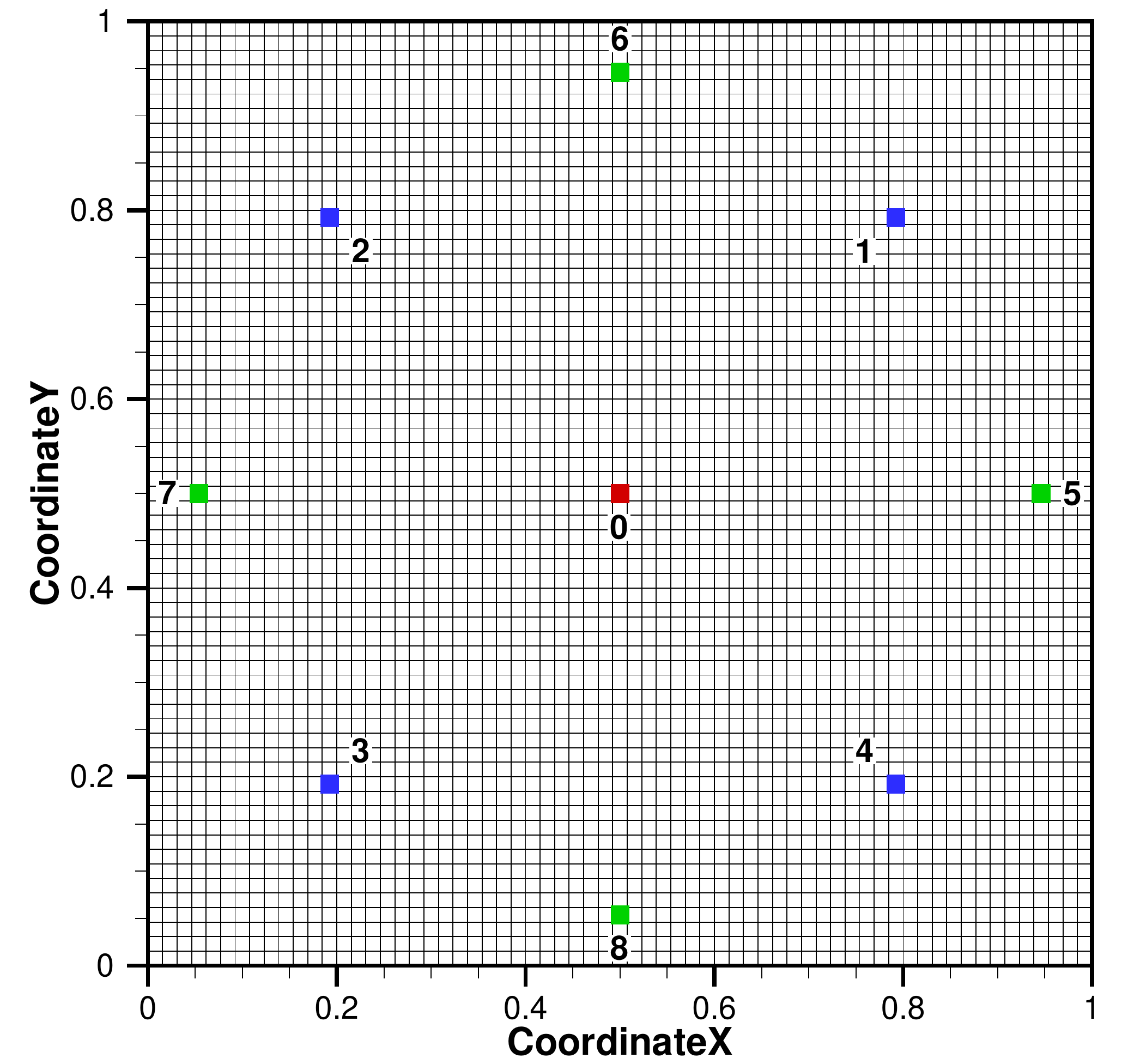}}
	\subfigure[Point $0$]{\includegraphics[width=0.45\textwidth]{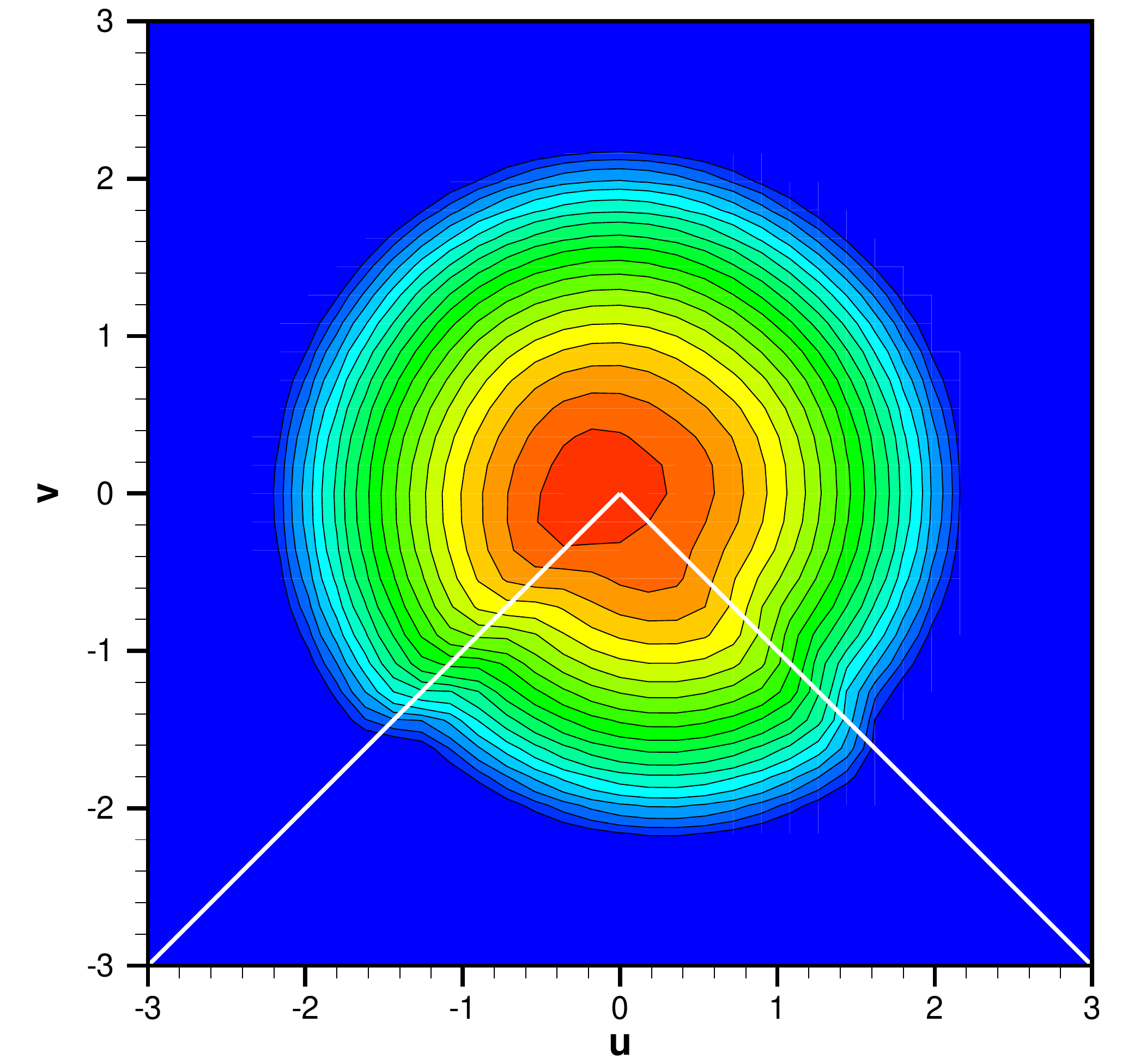}}\\
	\subfigure[Point $1$]{\includegraphics[width=0.24\textwidth]{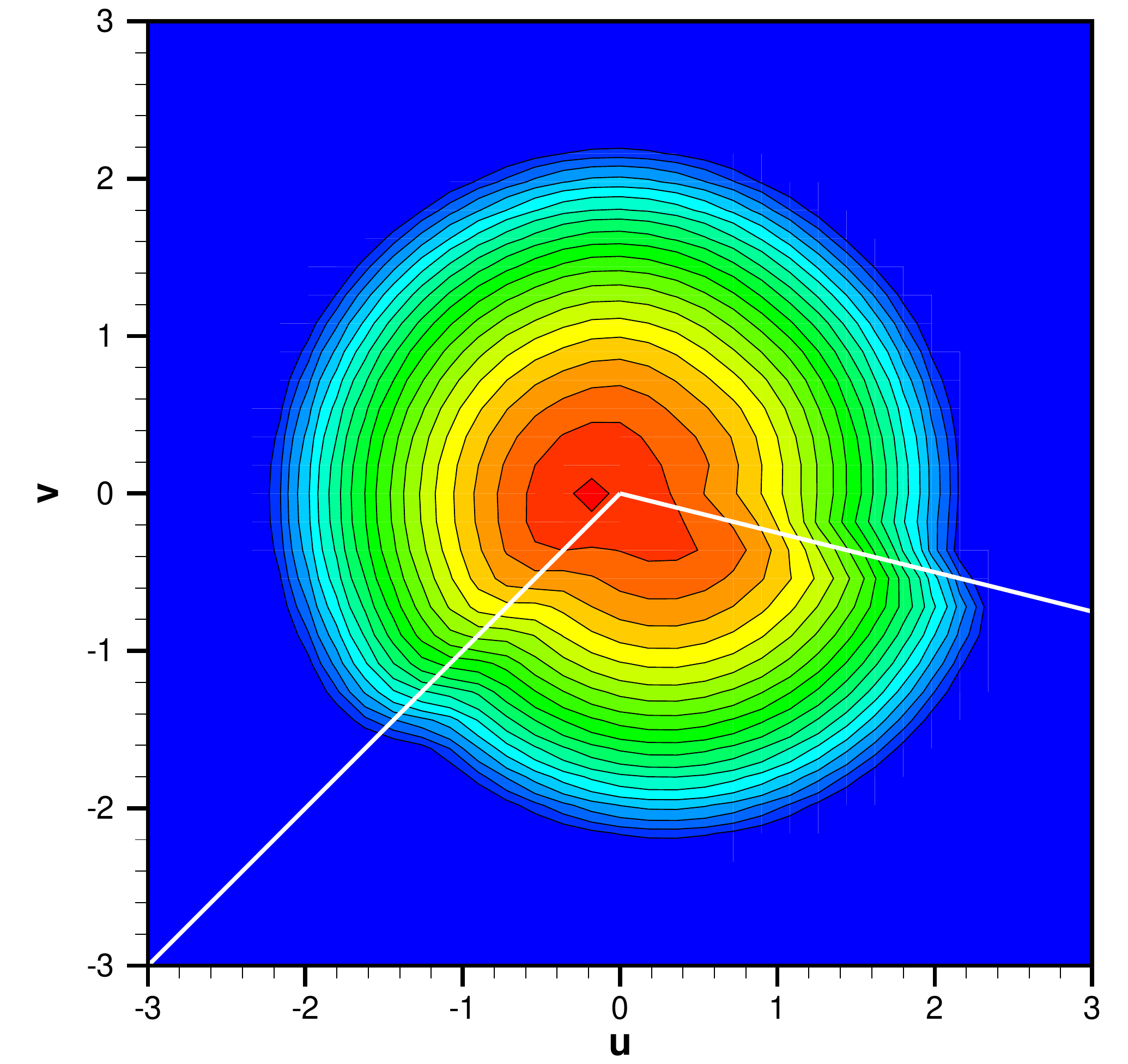}}
	\subfigure[Point $2$]{\includegraphics[width=0.24\textwidth]{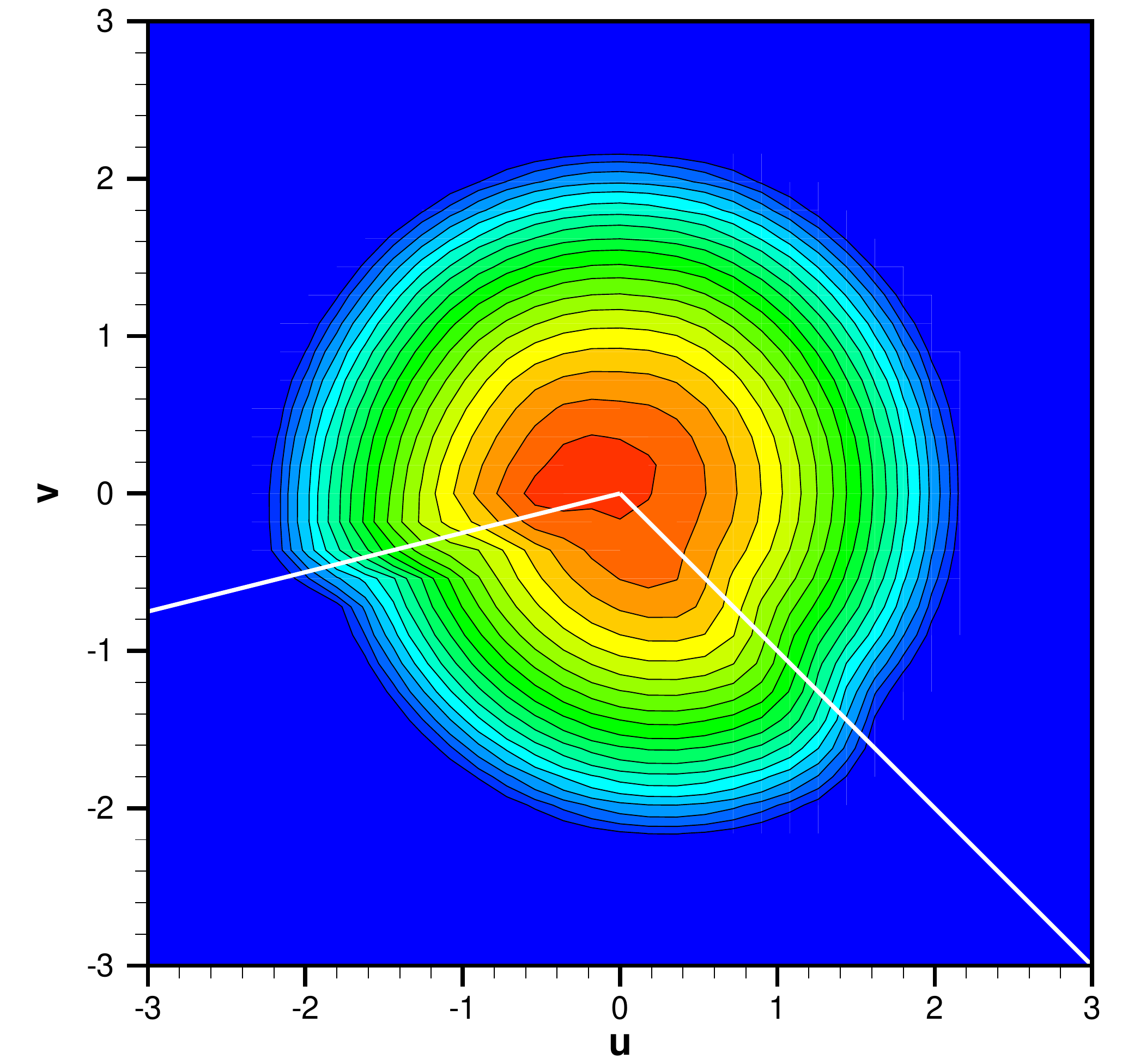}}
	\subfigure[Point $3$]{\includegraphics[width=0.24\textwidth]{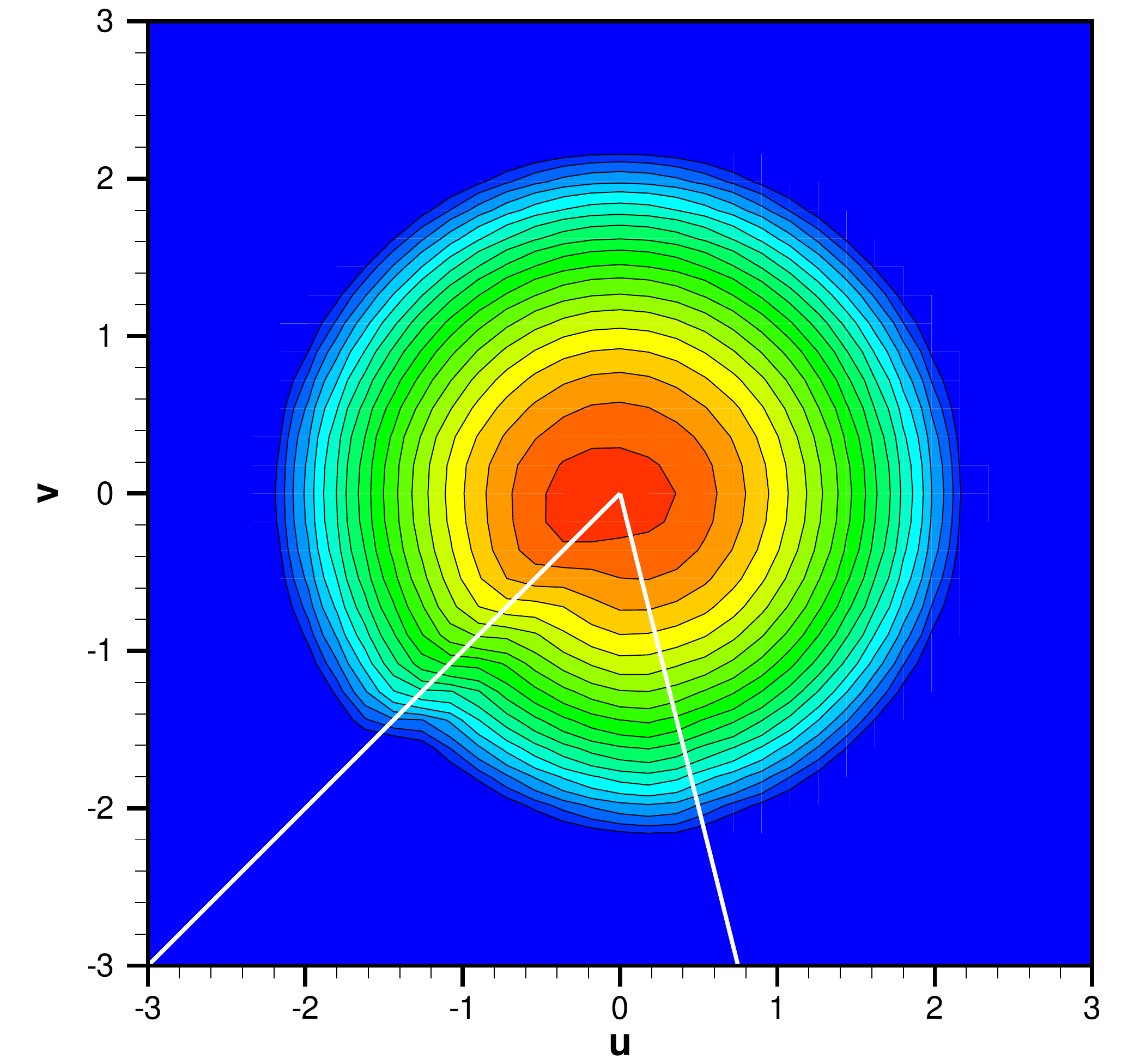}}
	\subfigure[Point $4$]{\includegraphics[width=0.24\textwidth]{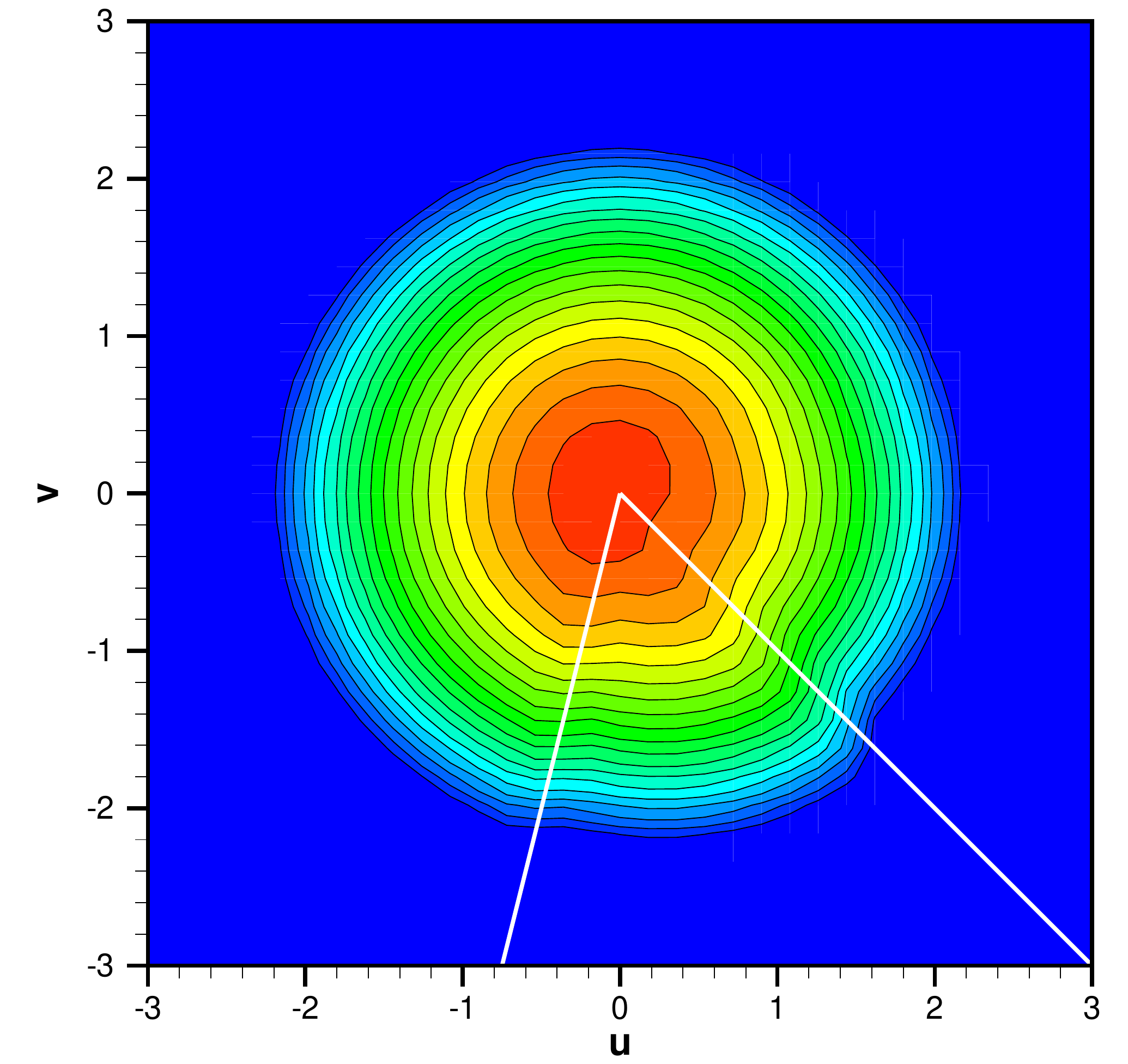}}\\
	\subfigure[Point $5$]{\includegraphics[width=0.24\textwidth]{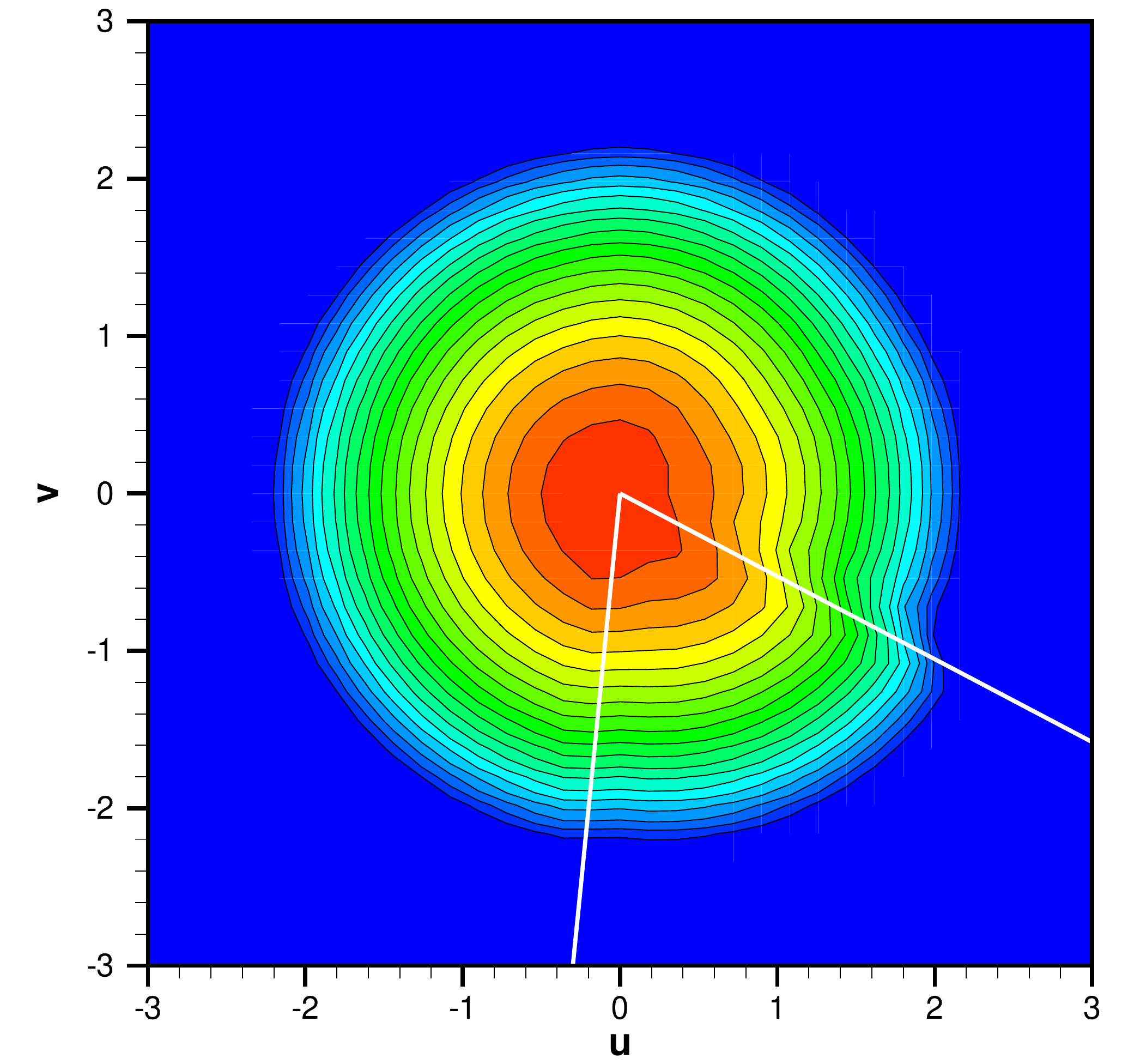}}
	\subfigure[Point $6$]{\includegraphics[width=0.24\textwidth]{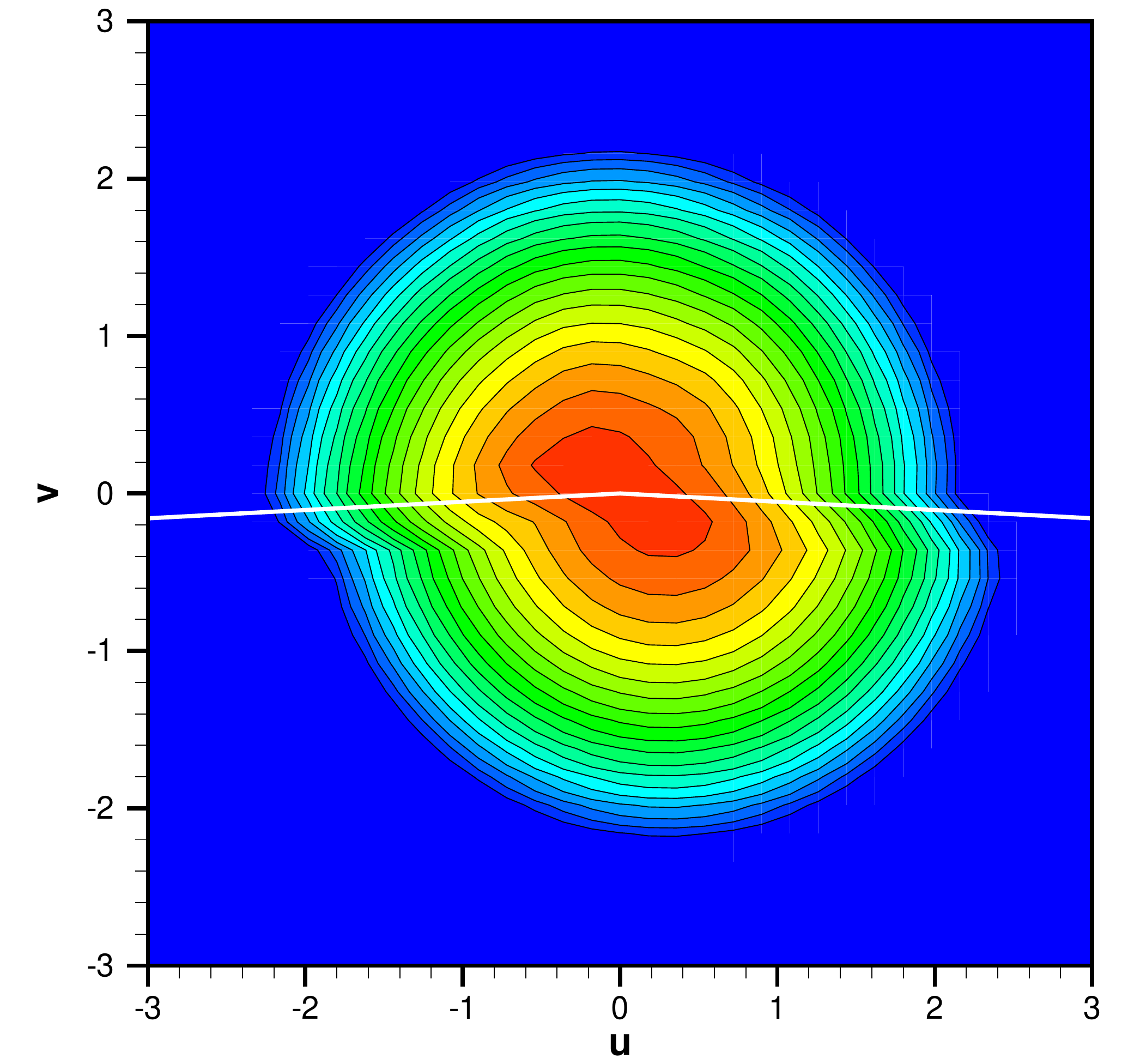}}
	\subfigure[Point $7$]{\includegraphics[width=0.24\textwidth]{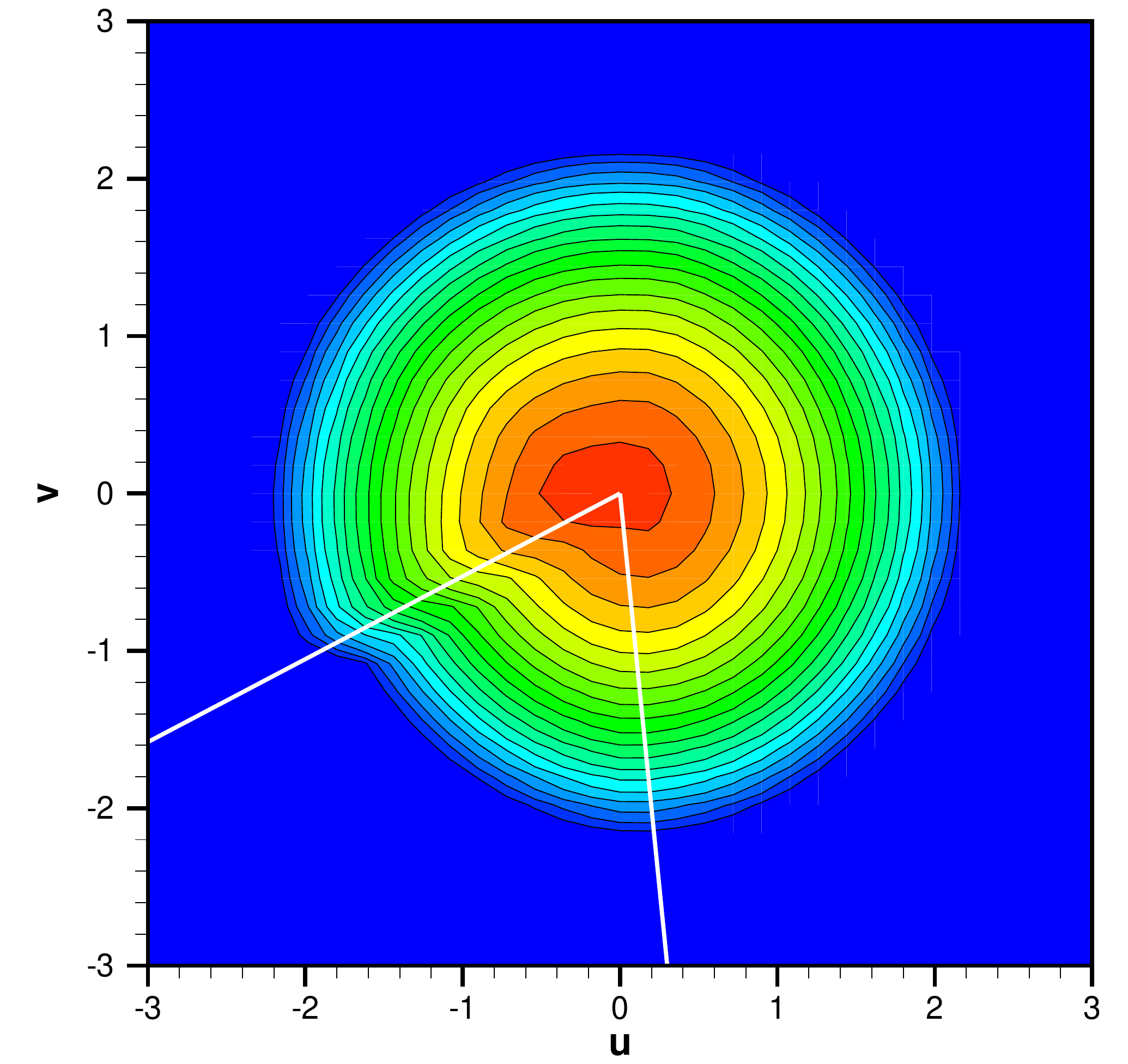}}
	\subfigure[Point $8$]{\includegraphics[width=0.24\textwidth]{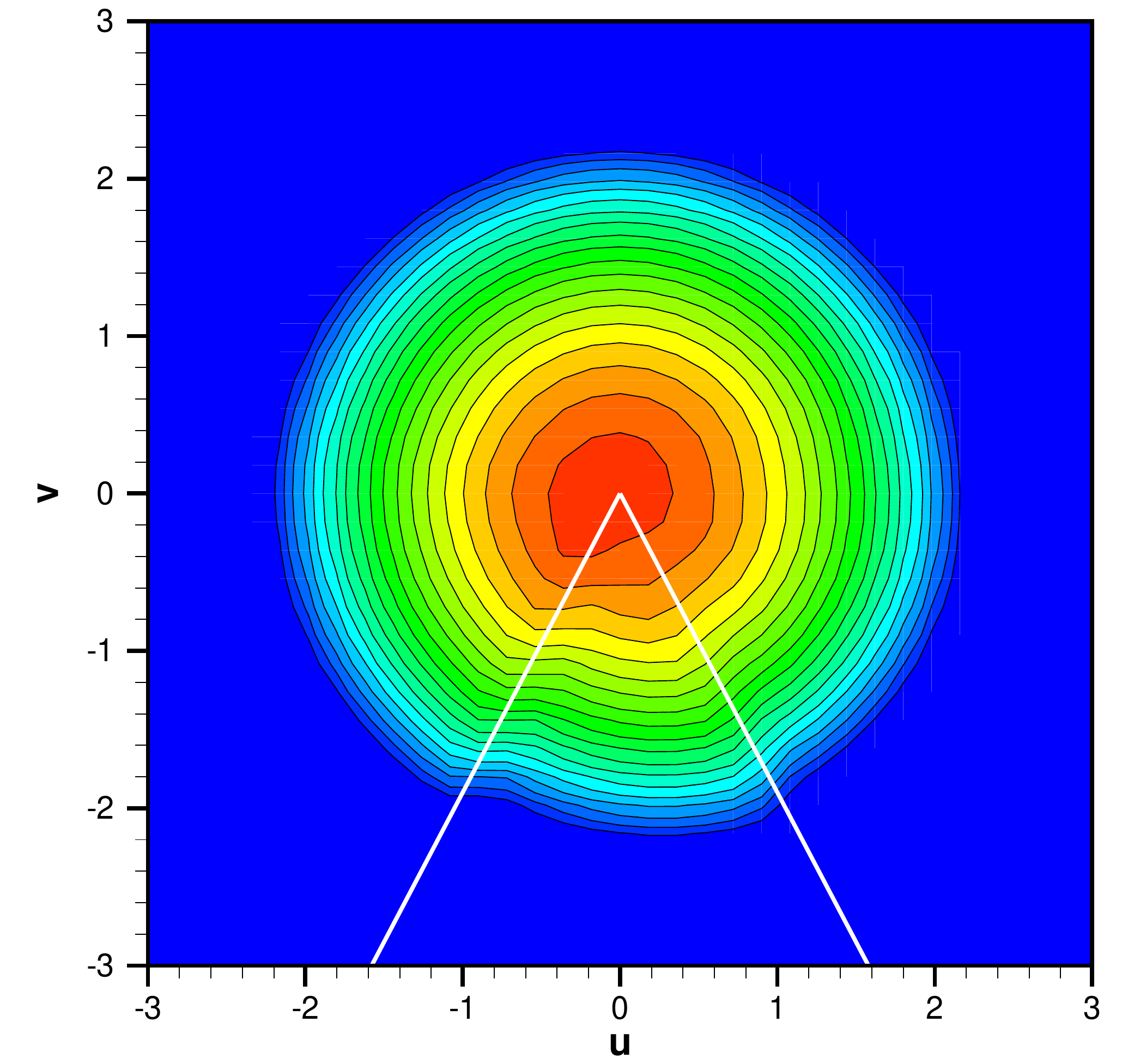}}
	\caption{\label{fig:cavity_vs}Distribution function in the velocity space for different location points. The dashed lines denotes the discontinuities induced by the top corners.}
\end{figure}

In Fig.~\ref{fig:cavity_phase}, we draw a digram to illustrate the distribution function at an arbitrary location point.
It can be found that the distribution function is formed by the reflected particles from the solid boundaries.
The velocity space is divided into four sections according to which wall the reflected particles come from.
Specifically, with the isothermal boundary condition, the distribution function is given by
\begin{equation}
	f(\theta) = g_M(\rho_w(\theta), U_w, T_w),
\end{equation}
where $g_M$ is the reflected Maxwellian distribution function, and $\rho_w(\theta)$ is obtained by the non-penetration solid boundary condition with zero mass flux.
Since the three stationary  walls $B_1$, $B_2$ and $B_3$ have the same temperature $T_w$ and velocity $U_w$, and the flow field variations near these boundaries are small, the reflected Maxwellian distribution from $B_1$, $B_2$ and $B_3$ would be almost the same.
Therefore, the discontinuities of the distribution function between the sections $S_1$, $S_2$ and $S_3$ in the velocity space would be mild.
However, the top boundary has a constant velocity and the density and temperature around the top two corners are quite different from those along the other three solid walls.
The reflected Maxwellian distribution function from the top corners especially the top right one will lead to strong discontinuities in the velocity space.
As shown in Fig.~\ref{fig:cavity_vs}, the distribution functions at different locations are plotted.
Since the discontinuity exists along the radial direction in the velocity space, the ray effect occurs due to the low resolution along circumferential direction, which results in the spatial distortions with radial patterns as observed in Fig.~\ref{fig:cavity_collisionless_temperature}.

\begin{figure}[H]
	\centering
	\subfigure[Unstructured velocity space\label{fig:cavity_unstructured_vs}]
	{\includegraphics[width=0.3\textwidth]{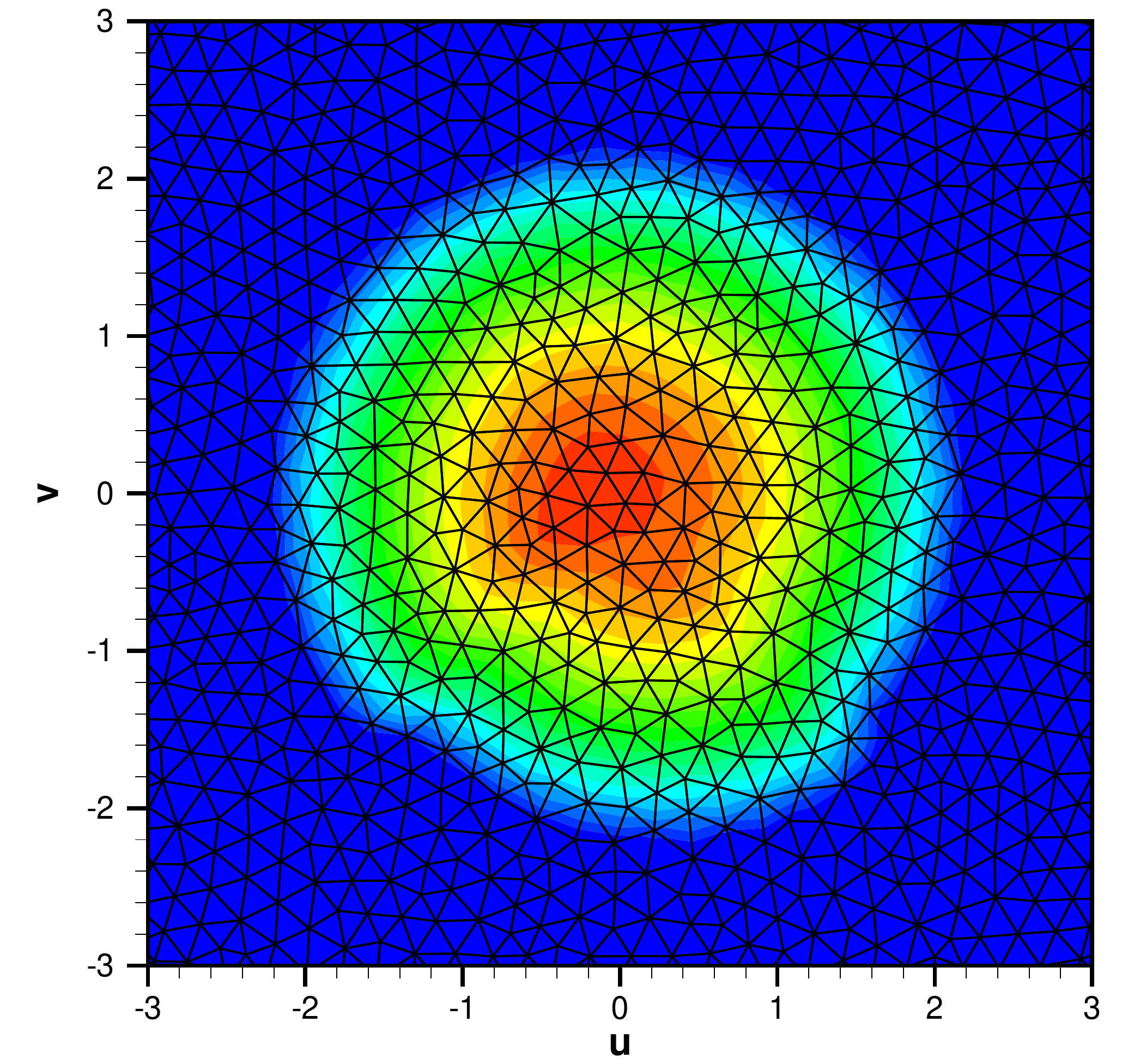}}
	\subfigure[Density\label{fig:cavity_unstructured_density}]
	{\includegraphics[width=0.32\textwidth]{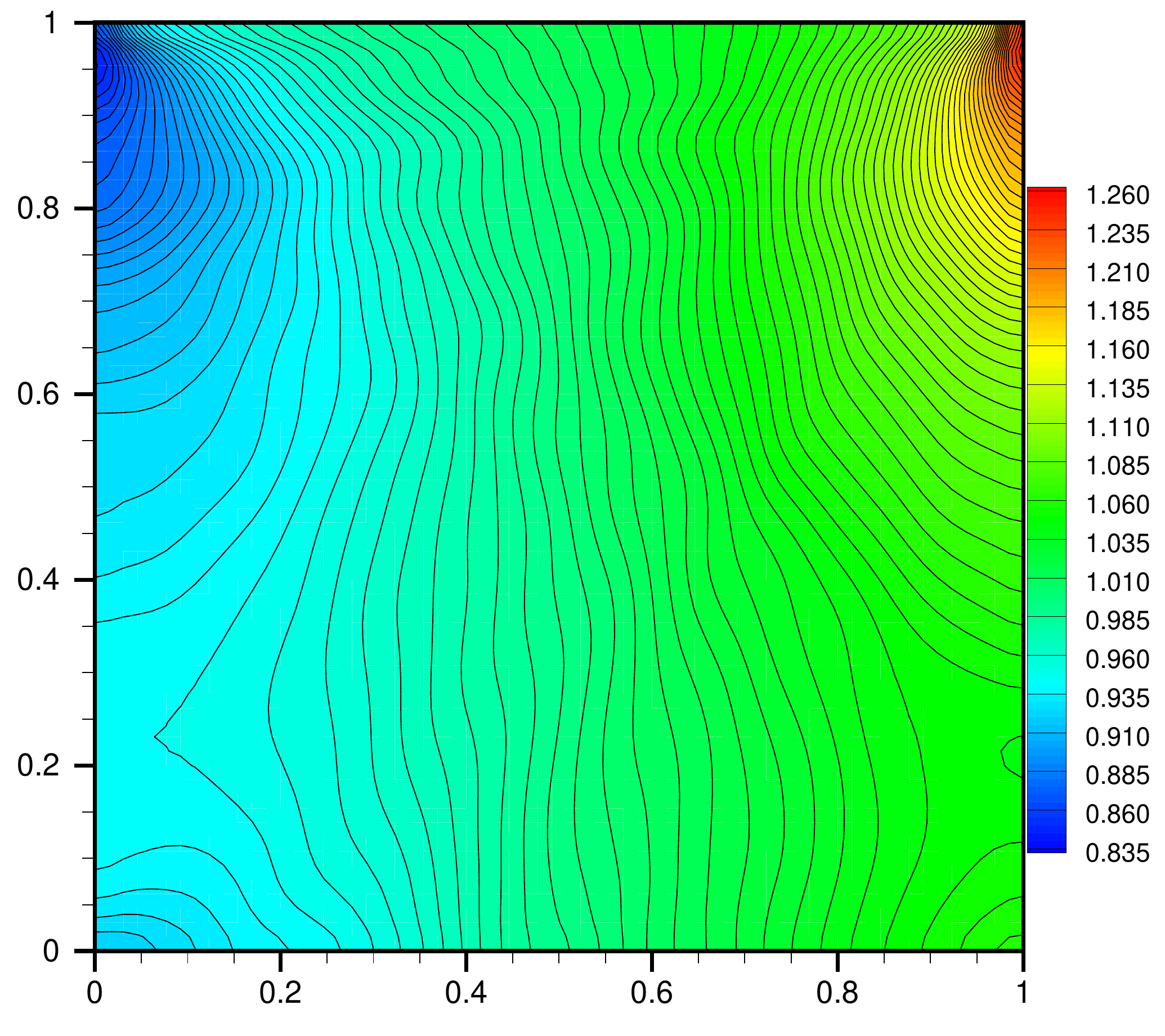}}
	\subfigure[Temperature\label{fig:cavity_unstructured_temperature}]
	{\includegraphics[width=0.32\textwidth]{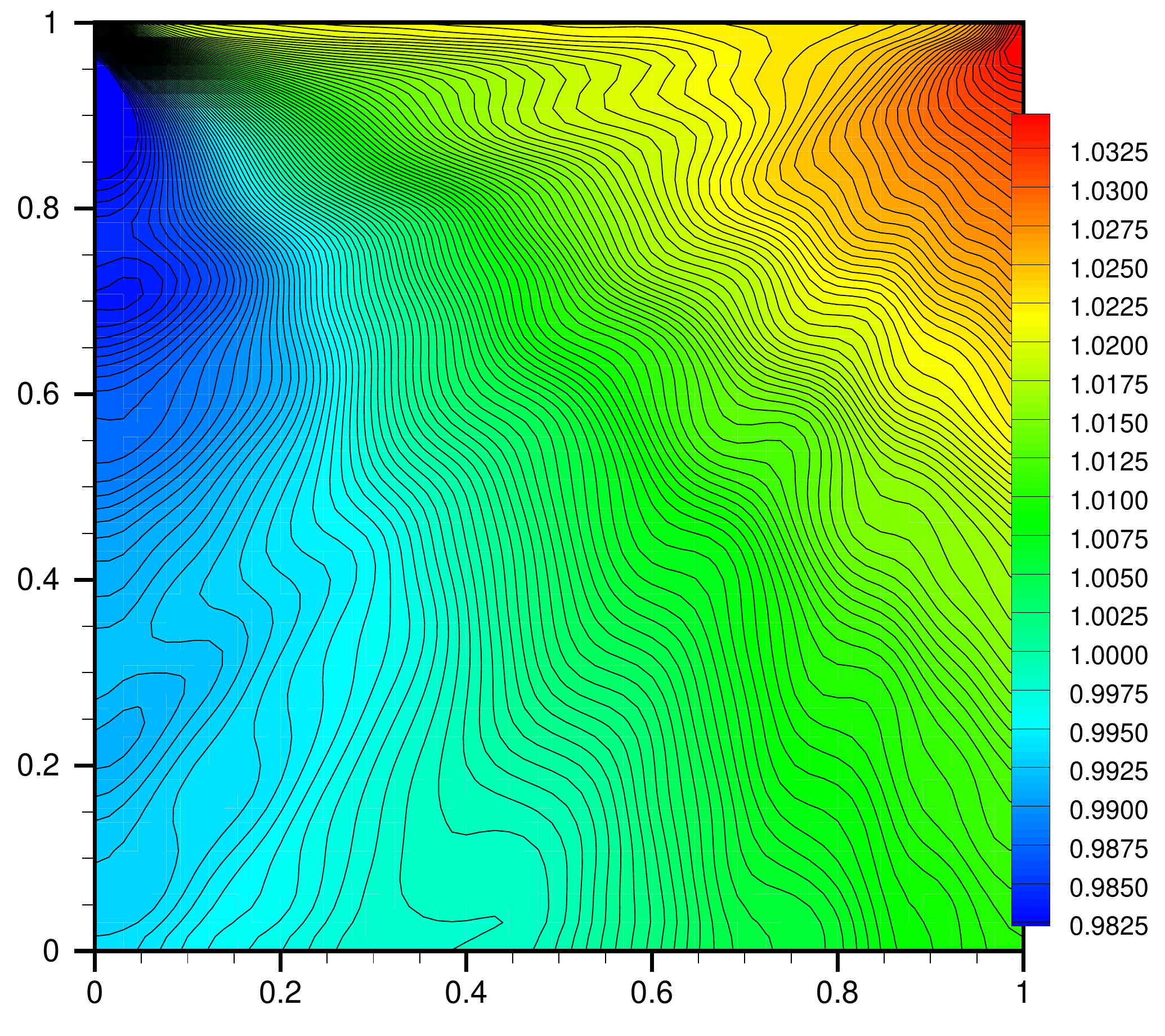}}
	\caption{\label{fig:cavity_unstructured} Cavity flow using unstructured mesh in velocity space. (a) Distribution function at $(x,y) =(0.5,0.5)$; (b) density; (c) temperature.}	
\end{figure}

Eq.~(\ref{eq:discrete_variation}) shows that each jump in the ray effect is due to the difference in the distribution function at the velocity points located along the line with discontinuities.
In the one dimensional case, when the discontinuity line moves along $u$ direction as shown in Fig~\ref{fig:rayleigh_vs}, the same number of discrete velocity points (the number of velocity discretization in $v$ direction) go across the discontinuity line for each jump.
However, for the two dimensional cavity flow, since the discontinuity lines are along the radial direction $\theta$, there are different numbers of discrete velocity points lying on the discontinuity line along different $\theta$ directions for the case with rectangular grid points in the velocity space.
There are more points lying in the directions of $\theta = 0\degree$, $45\degree$ and $90\degree$ with jumps than those on the other directions.
As a result, the ray effect will appear severely along the diagonal, horizontal and vertical directions, which can be observed in Fig.~\ref{fig:cavity_collisionless_temperature} in the central domain and the region near the top boundary.
Actually, this is due to the fact that with regular discrete velocity points along the directions of $0\degree$, $45\degree$ and $90\degree$ the local angular resolution $\Delta \theta$ in the velocity space is lower than that in the other irregular directions.
For comparative study, the cavity flow with an unstructured discretization in velocity space is computed.
As shown in Fig.~\ref{fig:cavity_unstructured_vs}, the distribution function at the location $(0.5,0.5)$ is plotted for the case with unstructured mesh points in the velocity space, where the areas of the triangles are basically equal to those in the previous uniform mesh and the integration weight $\omega$ is almost the same.
Since the angular resolution is approximately homogeneous for the unstructured velocity space, the enhanced ray effect along the diagonal directions and near the top boundary disappear, i.e., the results in Figs.~\ref{fig:cavity_unstructured_density} and \ref{fig:cavity_unstructured_temperature}.
In addition, in order to get rid of the influence of the geometry in the studies,
we present the solutions for cavities with different widths and heights for comparisons.
It can be found that the ray effect is still enhanced in the directions of $45\degree$, $0\degree$ and $90\degree$ instead of along the geometric diagonal direction.
This observation confirms the enhancement of the ray effect in specific directions is due to the rectangular discretization of the velocity space instead of the geometric configuration of computational domain in physical space.

\begin{figure}[H]
	\subfigure[Density]
	{\includegraphics[width=0.32\textwidth]{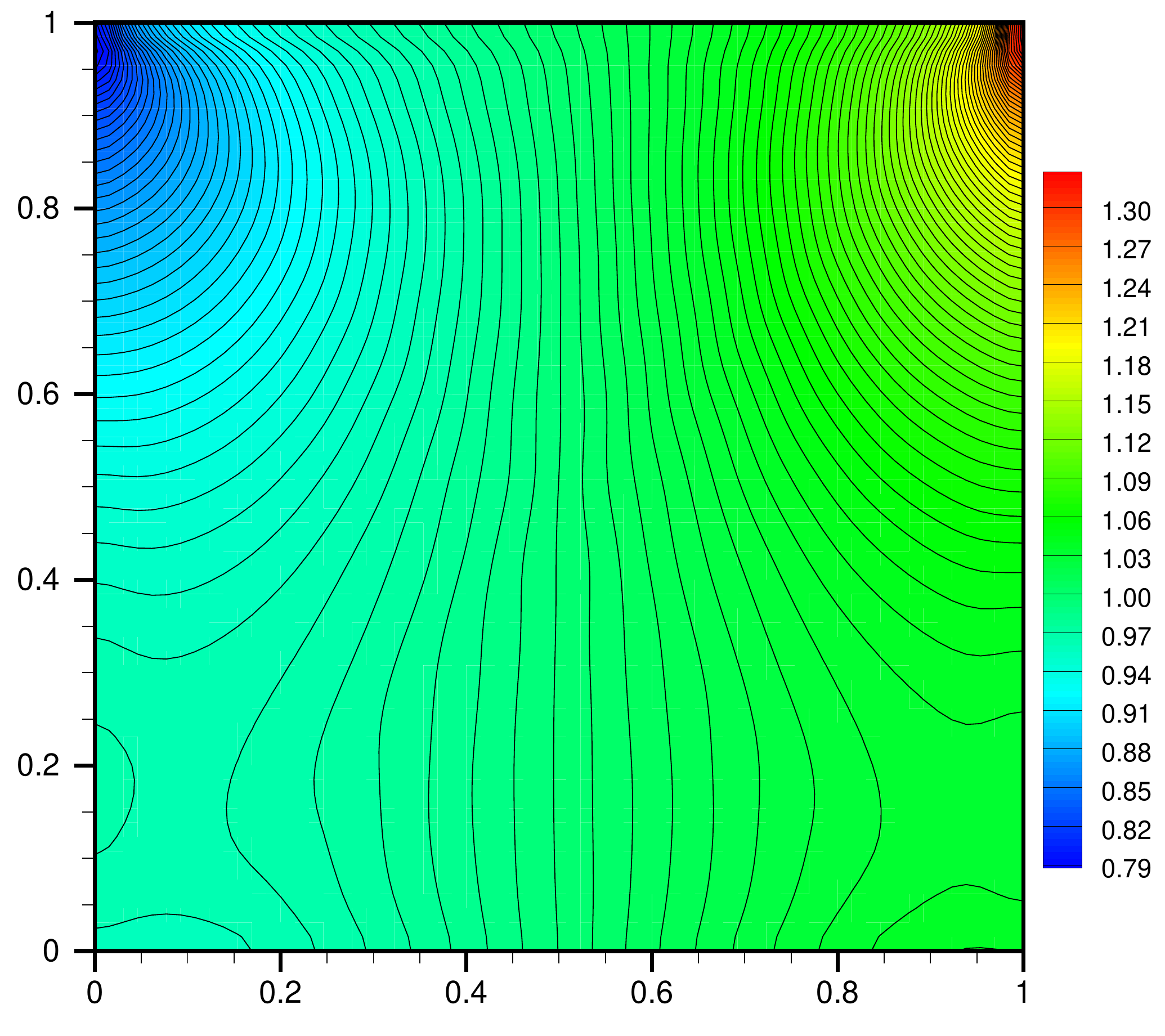}}
	\subfigure[Temperature]
	{\includegraphics[width=0.32\textwidth]{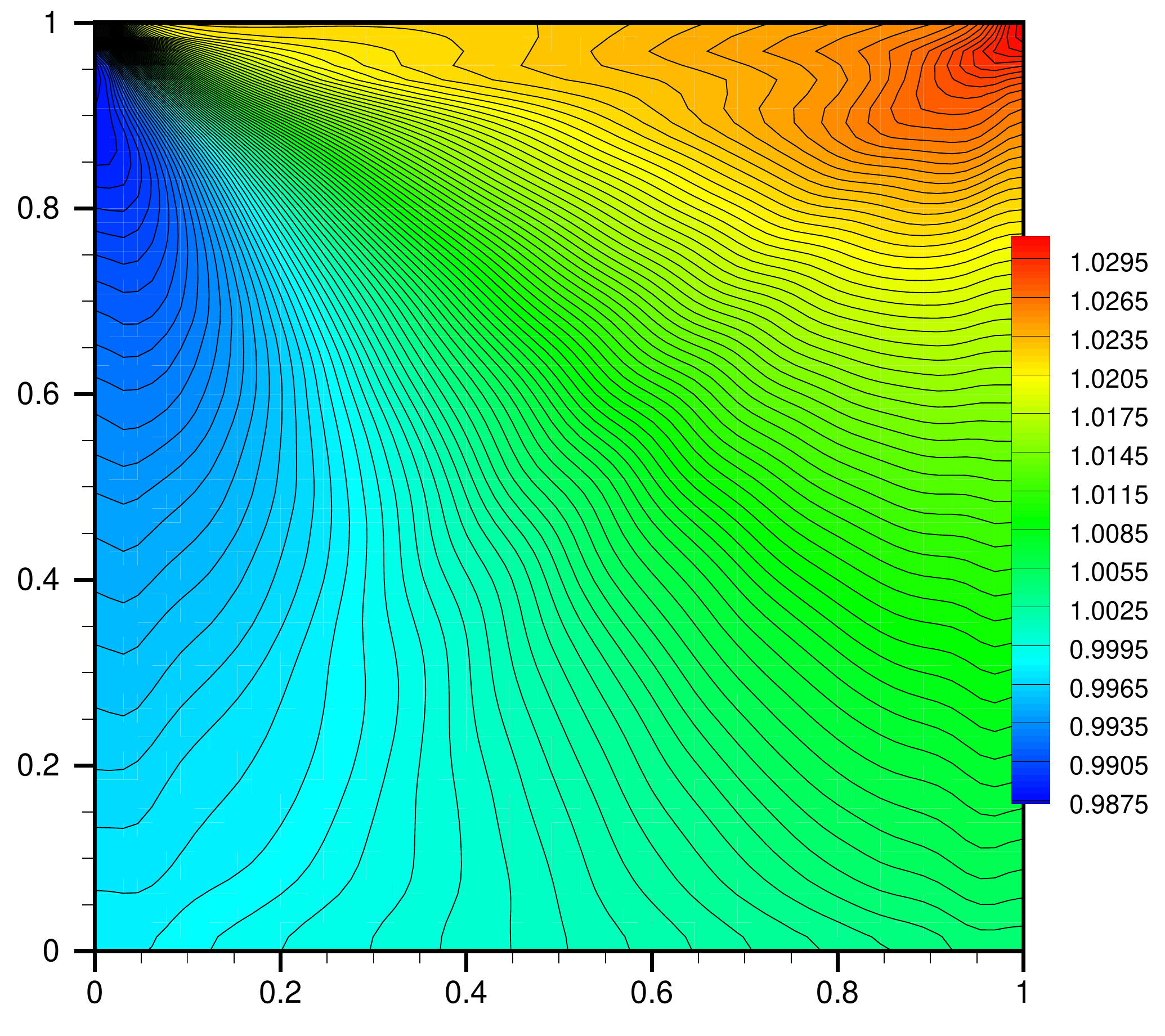}}
	\subfigure[Velocity]
	{\includegraphics[width=0.32\textwidth]{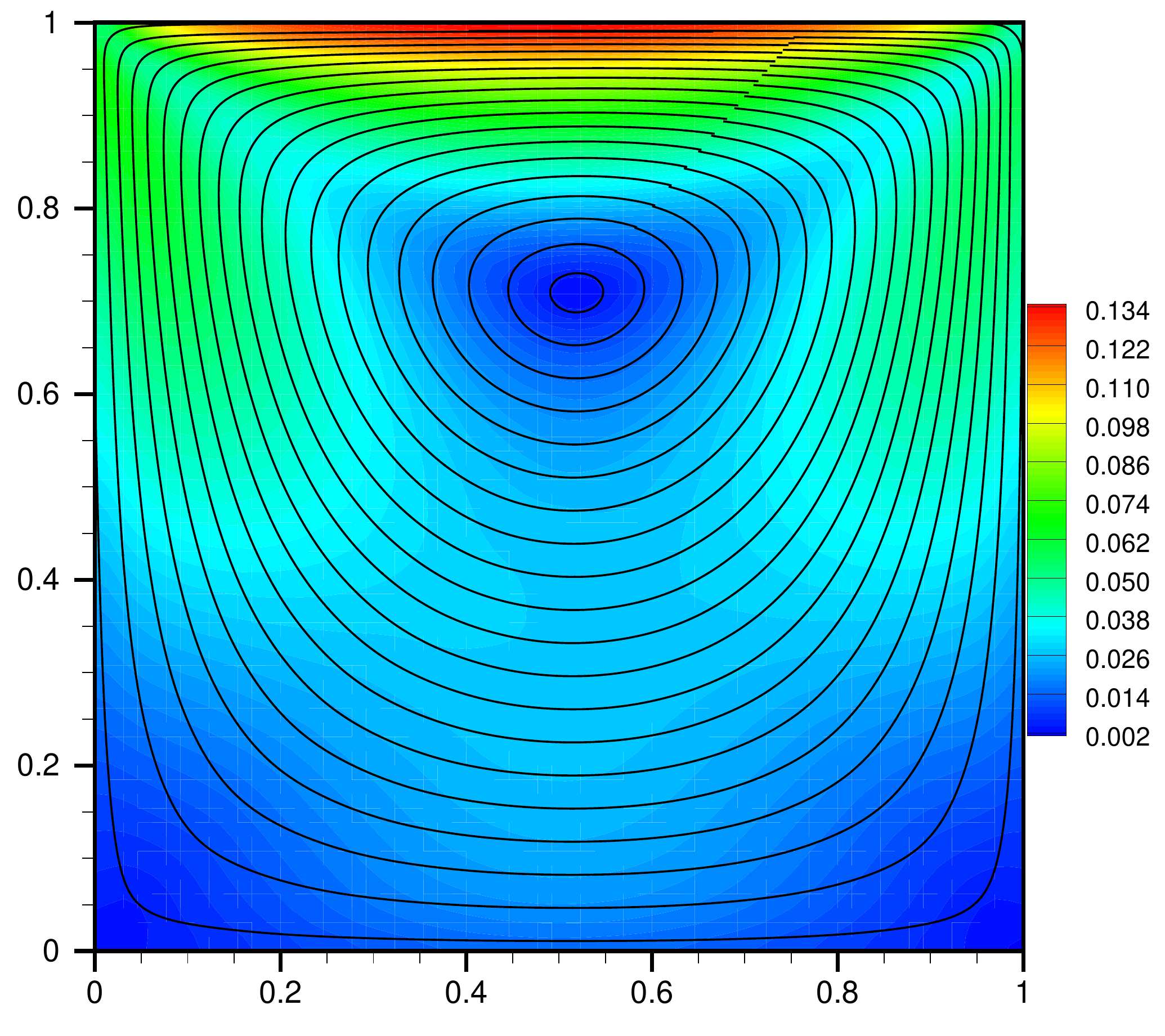}}
	\caption{\label{fig:cavity_collisional} Cavity flow at ${\rm Kn}=0.5$.}
\end{figure}

At smaller Knudsen numbers with increased particle collisions, the cavity flow at ${\rm Kn}=0.5$ is computed.
In Fig.~\ref{fig:cavity_collisional}, it can be seen that the ray effect is much reduced due to particles' collision which gradually smears the   reflected discontinuities from the wall.

\section{Mitigation solutions}\label{sec:mitigation}
Here, we rewrite the formula to compute the jump of ray effect, i.e., the numerical integration of differences in the distribution function along the corresponding discontinuity line,
\begin{equation}\label{eq:ray_effect_jump}
	\vec{w}(\vec{x}_m)= \sum_{h_m(\vec{u}_k)=0} {\vec{\psi}_k \Delta g_k \omega_k},
\end{equation}
where $h_m(\vec{u})=0$ denotes the $m$-th discontinuity line in the velocity space, for example, $u = u_m$ for Sod and Rayleigh cases and $\theta(\vec{u}) = \theta_m$ for cavity flows.
$\vec{x}_m$ is the physical location where the jump in the ray effect occurs due to the sudden change of the distribution function $\Delta g_k$ on the $m$-th discontinuity line.

From the previous theoretical analysis and numerical validation, we know that the cause of ray effect is due to low resolution in velocity space.
According to Eq.~(\ref{eq:ray_effect_jump}), we could obtain several principles for adjusting the discretization of velocity space to effectively mitigate the ray effect.
These are summarized as follows.
\begin{description}
	\item[A.] Refine the velocity point to reduce the integration weights $\omega_k$, especially for the regions where the moments of distribution function difference $\Delta g$ are large.
	\item[B.] Add more discrete velocity points in the direction across the discontinuity line in the velocity space, i.e., perpendicular to the discontinuity line $h_m(\vec{u})=0$, which is to reduce the step of ray effect and meanwhile to decrease the jumps by increasing the resolution corresponding to a smaller $\omega_k$.
	\item[C.] Reduce the number of discrete velocity points lying simultaneously on (across) the discontinuity line $h_m(\vec{u})=0$ in the velocity space, to diminish their contributions to the macroscopic flow variable jumps in Eq.~(\ref{eq:ray_effect_jump}).
\end{description}
Similar discussions can be found in radiative transfer in the discretization of angular variables \cite{li2003mitigation}.
In addition, the mitigation methods which are equivalent to obtaining the solutions with more discrete velocities can be employed as well, such as taking averaged solutions over multiple times of computations on different sets of velocity space discretization \cite{cumber2000jet,tencer2016ray,camminady2019ray}.

Based on the above analysis, it is clear that the refinement of the velocity discretization can substantially reduce the ray effect.
However, considering practical numerical simulations, there are other factors which could influence the ray effect as well,
such as the spatial discretization, accuracy of numerical scheme, particle collision, and the integration method.
On the discretized space, visual observation of ray effect can be attributed to the incompatibility between low resolution in the velocity space and high resolution in the physical space.
Therefore, if the requirement for spatial resolution is not too high, employing coarser mesh would be a very effective way to avoid the spatial distortion in the solution induced by inadequate resolution in the velocity space.
Similarly, low-order accurate numerical scheme can be used to reduce the spatial resolution, smear the discontinuity in the phase space, and to get  a smoother result.

Since the ray effect is more severe in the highly rarefied regime and the particle collision always reduces the ray effect, only the collisionless case will be considered in the following discussions.
Different integration methods, such as trapezoidal rule and Newton--Cotes method,  just provide different jumps through their different integration coefficients, and they will present the similar ray effect.
Therefore, we will not further consider the difference between integration methods.

In the following, we will give numerical results obtained by DVM with the inclusion of the above mitigation techniques.
The solutions obtained by stochastic particle method will be discussed as well.
Since only collisionless cases are considered, the numerical schemes based on discrete velocities, such as the DVM \cite{yang1995rarefied,mieussens2000discrete}, GKUA \cite{li2004gkua}, UGKS \cite{xu2010unified,xu2015direct} and DUGKS \cite{guo2013discrete,guo2015discrete} will give the same results; and all the stochastic particle methods such as DSMC \cite{bird1994book,pareschi2000asymptotic}, USP \cite{fei2018particle}, UGKP \cite{li2018ugkp} and UGKWP \cite{liu2018ugkwp,zhu2019ugkwp,li2019ugkwp} will become the same particle tracking schemes.

\subsection{Sod test case}

For the Sod test case, the mitigation techniques for the ray effect are employed to give better numerical solutions.
For this case, a uniform mesh with $1000$ cells is used in the physical space and $61$ discrete velocity points are used to cover the velocity space $\left[-10, 10\right]$ and  take trapezoidal integration method.
The time step is $5\times10^{-5}$ and the output time is $t_e = 0.15$.

\begin{figure}[H]
	\centering
	\subfigure[Density]{\includegraphics[width=0.32\textwidth]{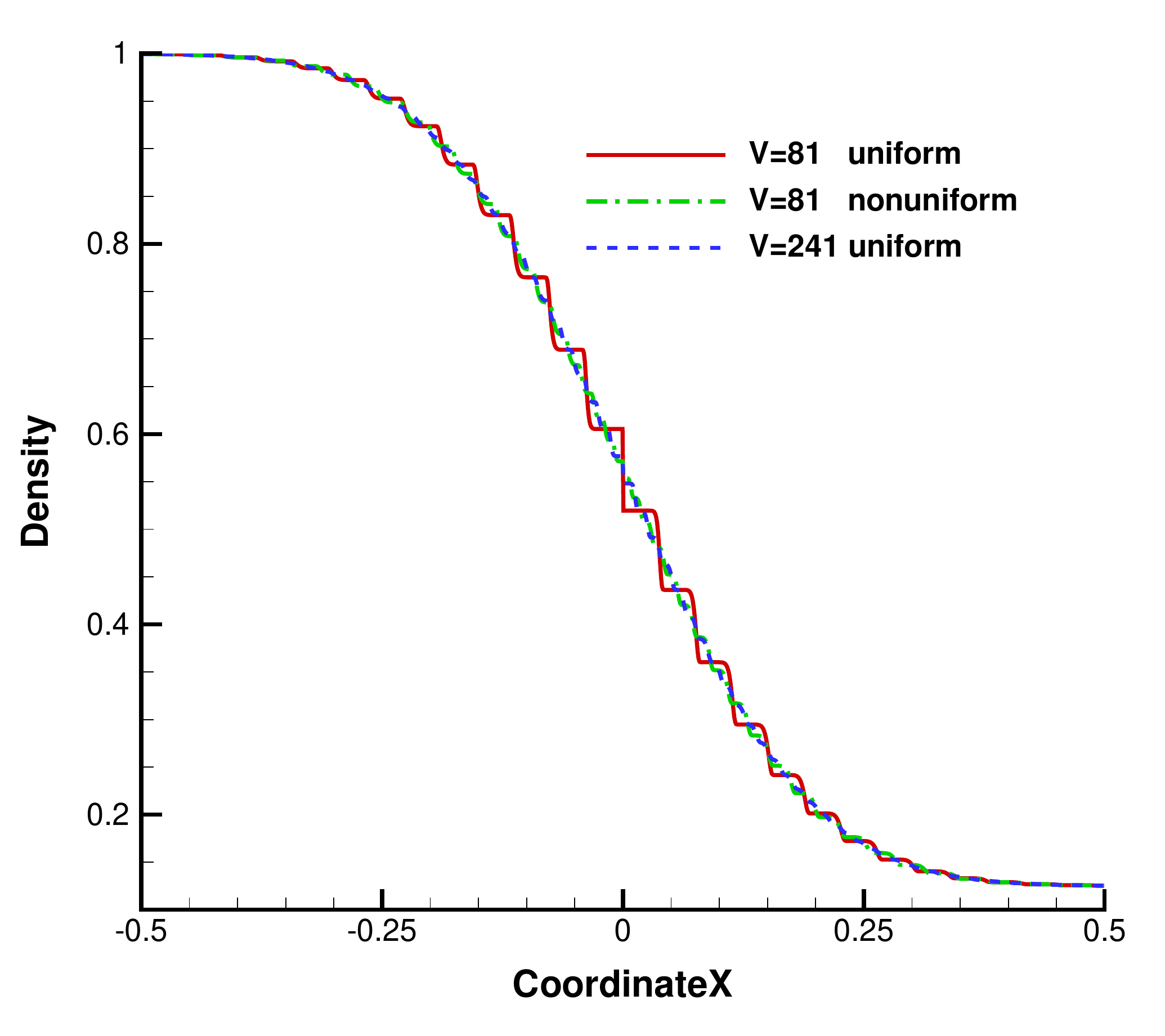}}
	\subfigure[Velocity]{\includegraphics[width=0.32\textwidth]{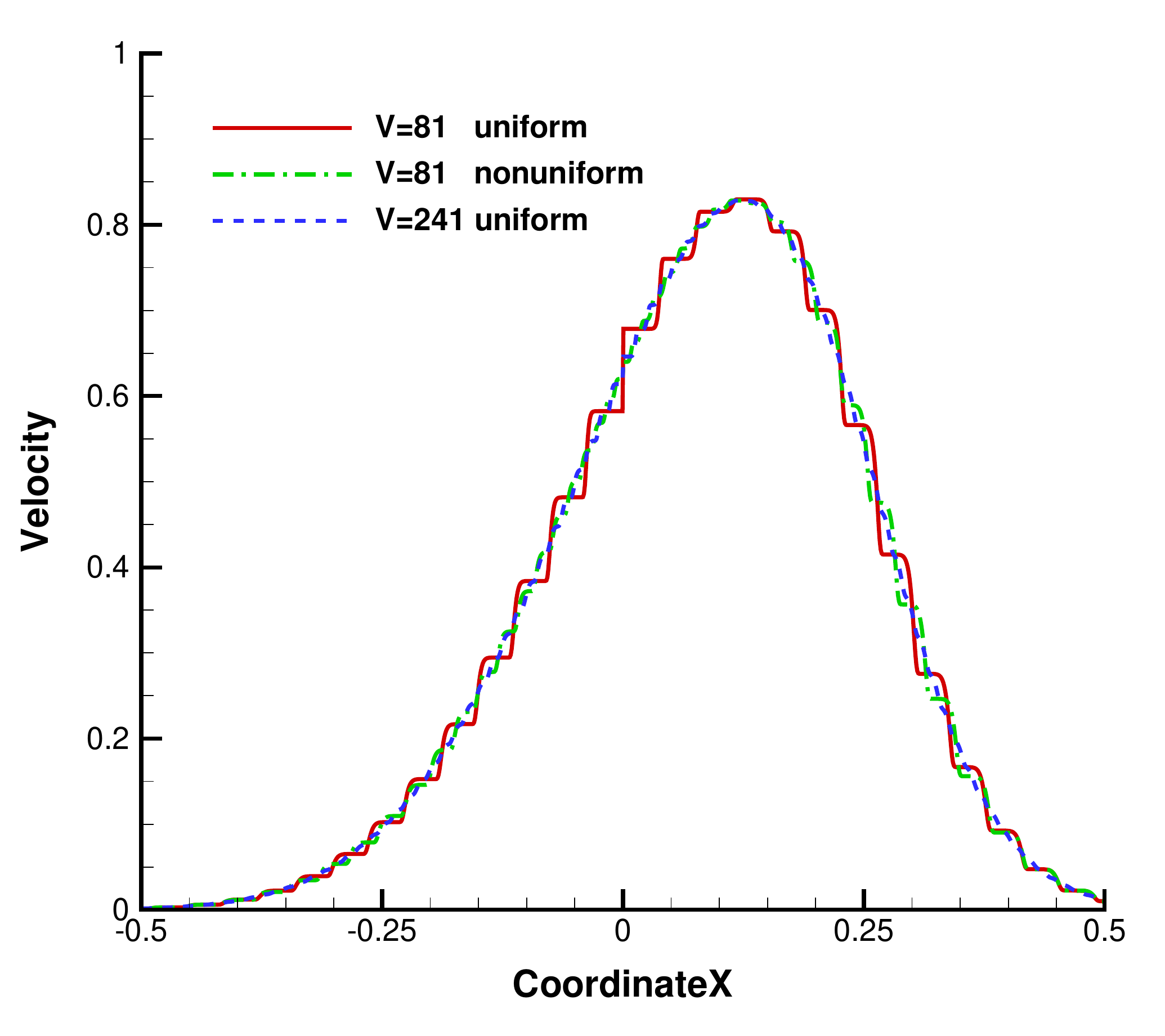}}
	\subfigure[Temperature]{\includegraphics[width=0.32\textwidth]{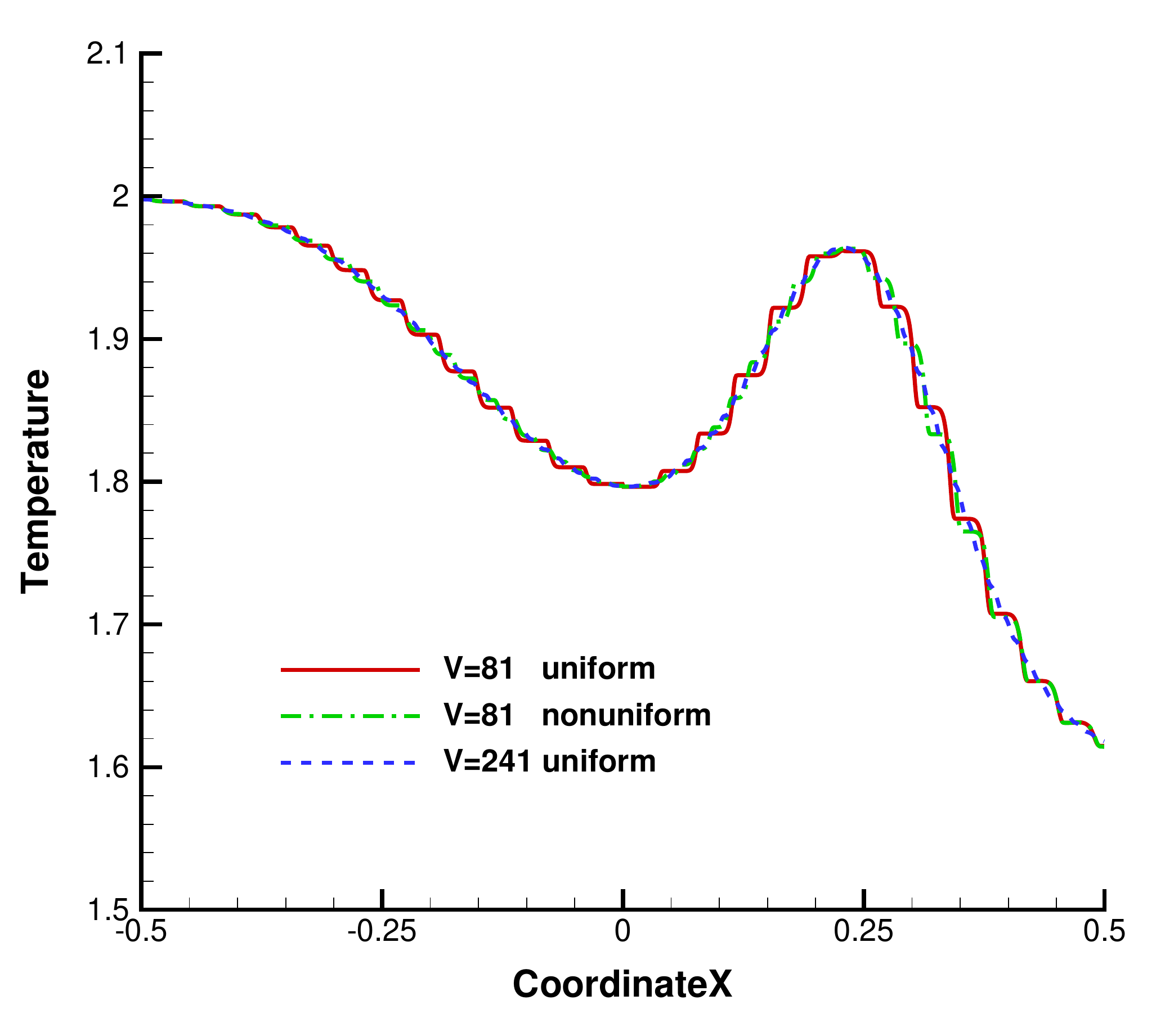}}
	\caption{\label{fig:sod_vs}The Sod test cases with refined mesh in the velocity space.}
\end{figure}

In Fig.~\ref{fig:sod_vs}, we show the results of the Sod test at ${\rm Kn}\to \infty$ on a refined velocity mesh,
where  uniform meshes with $81$ and $241$ discrete velocity points and a nonuniform mesh with $81$ points refined near $u_k = 0$  are employed.
It shows that  the ray effect are effectively reduced by a fine mesh with $241$ discrete points in the velocity space.
Since the difference in the distribution function has larger values around the zero velocity point, the local refinement around $u_k=0$ can mitigate the density jump as well.
However, it should be noted that for higher order moments of the distribution function, such as the velocity and temperature, the moments of the distribution function are different from the density jump evaluation, so the local refinement around the zero velocity point is less effective for the high order quantities in the ray effect.

\begin{figure}[H]
	\centering
	\subfigure[Density]{\includegraphics[width=0.32\textwidth]{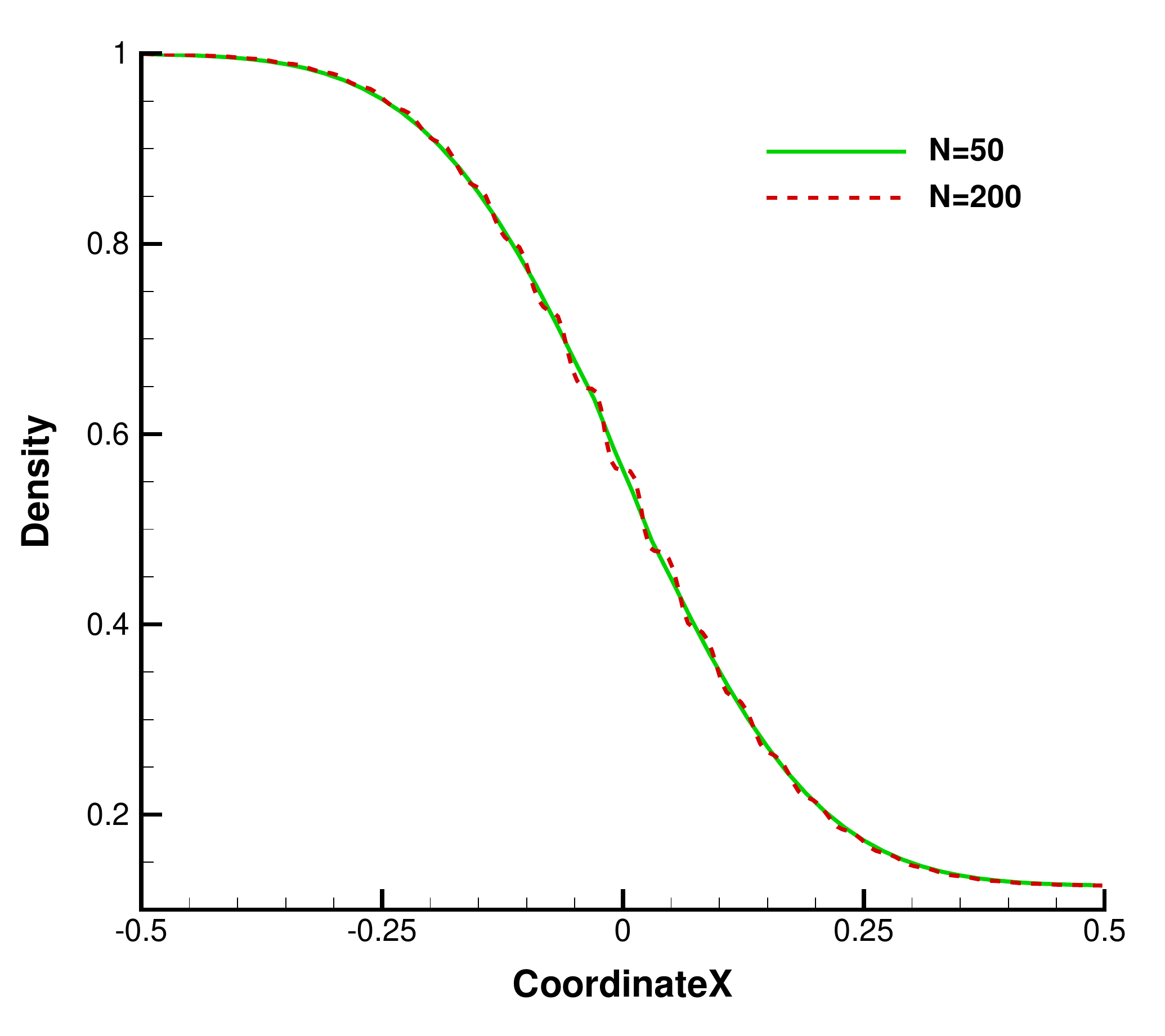}}
	\subfigure[Velocity]{\includegraphics[width=0.32\textwidth]{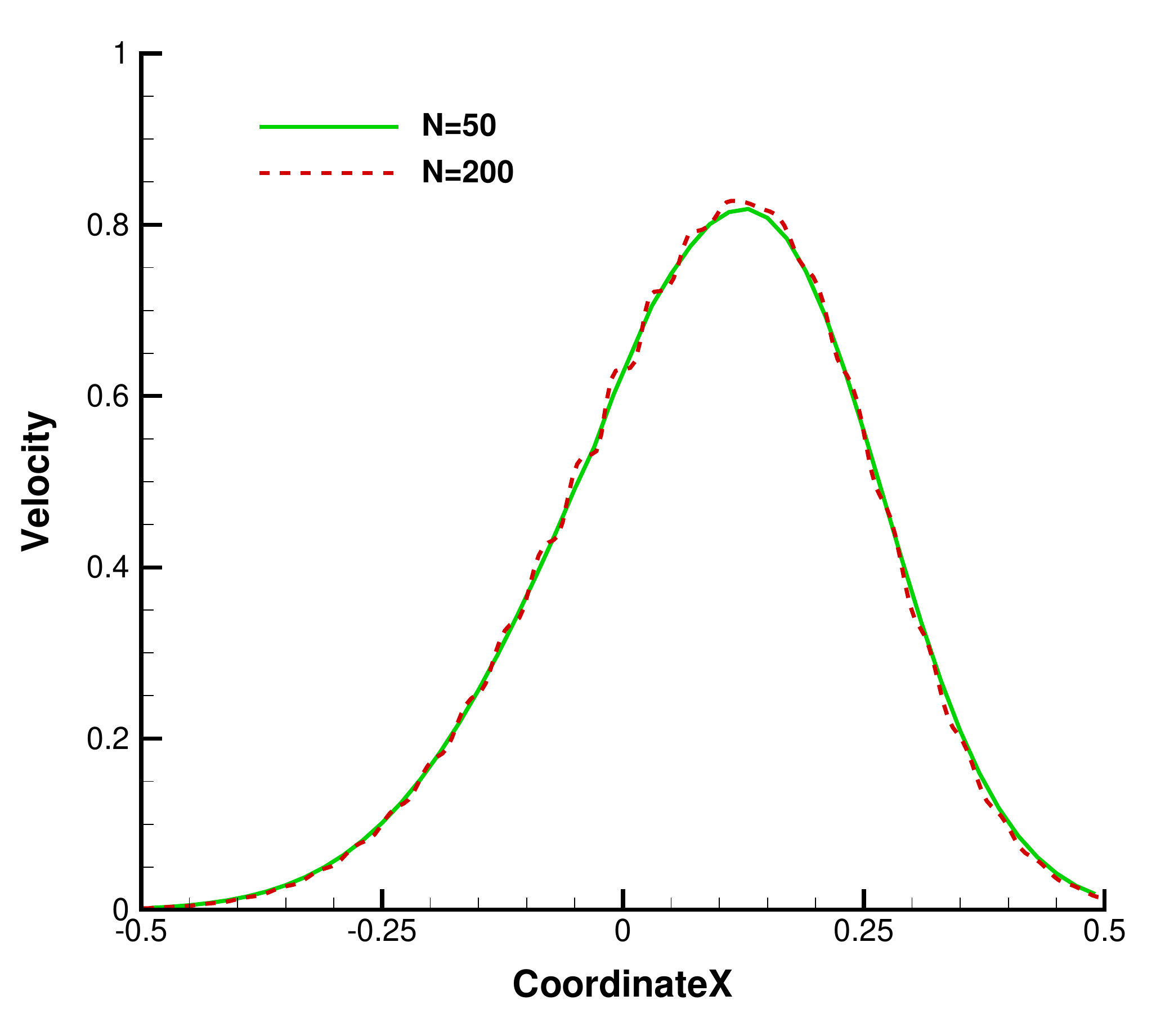}}
	\subfigure[Temperature]{\includegraphics[width=0.32\textwidth]{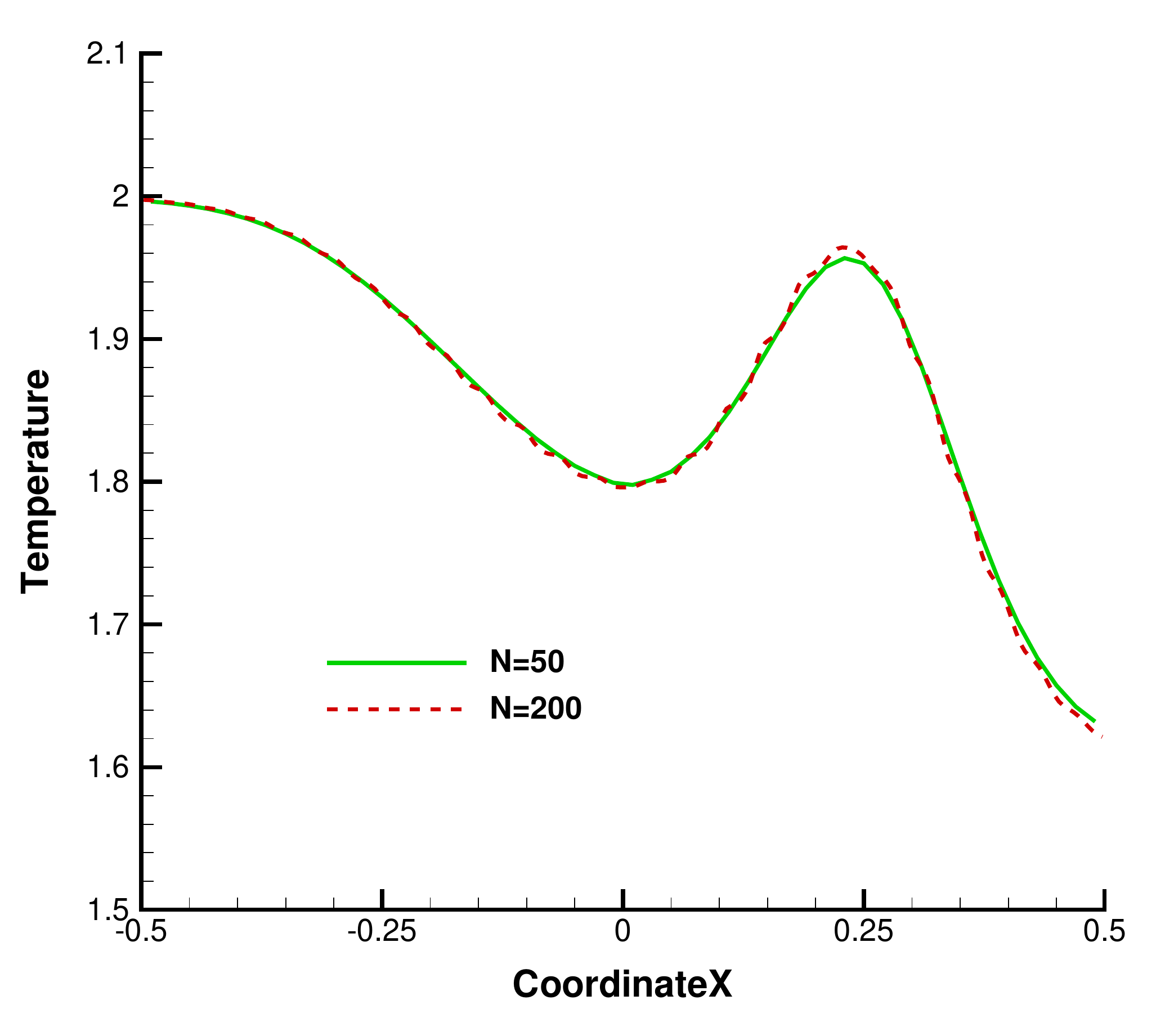}}
	\caption{\label{fig:sod_mesh}Numerical solutions for Sod test case at ${\rm Kn}\to \infty$ on different meshes.}
\end{figure}

With lower numerical resolution in the physical space, the ray effect can be mitigated visually on the coarse meshes with $50$ and $200$ cells as shown in Fig.~\ref{fig:sod_mesh}, where the $80$ velocity points are employed in the velocity space.
Theoretically, using a low spatial resolution is a passive way to mitigate the ray effect because it does not solve the problem essentially,
but unresolves the detailed distorted flow structure.
However, in the aspect of numerical simulations, as long as the numerical solutions with low spatial resolution are acceptable,  using coarser mesh is a very effective way to reduce the ray effect and the computational efficiency can be improved.
As pointed out, the limited resolution on the sub-scale between grid points in the velocity space will result in uncertainty in the physical space on the scale of $t\Delta u$.
The detailed flow structure within $t\Delta u$ cannot be resolved even with a fine mesh in the physical space.
Thus, for the numerical simulations of highly rarefied flow,   compatible discretization of velocity space and physical space are required, such as
$\Delta x \ge t_e\Delta u$ or $\Delta u \le \Delta x / t_e$.

\begin{figure}[H]
	\centering
	\subfigure[Density]{\includegraphics[width=0.32\textwidth]{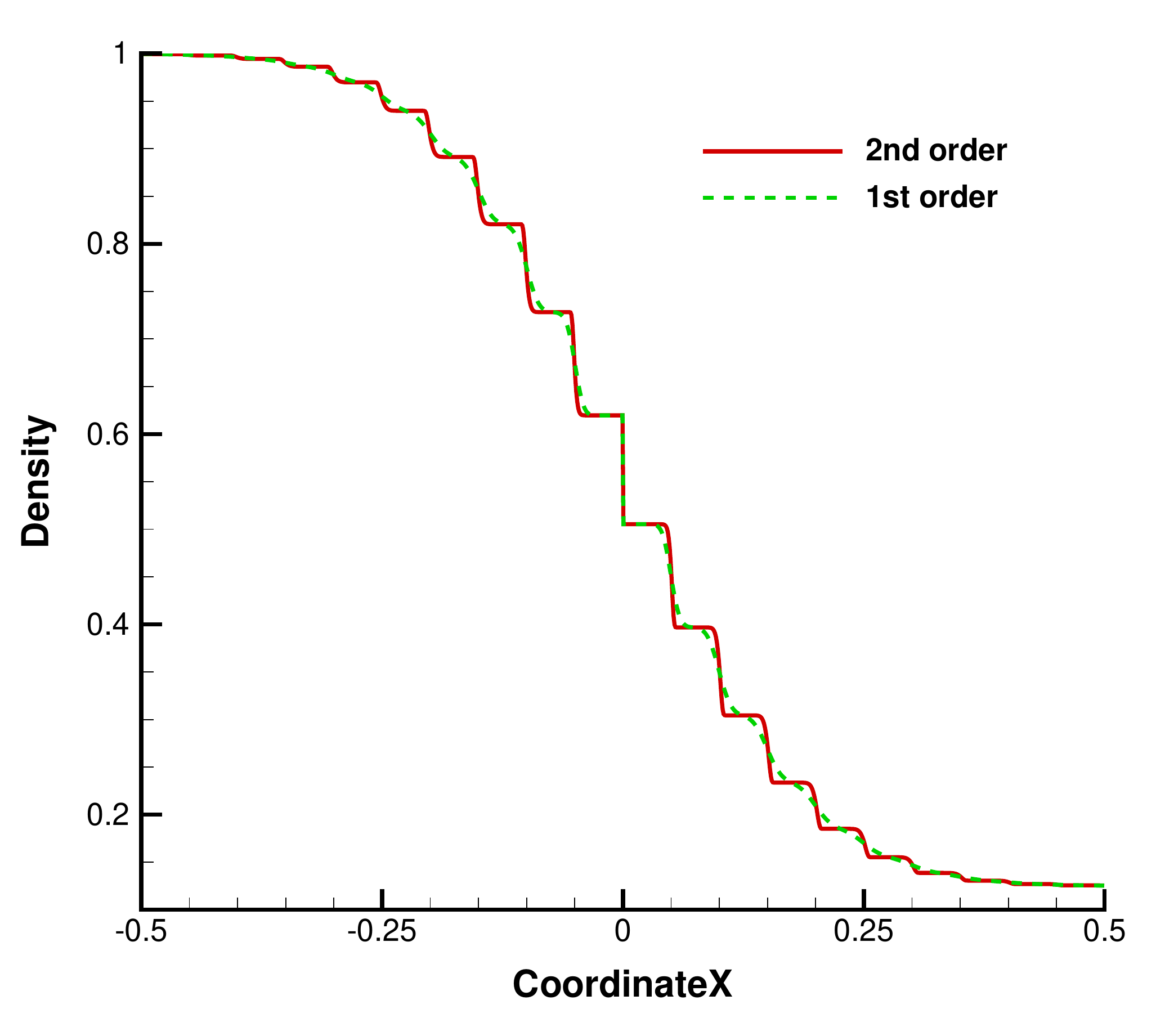}}
	\subfigure[Velocity]{\includegraphics[width=0.32\textwidth]{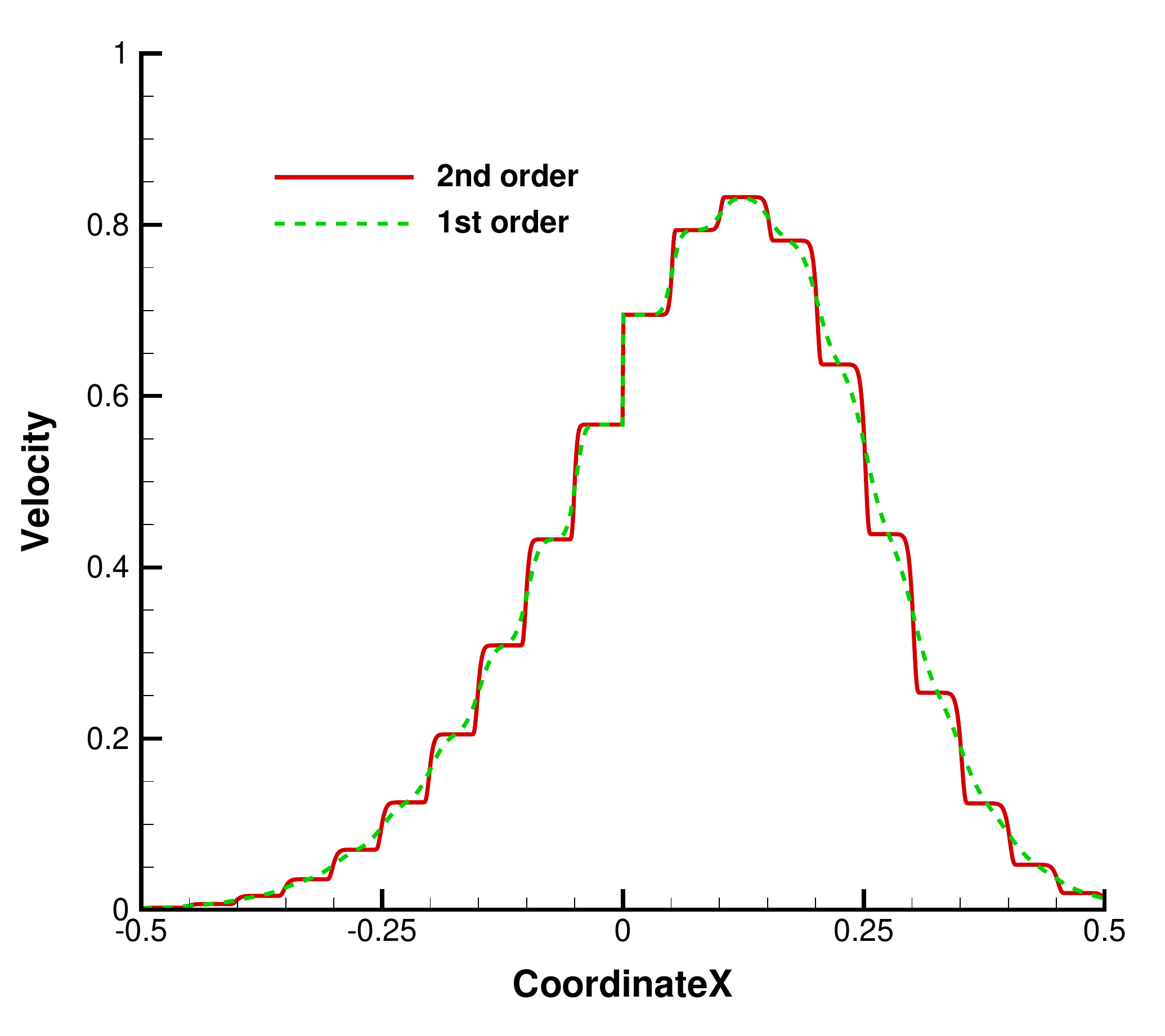}}
	\subfigure[Temperature]{\includegraphics[width=0.32\textwidth]{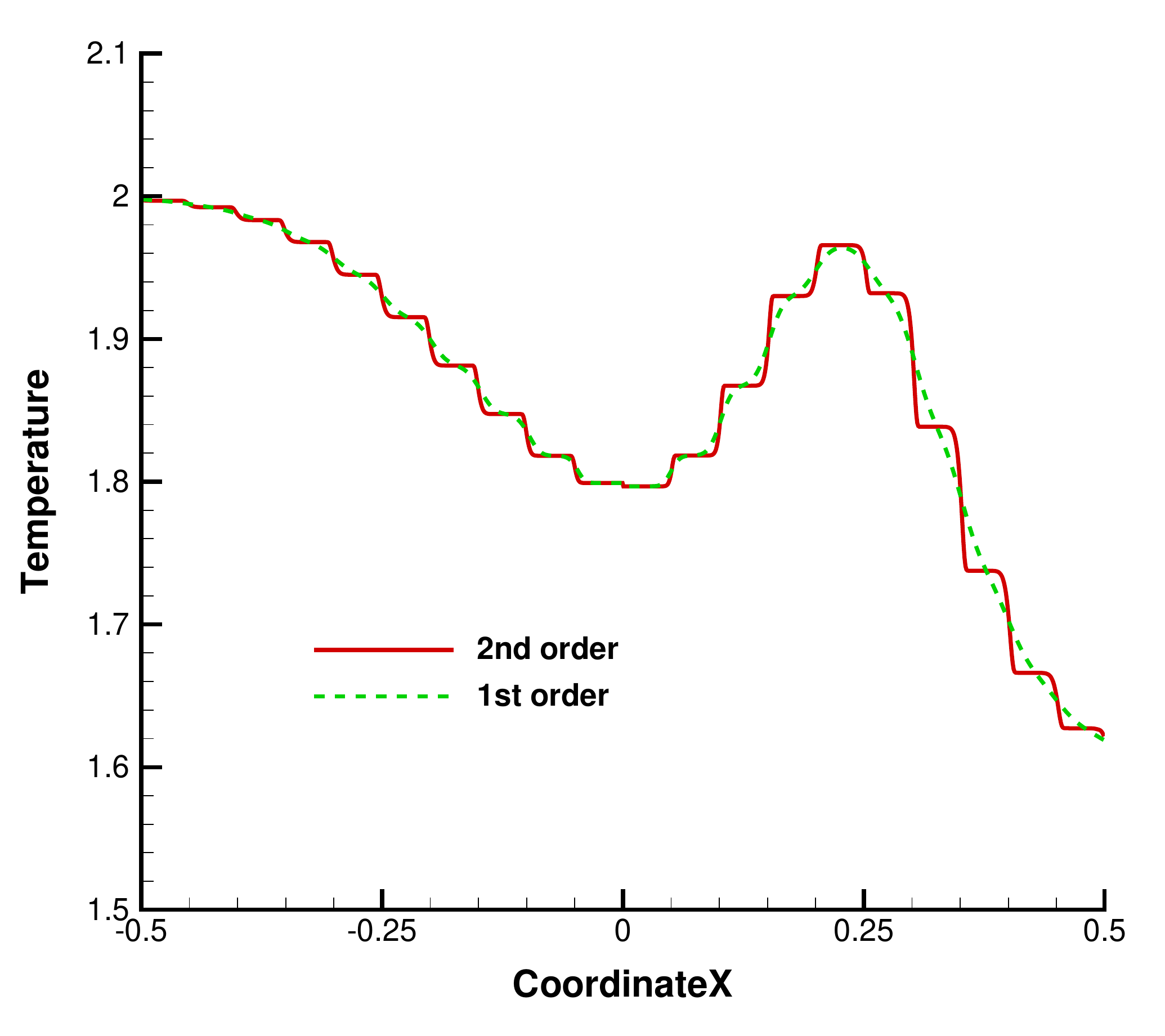}}
	\caption{\label{fig:sod_1st}Numerical solutions for Sod test case at ${\rm Kn}\to \infty$ obtained by the first-order and second order schemes.}
\end{figure}

Similar to using a coarse mesh in the physical space, it is helpful to reduce the ray effect as well by adopting a lower order  numerical scheme without resolving the discontinuities due to the enhanced numerical dissipation.
The solutions obtained by the first-order and second-order numerical schemes are plotted in Fig.~\ref{fig:sod_1st}.
It can be found that the first-order scheme obtains smoother results than the second-order one,
especially for the jumps away from the initial discontinuity,
because the discontinuity line in the phase space is widened by the numerical dissipation and finite volume average.
This is more obvious for the distribution function at large discrete velocities, which have crossed more cells.
It should be pointed out that this remedy doesn't work at the point $u_k=0$, i.e., the jump at $x=0$, because the distribution function discretized at $u_k=0$ has no contribution in the time evolution of numerical solutions,
and the jump caused by the discontinuity at $u_k=0$ remains.
Therefore, for numerical simulations with discrete distribution function, it is better not to include the zero velocity points, not only because of the reason given above, but also due to the fact that the transport property of the low speed particles with both positive and negative microscopic velocities cannot be represented on the average by the evolution of the distribution function exactly at $u_k=0$.

\subsection{Rayleigh flow}\label{sec:mitigation_rayleigh}

The Rayleigh flow in the collisionless limit is re-calculated by implementing the mitigation techniques for the ray effect.
The default setup is the same as that given in Section \ref{sec:numerical_rayleigh}.

\begin{figure}[H]
	\centering
	\subfigure[Density]{\includegraphics[width=0.32\textwidth]{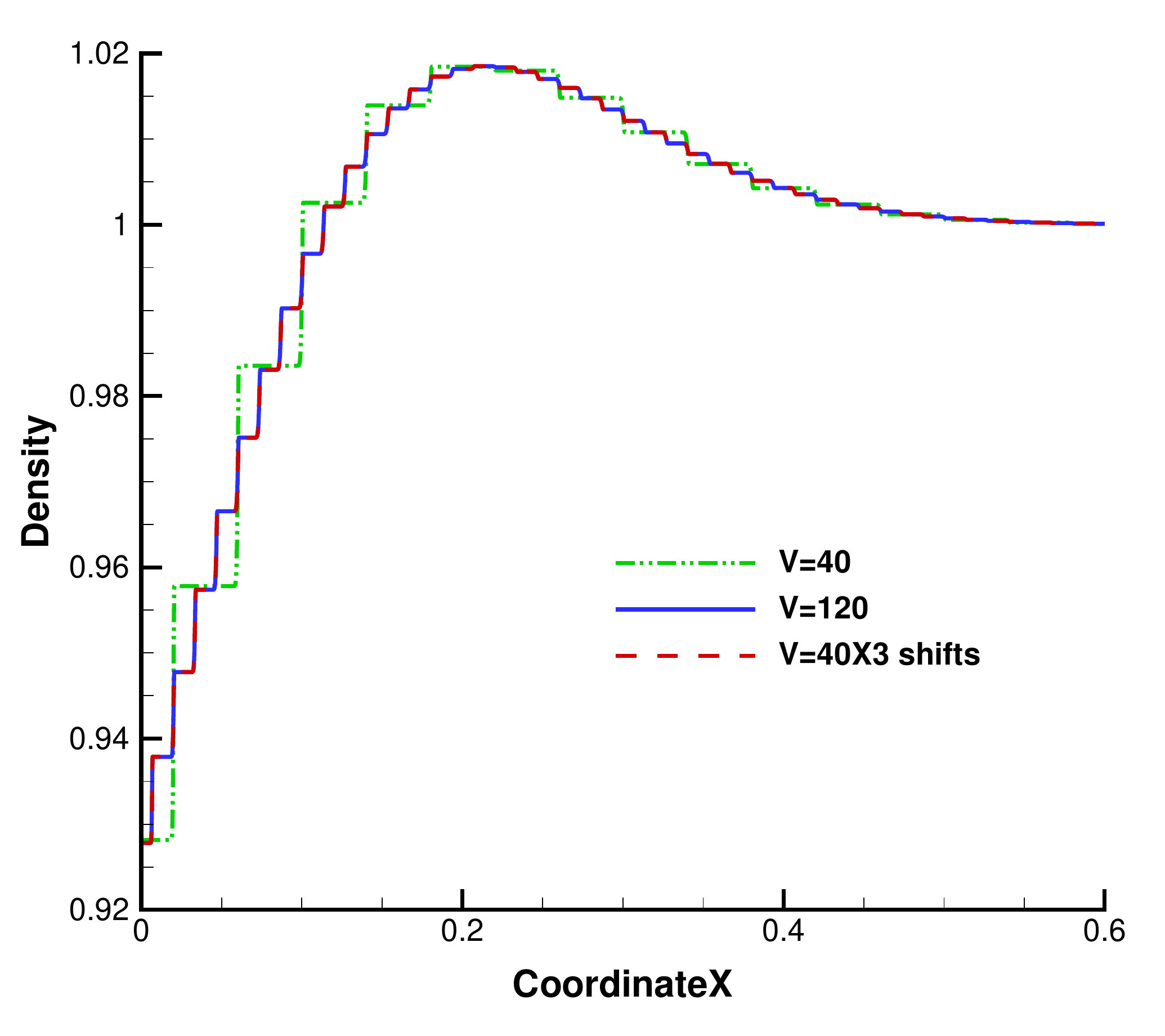}}
	\subfigure[Velocity]{\includegraphics[width=0.32\textwidth]{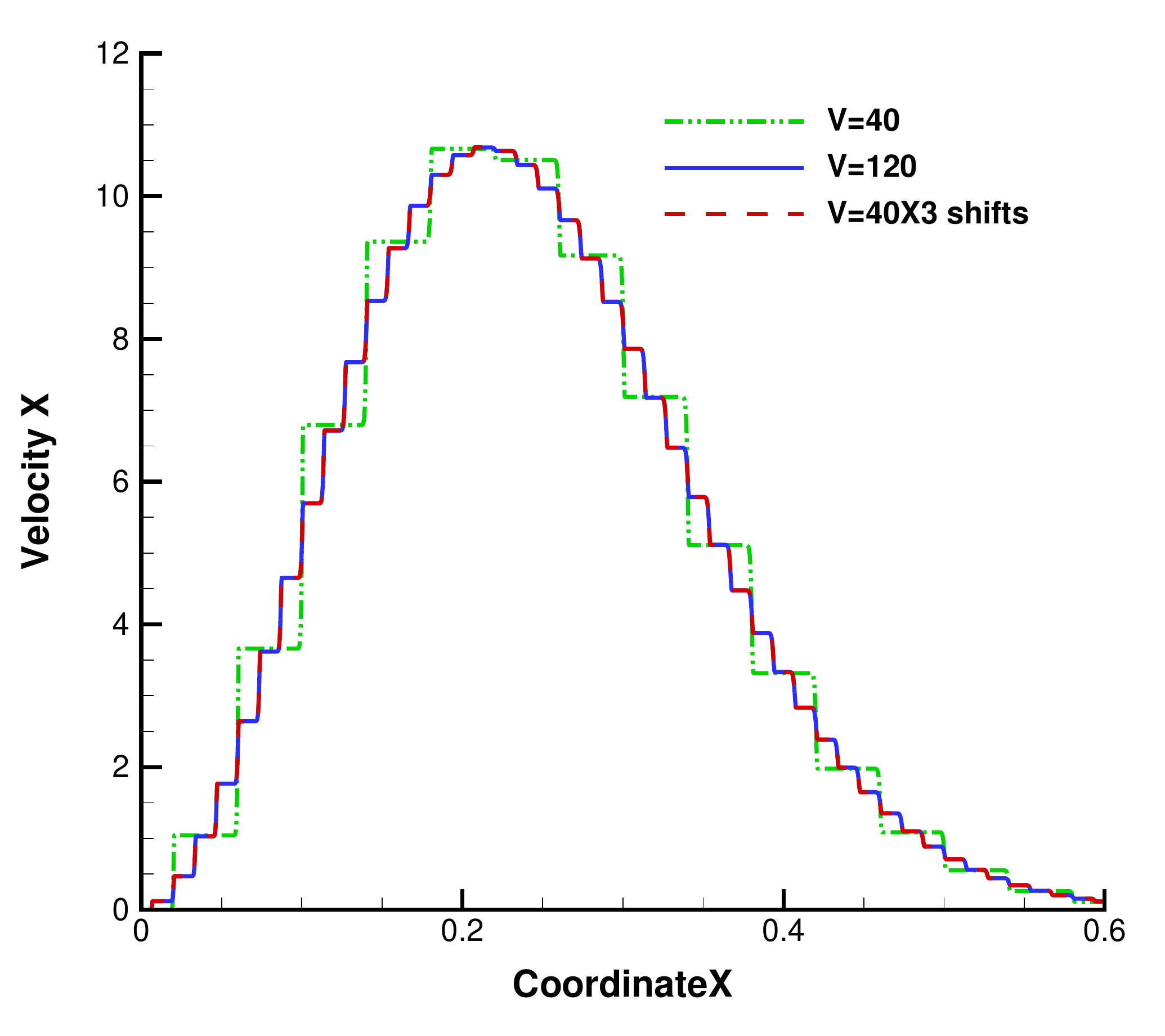}}
	\subfigure[Temperature]{\includegraphics[width=0.32\textwidth]{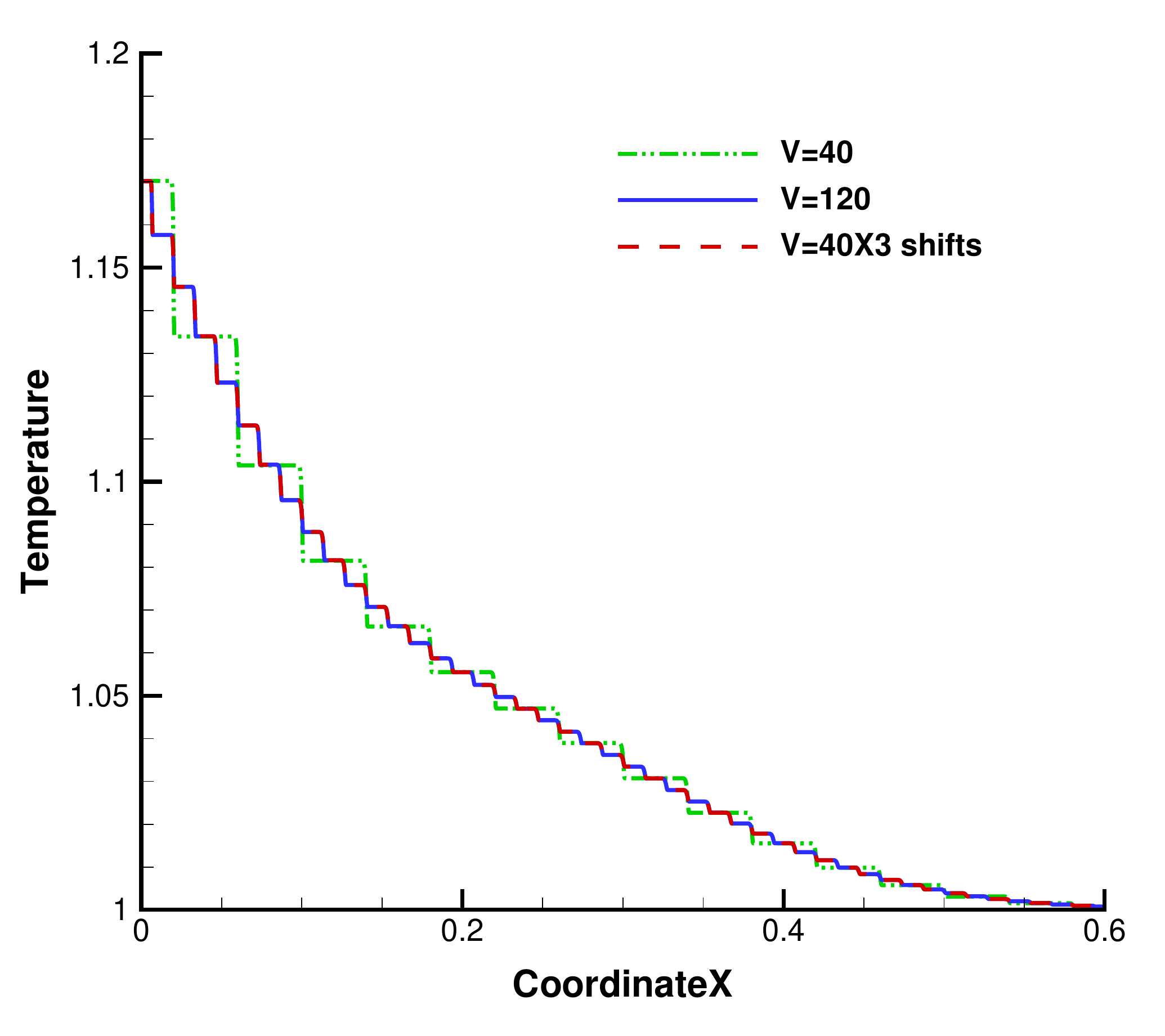}}
	\caption{\label{fig:rayleigh_vs_shift} Rayleigh flow at ${\rm Kn}\to\infty$ with refined velocity space. The red dashes line denotes the averaged solution on three different discretizations of the velocity space, which are obtained by shifting the discrete velocity points with $-\Delta u / 3$, $0$ and $\Delta u / 3$.}
\end{figure}

Figure \ref{fig:rayleigh_vs_shift} shows the result of the Rayleigh flow with refined velocity space.
The solutions obtained with $120\times40$ discrete velocity points seem better than that with $40\times40$ points.
We also shift the $40\times40$ discrete velocity points in $u$ direction by $-\Delta u / 3$, $0$ and $\Delta u / 3$.
It shows that the averaged solutions from the shifted discrete velocity points are identical to those with refined velocity space
because more discrete velocity points are equivalently considered in the final averaged solution.
In \cite{varoutis2008application}, it is claimed that the ray effect cannot be eliminated by simply increasing the number of discrete velocity points, since with the decreasing of the oscillation amplitude the frequency may be increased.
However, from the analysis and observation, we would like to regard the ray effect as spatial distortions rather than oscillations because the solution does not vibrate in space and time.
Therefore, the description and analysis using the frequency is not fully appropriate.
Ray effect is more like the discrete velocity induced mosaic in the physical space; 
therefore, the step in the ray effect will get smaller with the increase of discrete velocity points. 
The numerical solutions will approach to the analytic solution once a continuous velocity space is being recovered in the grid refinement process.

\begin{figure}[H]
	\centering
	\subfigure[Density]{\includegraphics[width=0.32\textwidth]{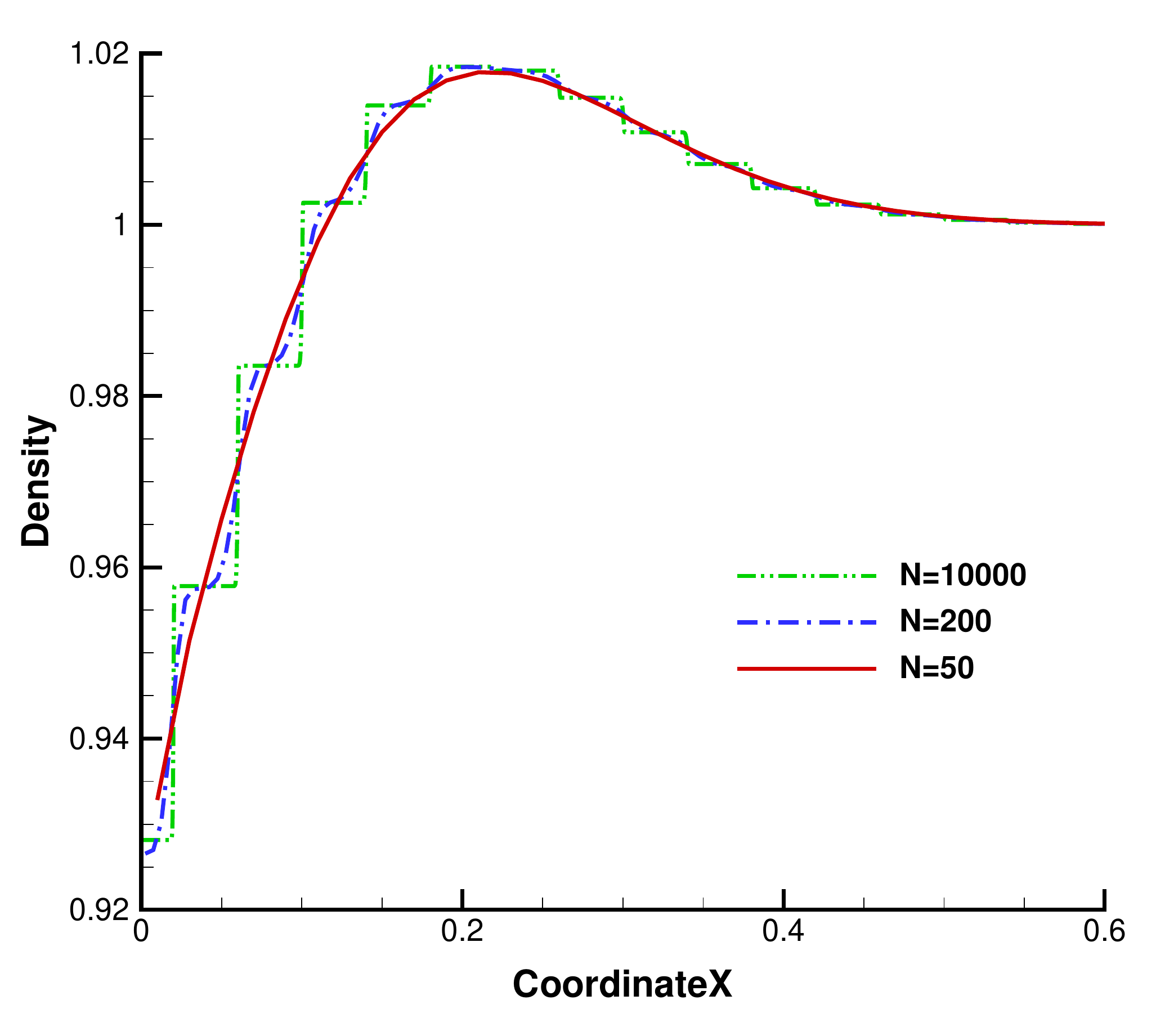}}
	\subfigure[Velocity]{\includegraphics[width=0.32\textwidth]{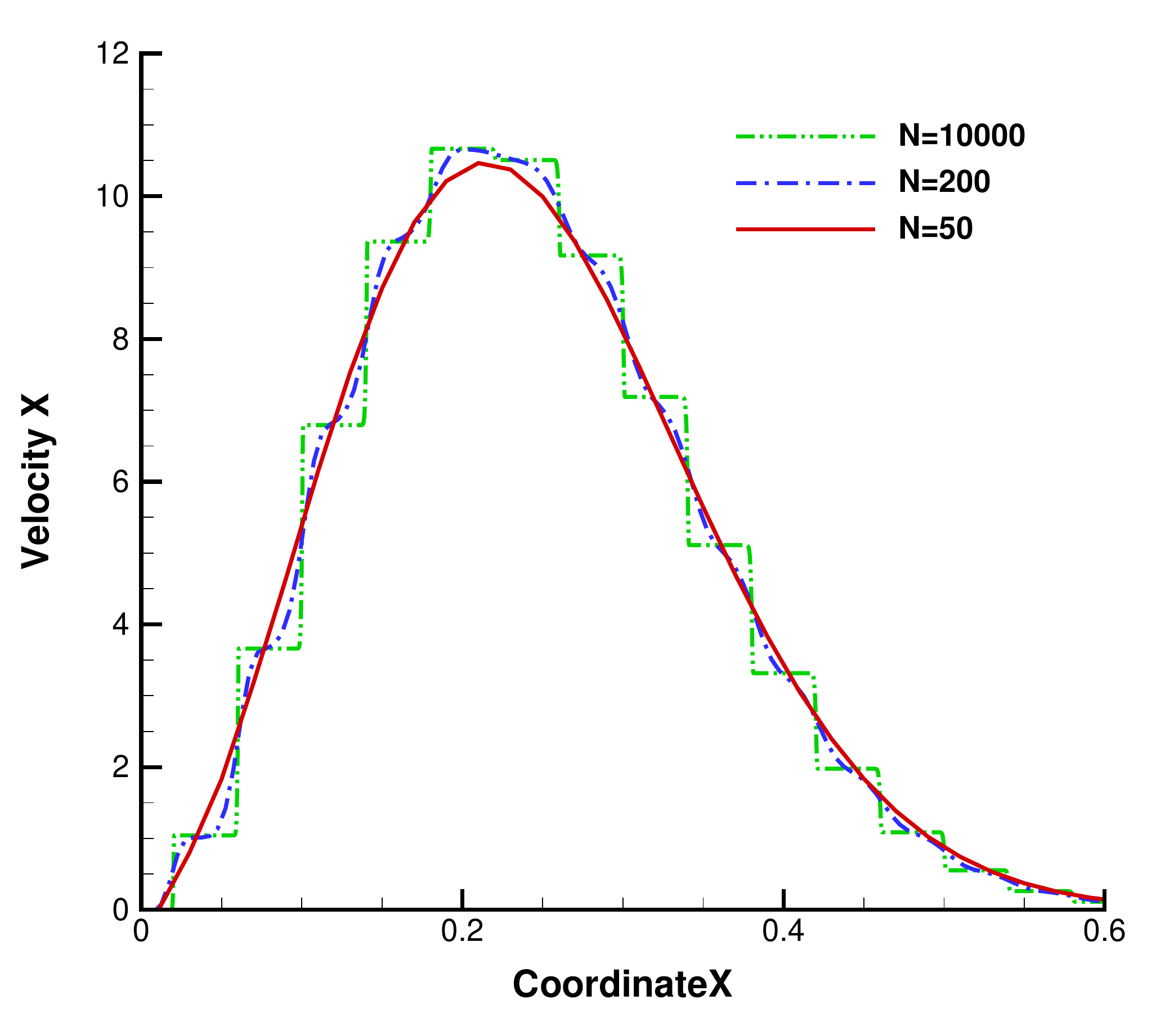}}
	\subfigure[Temperature]{\includegraphics[width=0.32\textwidth]{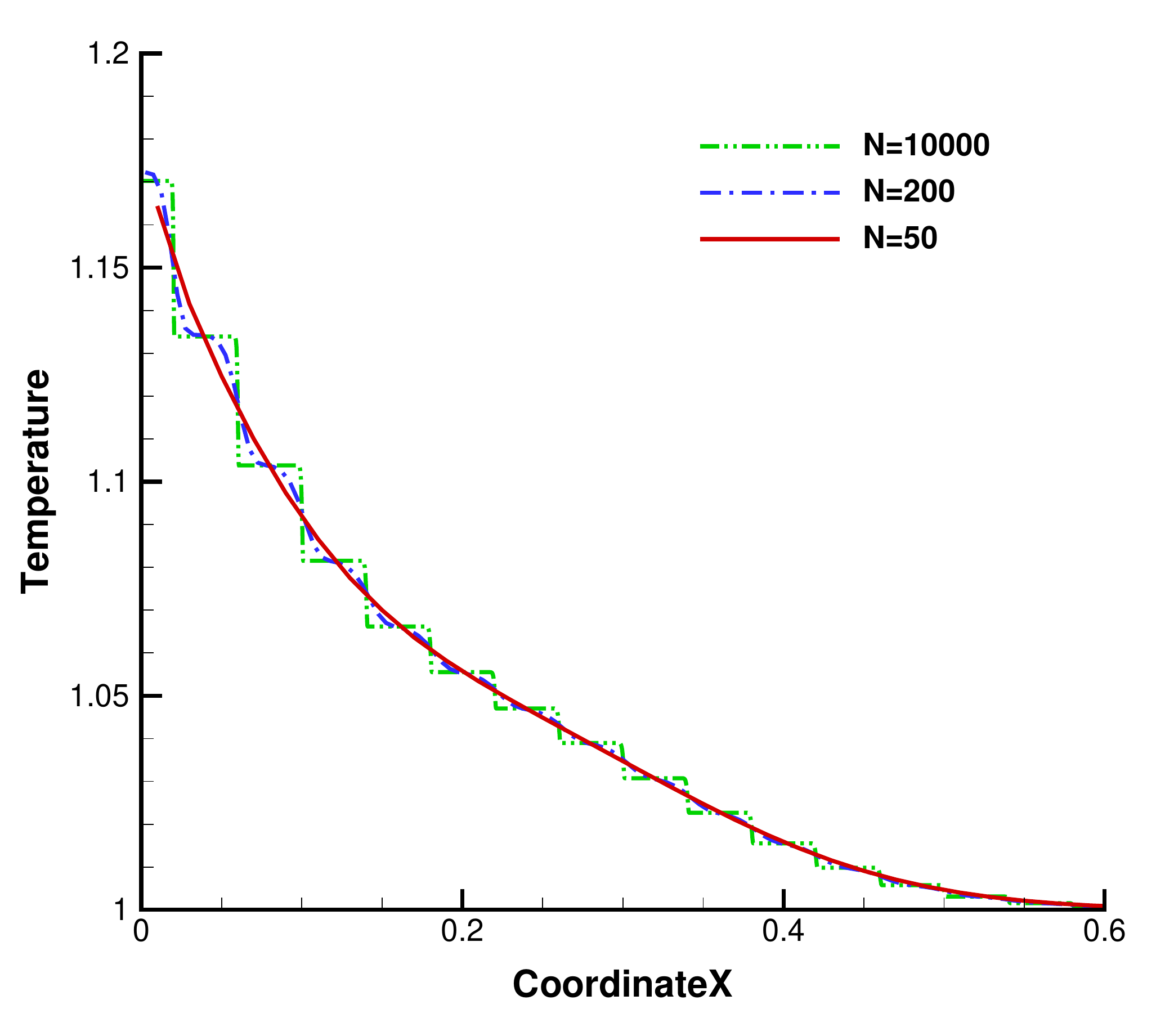}}
	\caption{\label{fig:rayleigh_mesh} Rayleigh flow at ${\rm Kn}\to \infty$ on coarser meshes.}
\end{figure}

The solutions on the coarse meshes with $50$ and $200$ cells are given in Fig.~\ref{fig:rayleigh_mesh}.
As expected, the ray effect can be merely observed on the coarse mesh with $50$ uniform cells.

\begin{figure}[H]
	\centering
	\subfigure[Density]
	{\includegraphics[width=0.32\textwidth]{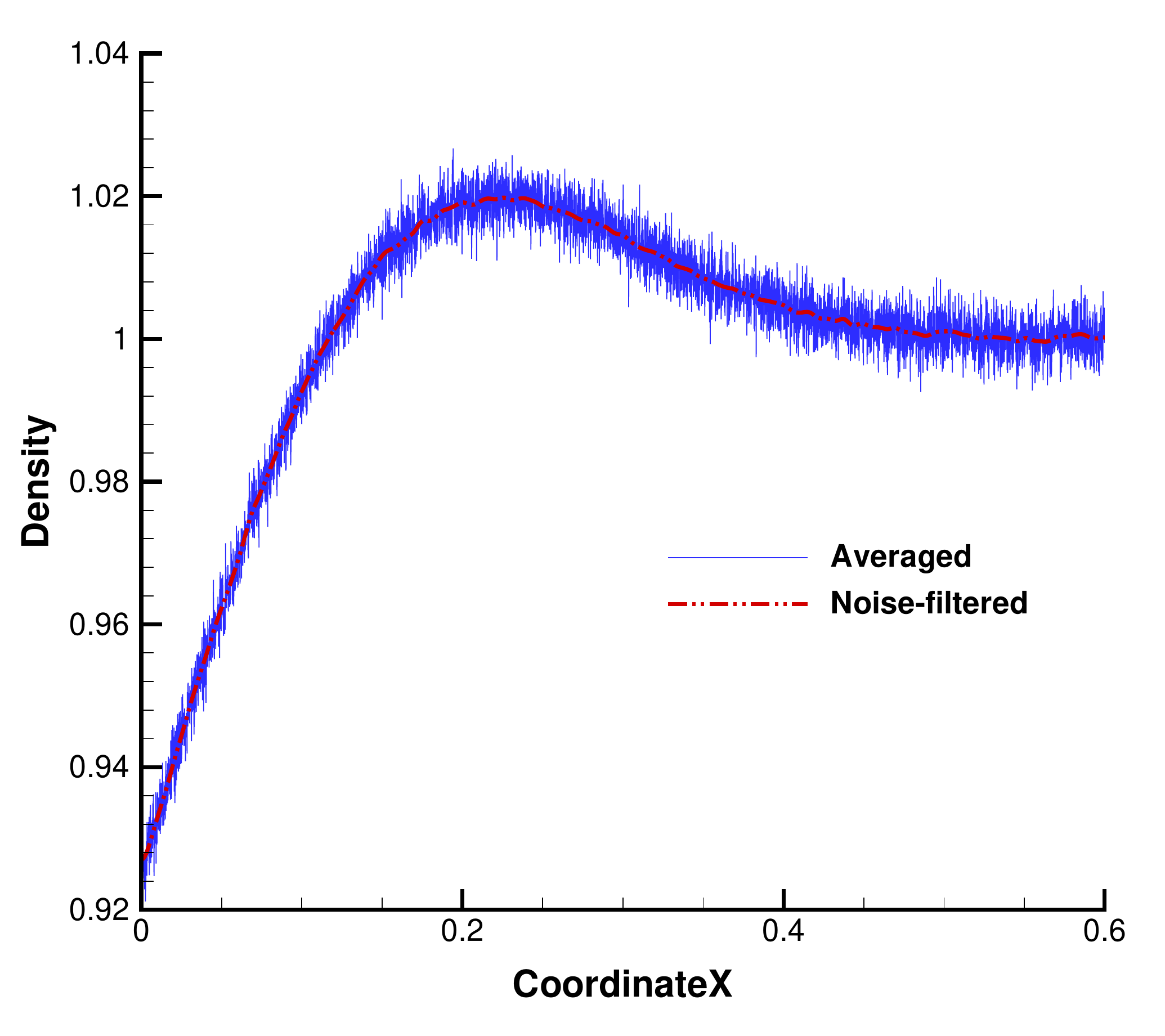}}
	\subfigure[Velocity]
	{\includegraphics[width=0.32\textwidth]{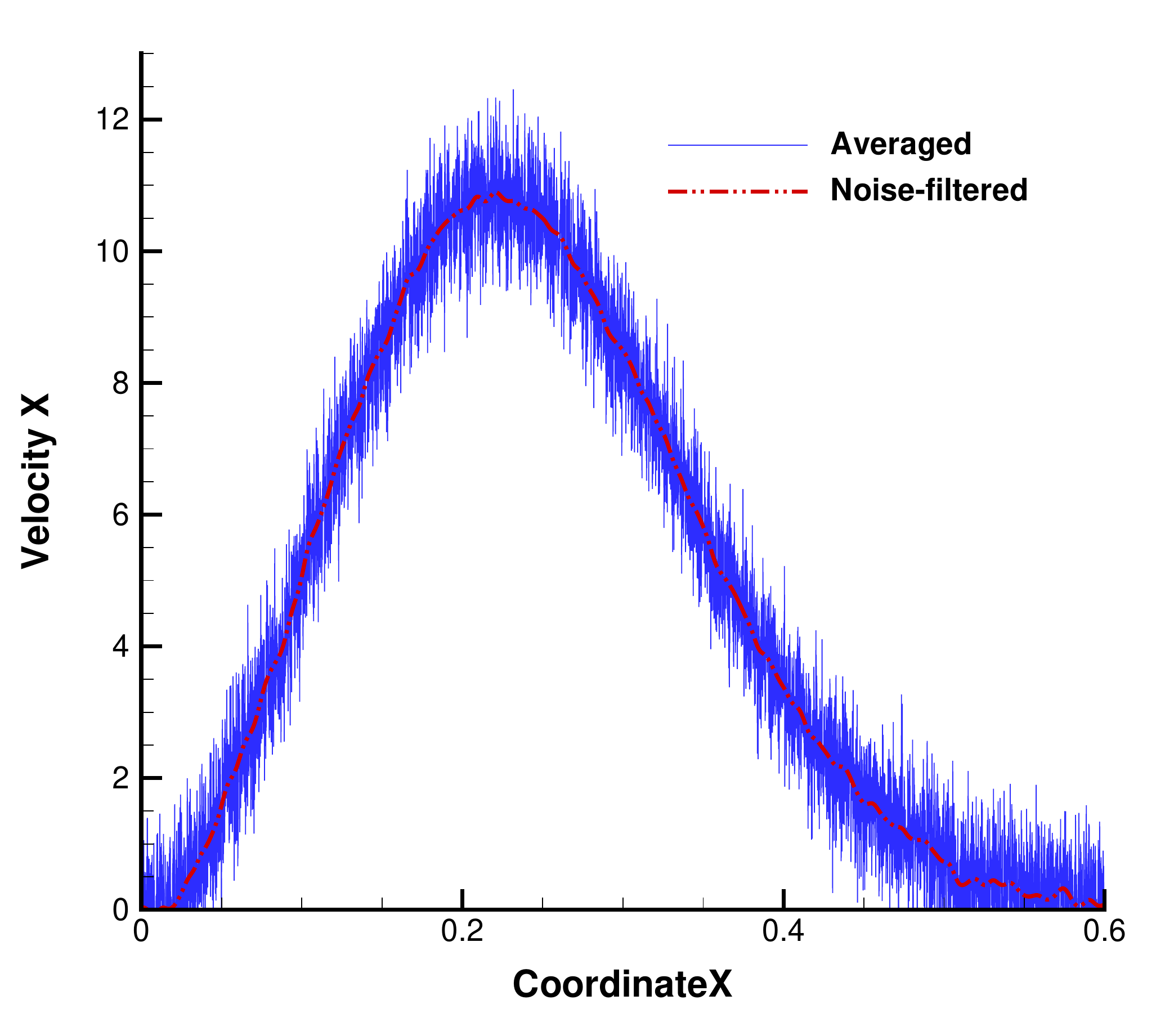}}
	\subfigure[Temperature]
	{\includegraphics[width=0.32\textwidth]{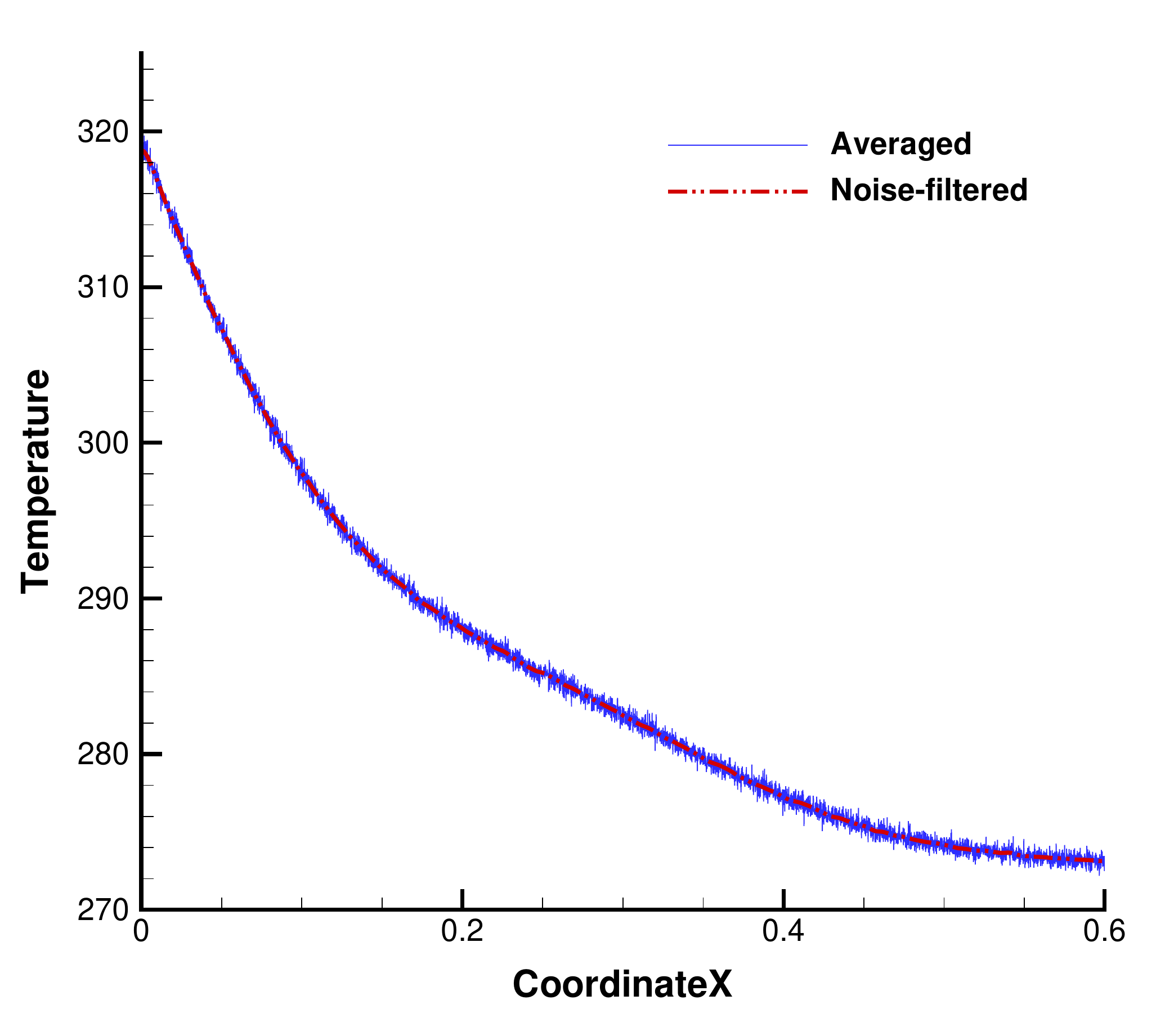}}
	\caption{\label{fig:rayleigh_particle}Rayleigh flow at ${\rm Kn}\to \infty$ obtained by particle method, where the blue solid line is the averaged solution over $100$ times of computations and the red dashed line denotes the noise-filtered solution.}	
\end{figure}

We also compute the numerical solution by the particle method.
Initially, the number of particles in each cell is $1600$, the same as the number of discrete velocities in previous calculations.
In the collisionless case, the particles can be accurately tracked.
Mostly, the ray effect happens more severely in the velocity space where the distribution functions and their differences have large values.
For particle method, more particles will be sampled in the velocity space at the place where the value of the distribution function is high.
Therefore, the principle A is satisfied for stochastic method automatically.
The random discretization in the velocity space from the particle method could give more discrete velocity points in the direction across the discontinuity line, and principles B and C are satisfied as well.
In terms of the velocity discretization, the particle method provides an optimal strategy in velocity space adaptation to overcome the ray effect.
However, in the numerical simulation of unsteady flows, the statistical noise becomes dominant and the influence of ray effect can be neglected in this case.
The numerical solutions averaged over $100$ times of computations are plotted in Fig.~\ref{fig:rayleigh_particle}.
From the noise-filtered results, the typical characters of ray effect, such as stepped structures, have not been observed.

\subsection{Non-stationary initial discontinuity}
\begin{figure}[H]
	\centering
	\subfigure[Density]{\includegraphics[width=0.32\textwidth]{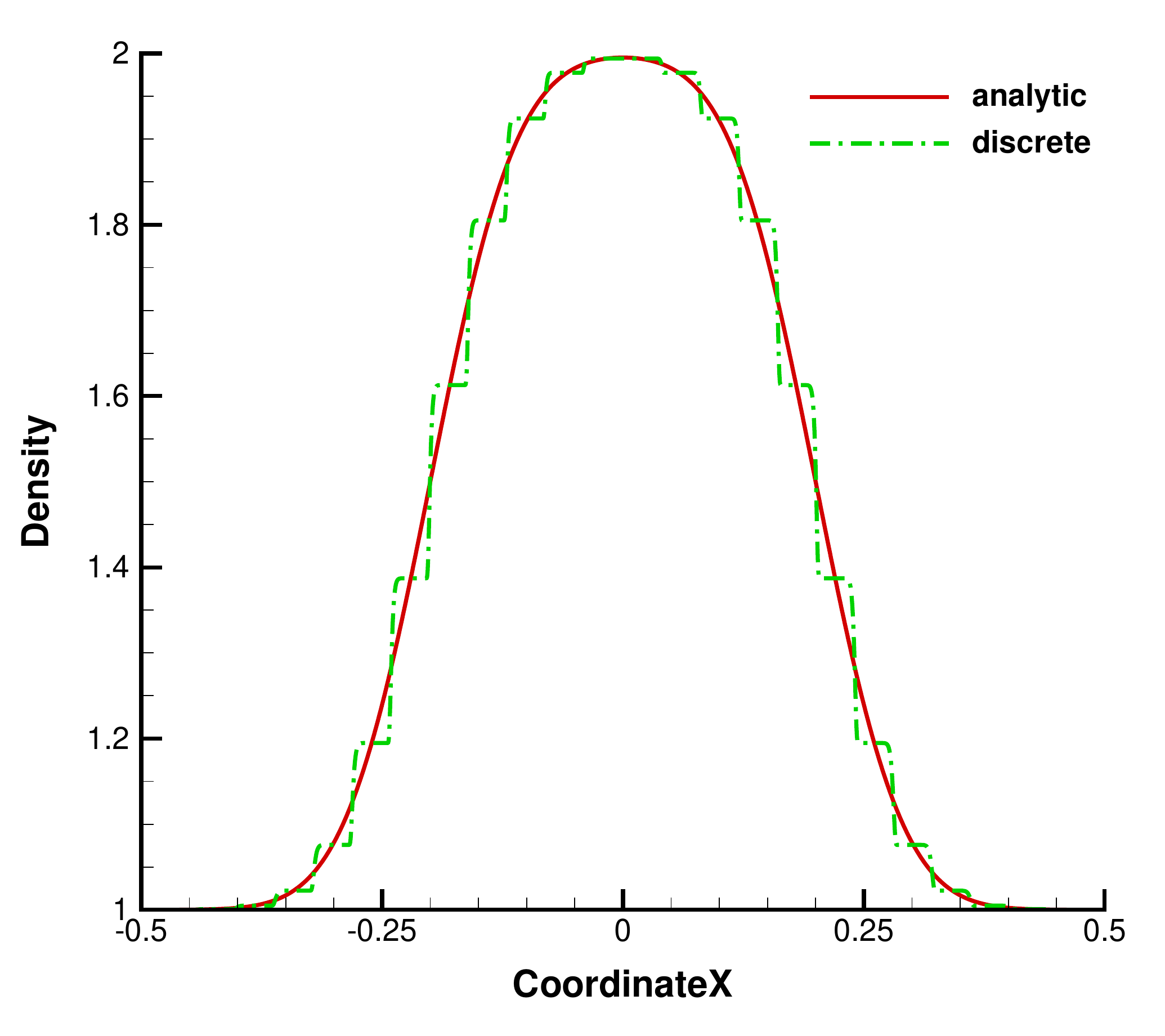}}
	\subfigure[Velocity]{\includegraphics[width=0.32\textwidth]{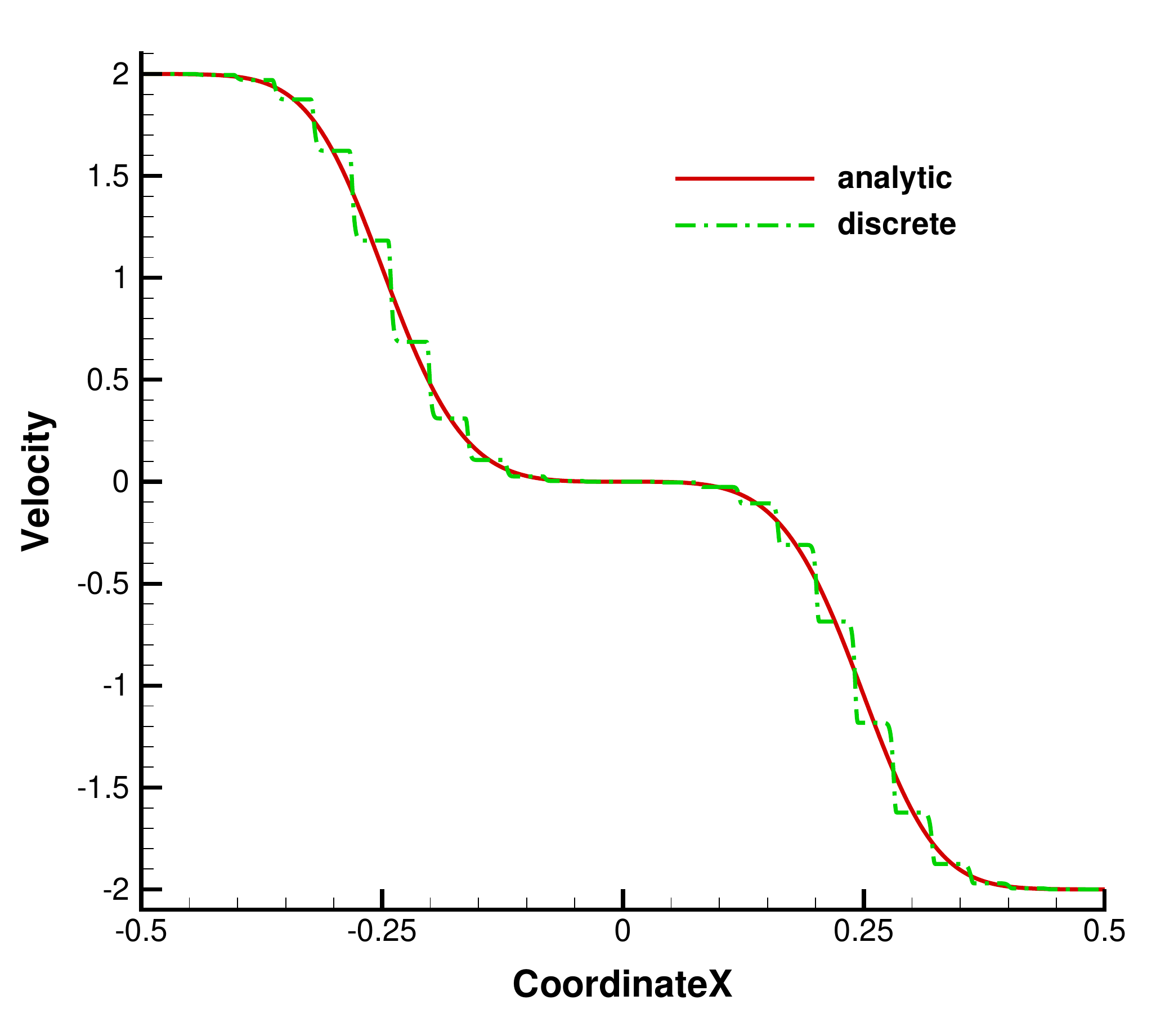}}
	\subfigure[Temperature]{\includegraphics[width=0.32\textwidth]{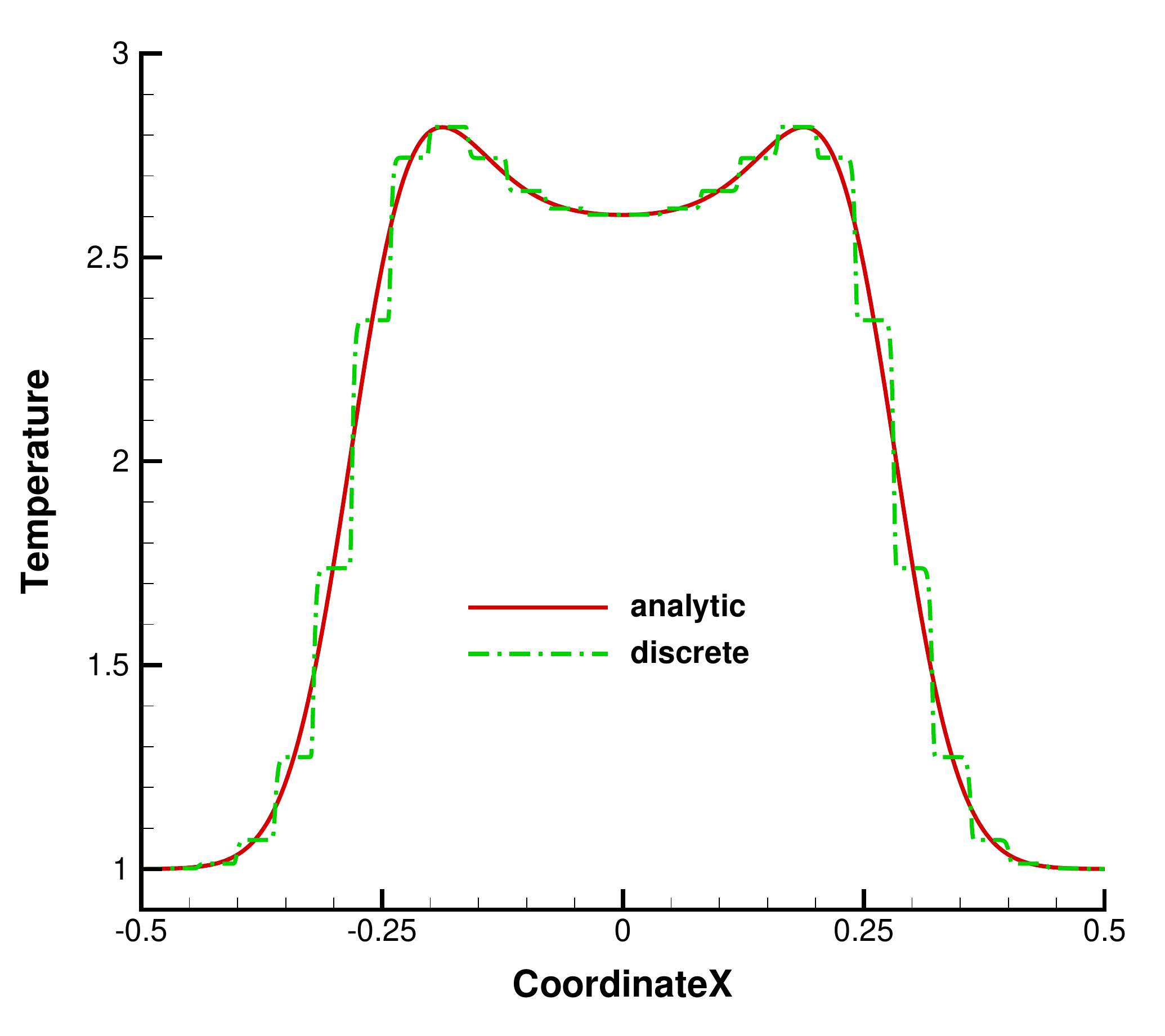}}
	\caption{\label{fig:hit_solution}Numerical solution obtained by using discrete velocity distribution function in comparison with the analytic data.}
\end{figure}

\begin{figure}[H]
	\centering
	\subfigure[Density]{\includegraphics[width=0.32\textwidth]{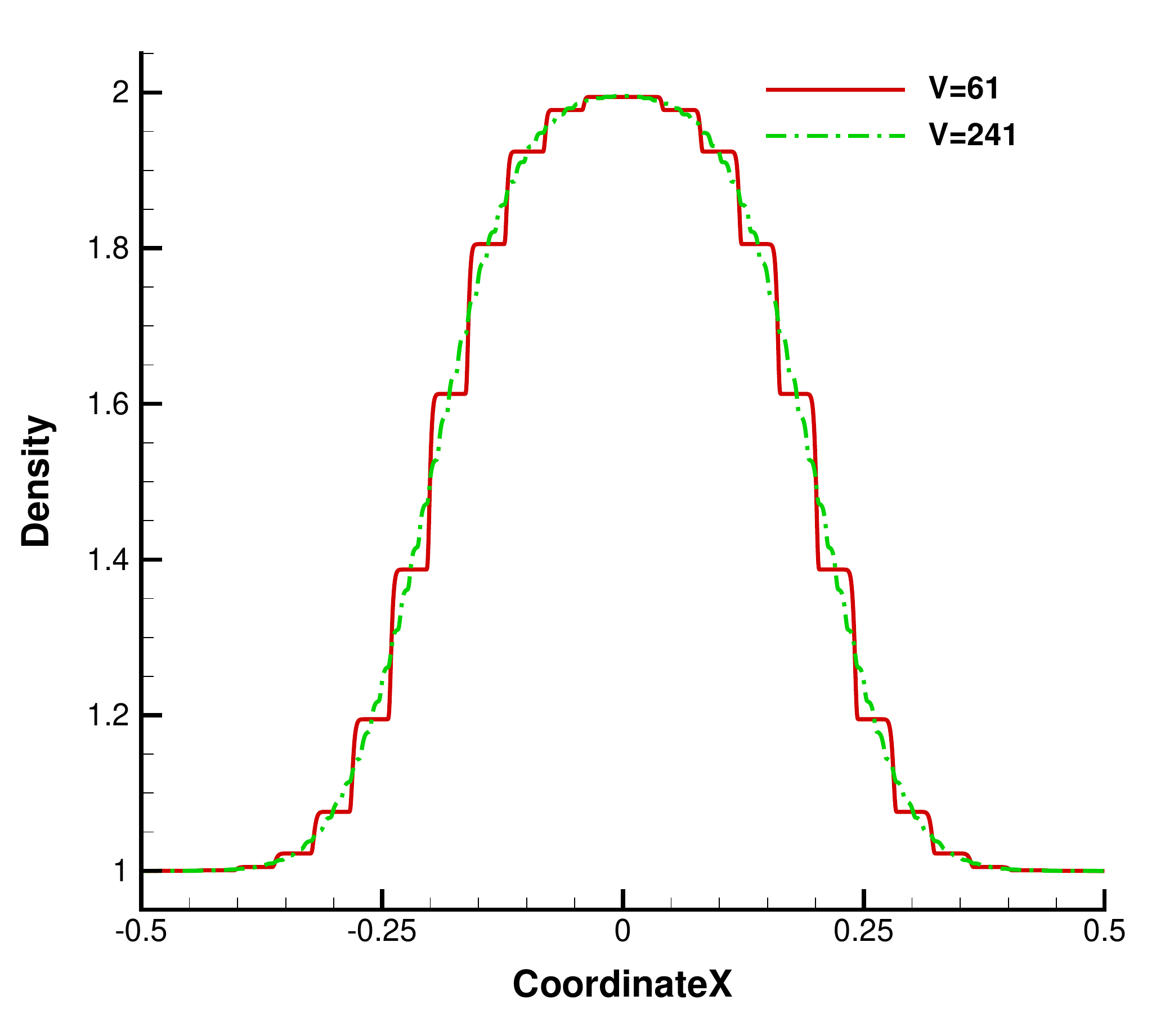}}
	\subfigure[Velocity]{\includegraphics[width=0.32\textwidth]{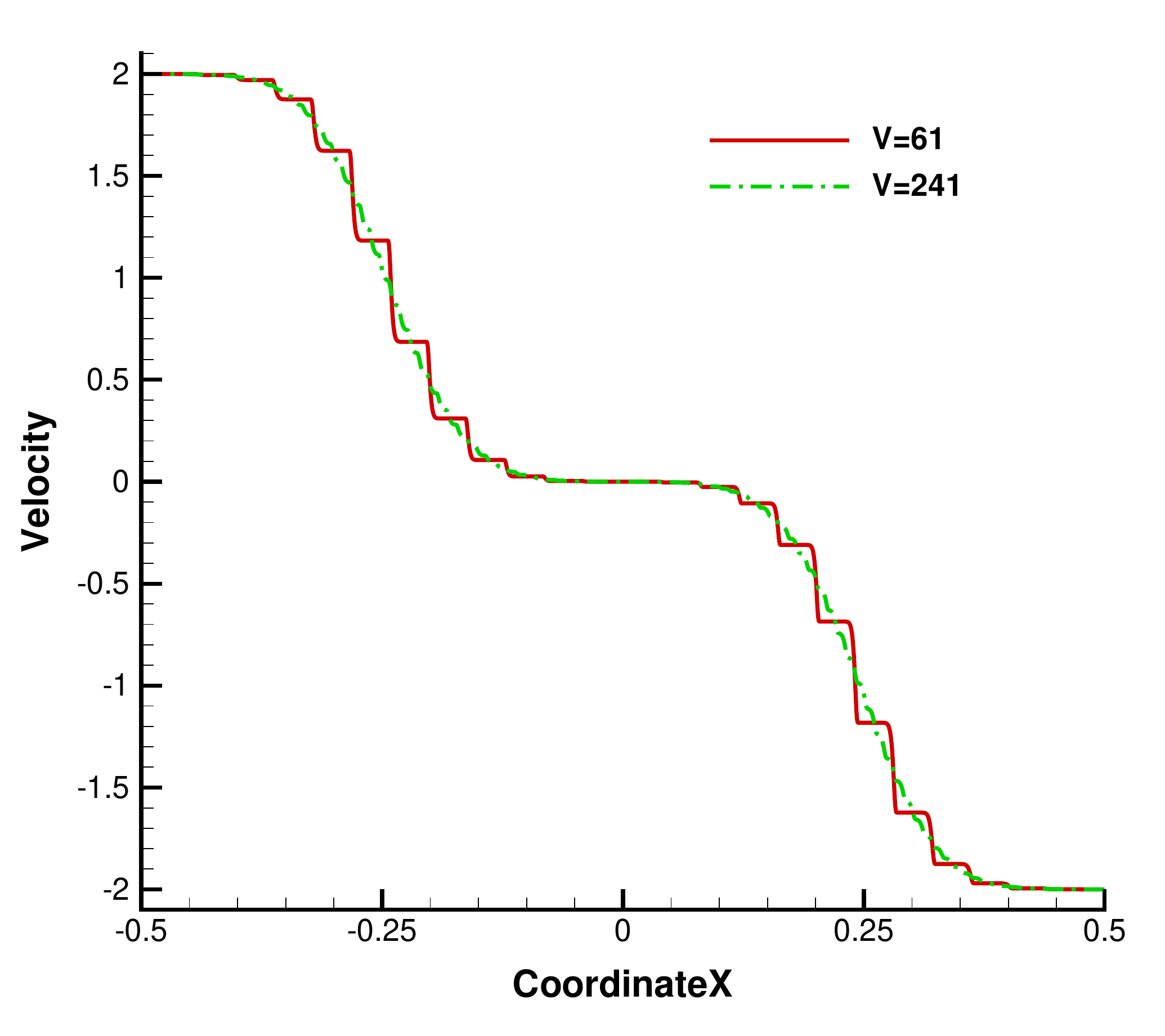}}
	\subfigure[Temperature]{\includegraphics[width=0.32\textwidth]{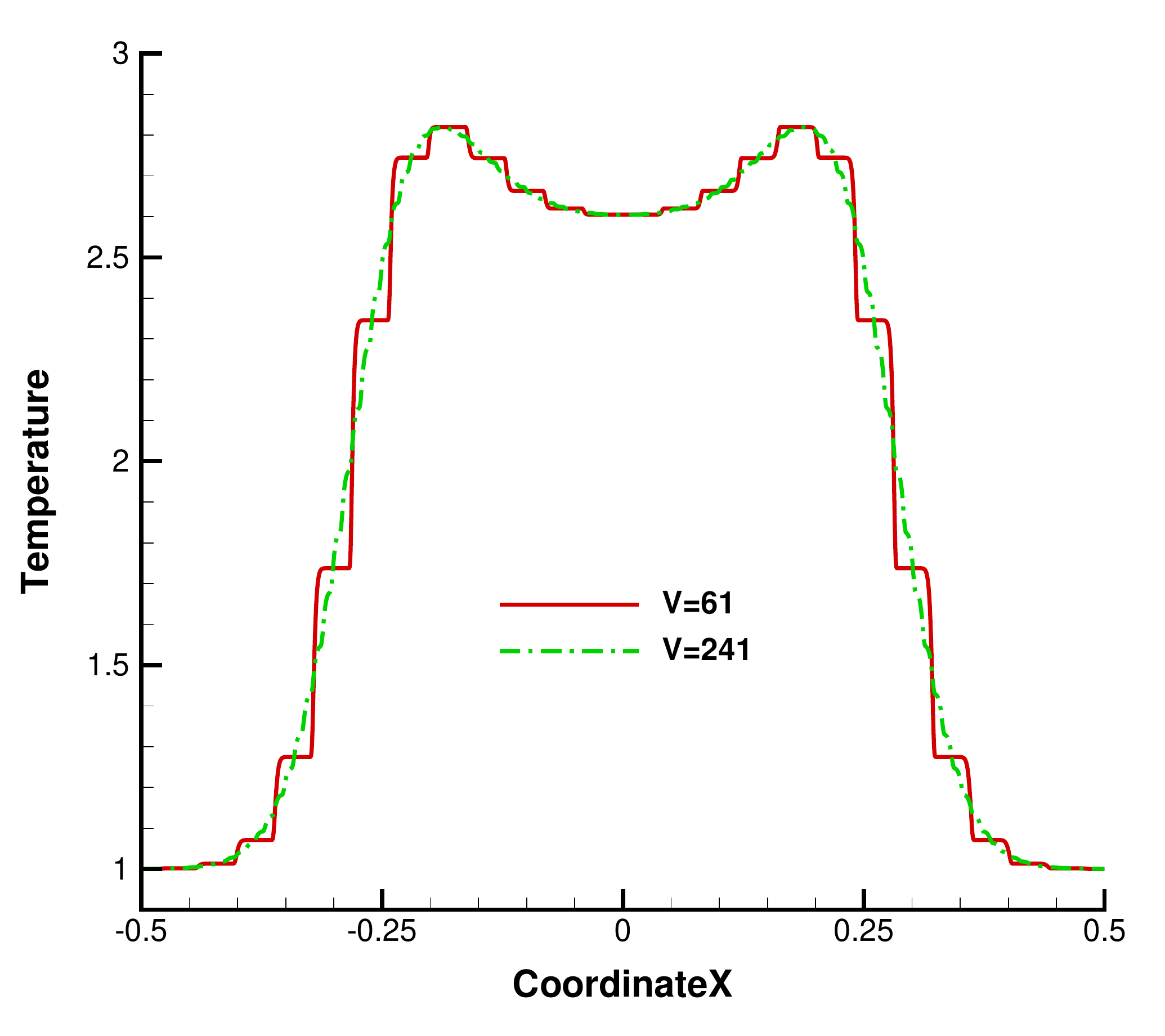}}
	\caption{\label{fig:hit_refine}Numerical solutions obtained by using different numbers of discrete velocity points.}
\end{figure}

\begin{figure}[H]
	\centering
	\subfigure[Density]{\includegraphics[width=0.32\textwidth]{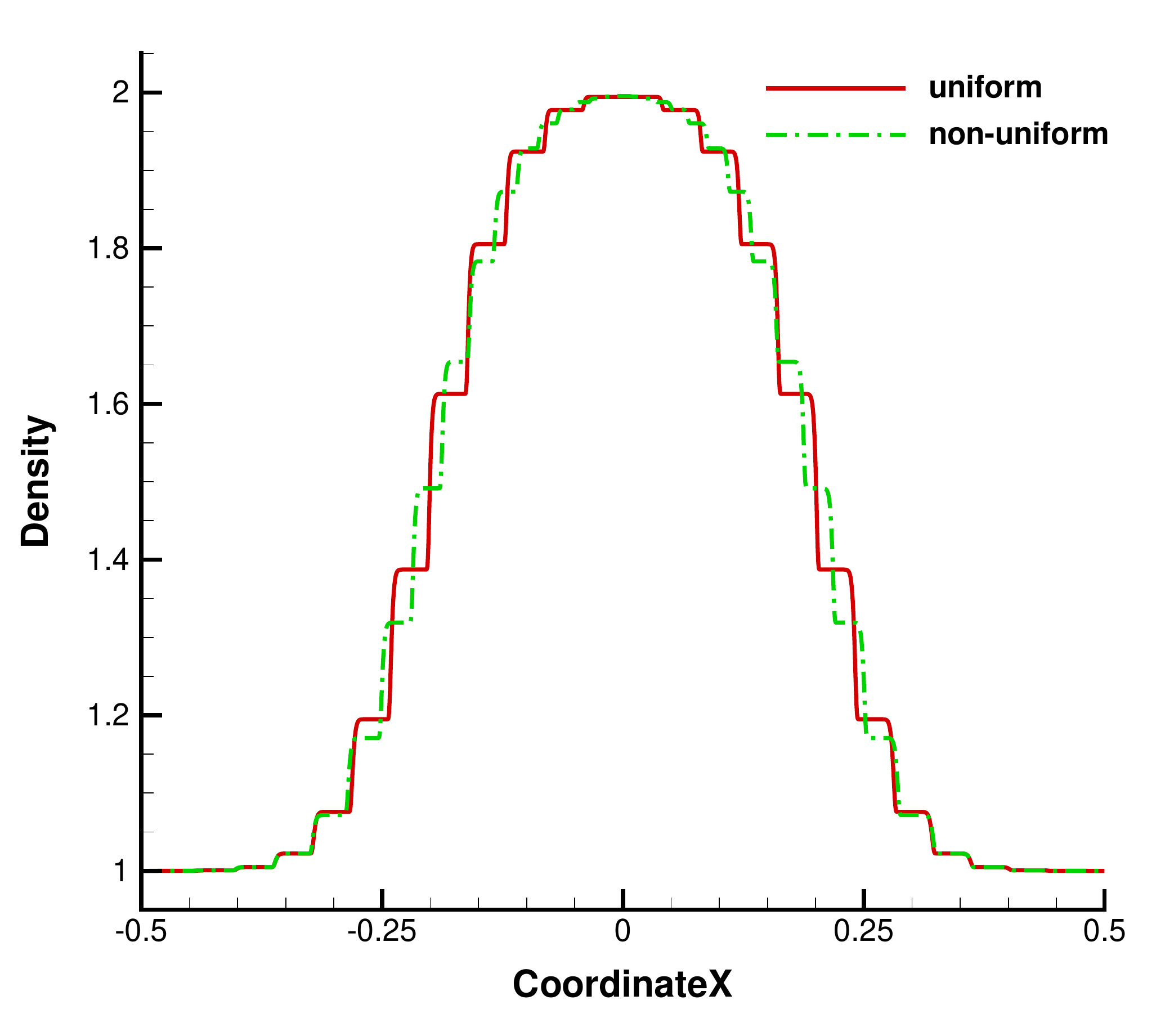}}
    \subfigure[Velocity]{\includegraphics[width=0.32\textwidth]{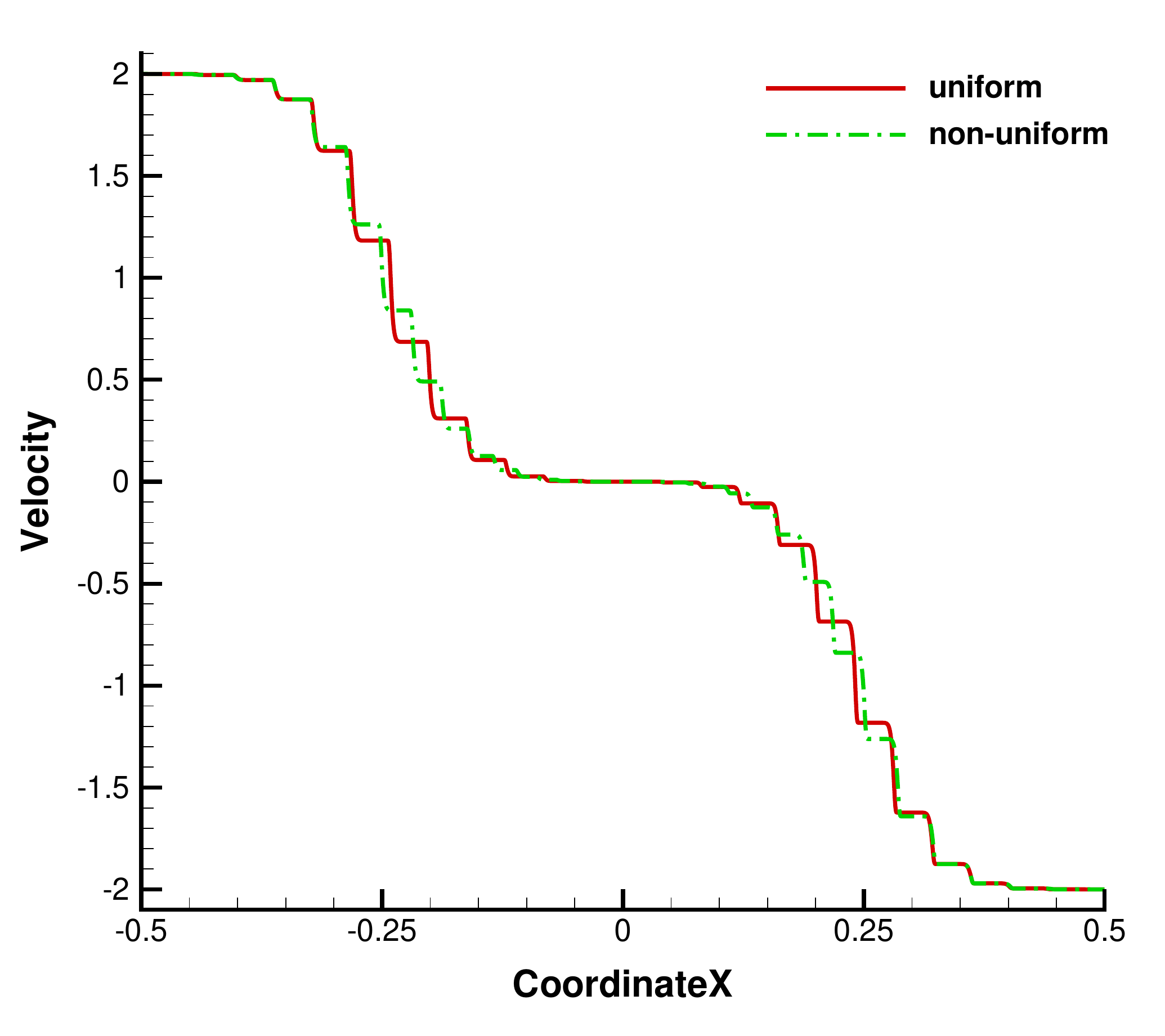}}
	\subfigure[Temperature]{\includegraphics[width=0.32\textwidth]{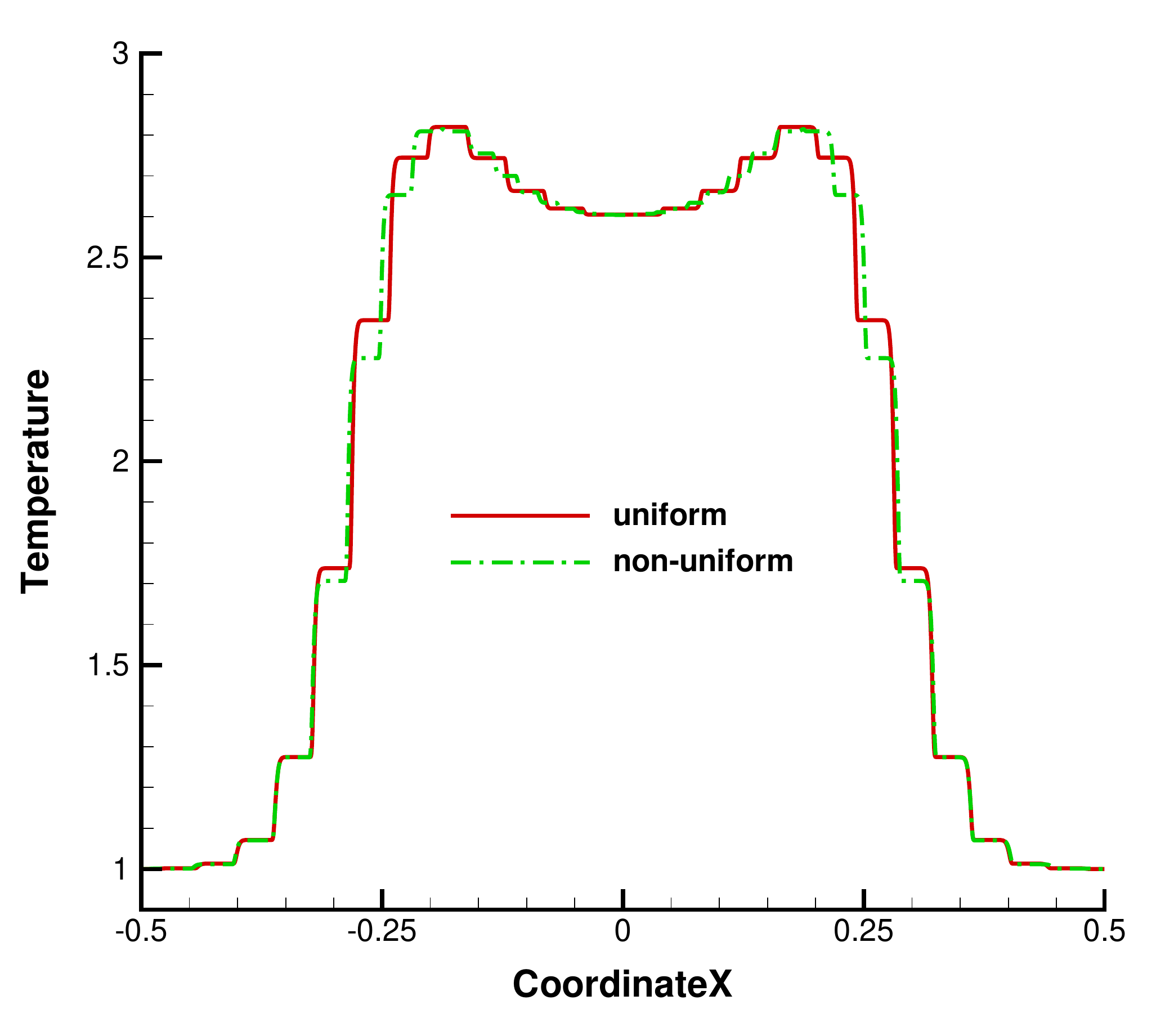}}
	\caption{\label{fig:hit_stretch}Numerical solutions obtained by using uniform and non-uniform meshes in the velocity space, where the non-uniform mesh is stretched near the zero point and has the same number of discrete velocities as the uniform mesh.}
\end{figure}

\begin{figure}[H]
	\centering
	\subfigure[$g_l(u,v)$]
	{\includegraphics[width=0.24\textwidth]{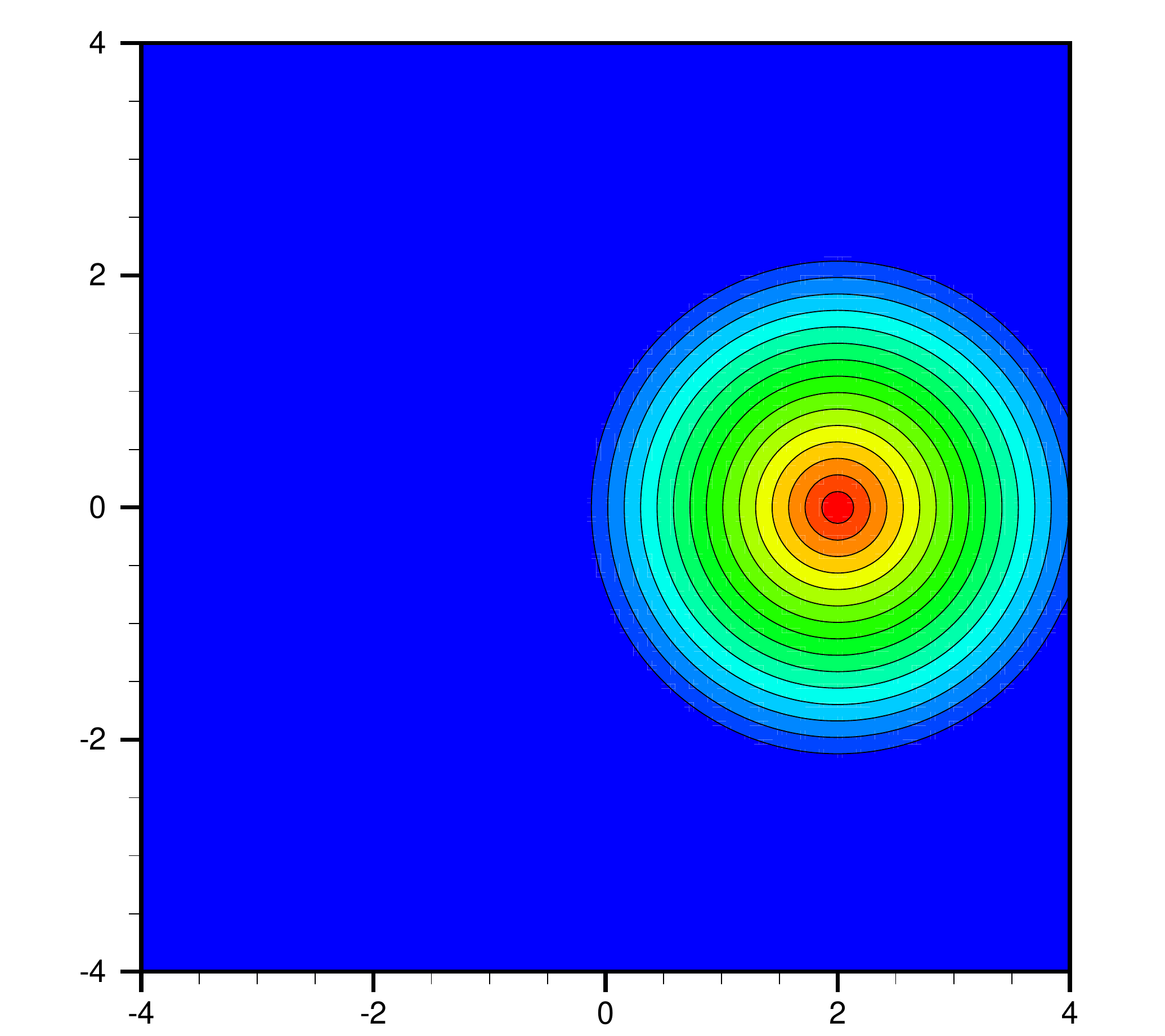}}
	\subfigure[$g_r(u,v)$]
	{\includegraphics[width=0.24\textwidth]{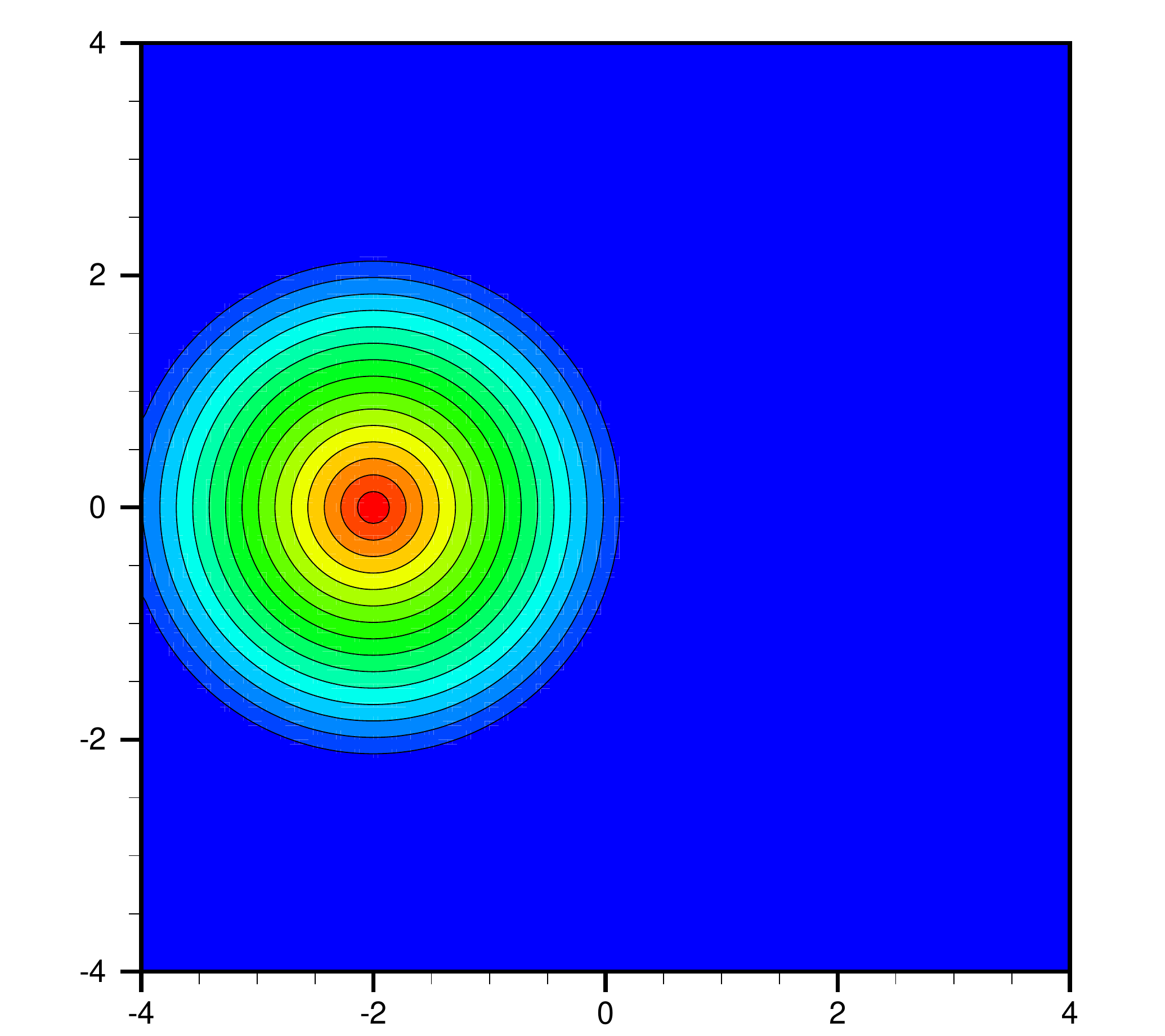}}
	\subfigure[$f(u,v)$]
	{\includegraphics[width=0.24\textwidth]{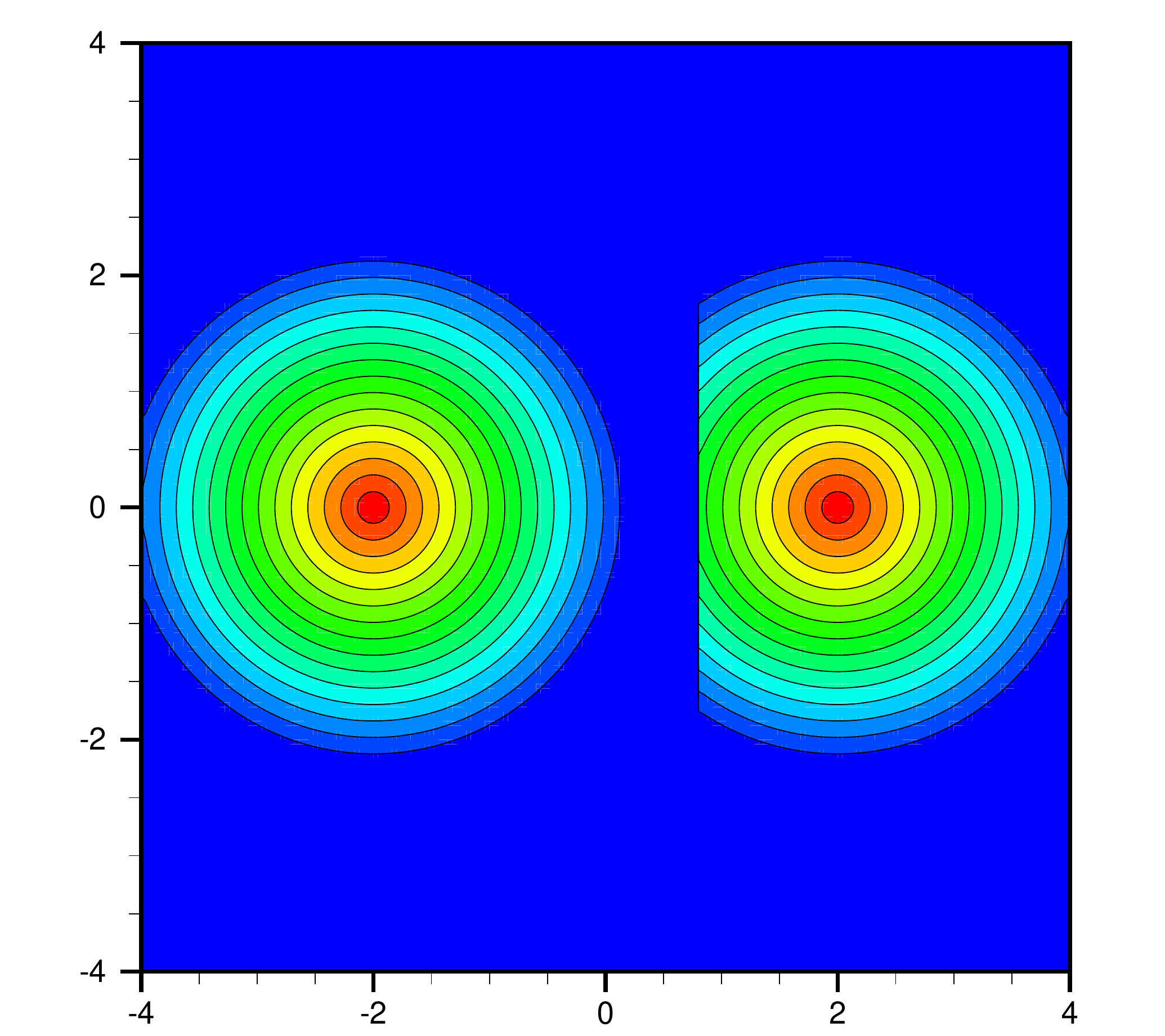}}
	\subfigure[$|g_l(u,v) - g_r(u,v)|$]
	{\includegraphics[width=0.24\textwidth]{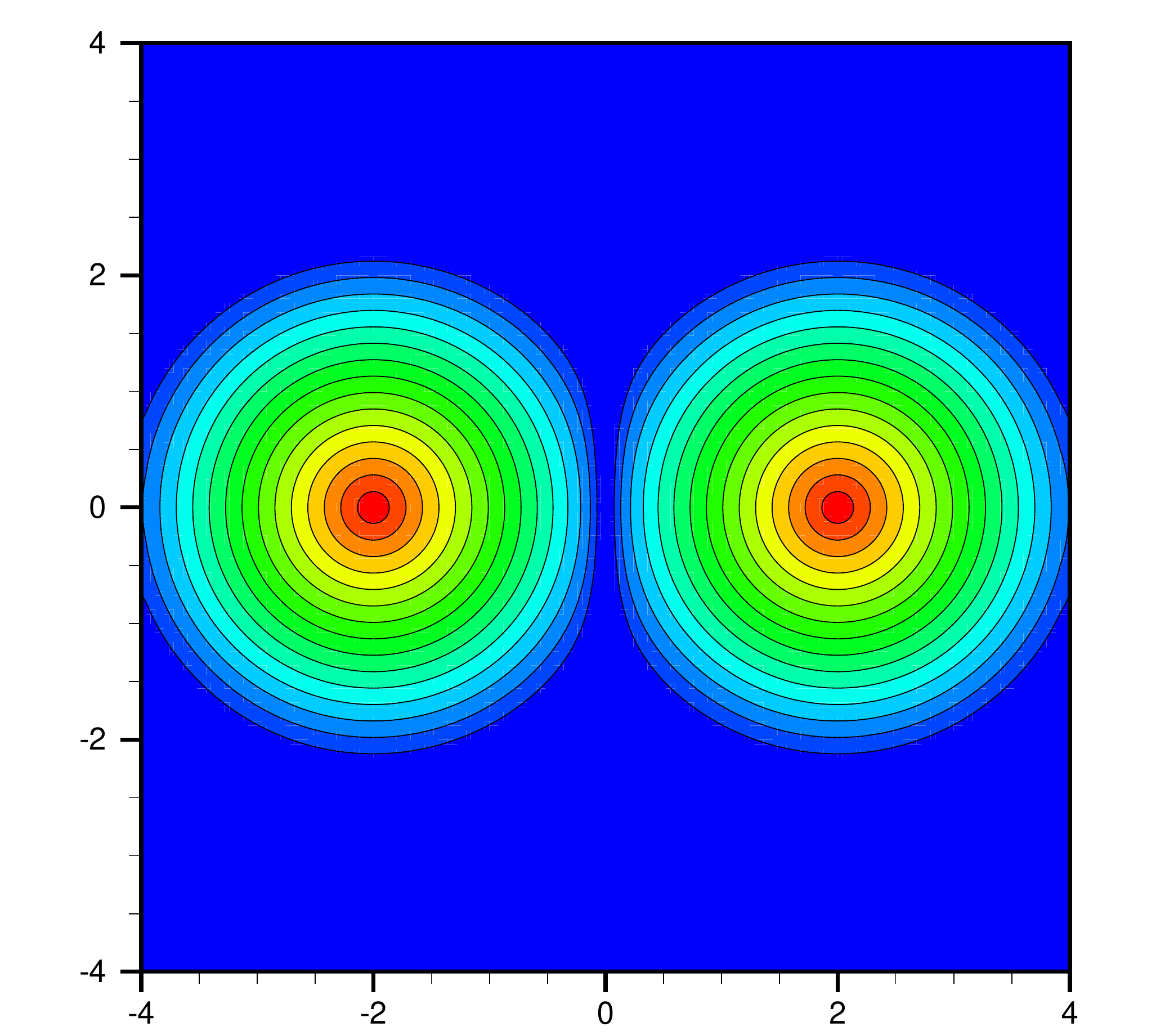}}
	\caption{\label{fig:hit_vs}Analytic distribution function in velocity space. (a) Initial Maxwellian distribution for left state; (b) initial Maxwellian distribution for right state; (c) distribution function at a specific location $x$; (d) distribution function difference between $g_l$ and $g_r$ for evaluation of ray effect jumps.}
\end{figure}

A one-dimensional case similar to the Sod test with a non-stationary initial discontinuity is computed here.
The main difference from the Sod case is that the initial velocity is not zero.
The initial condition gives
\begin{equation}\label{eq:initial_setup}
(\rho, U, T) =
\begin{cases}
\left(1, 2, 1 \right), &\quad x < 0, \\
\left(1, -2, 1 \right), &\quad x > 0,
\end{cases}
\end{equation}
For this case, the computational domain is discretized by a uniform mesh with $2000$ cells, and $61$ discrete velocity points for the range of $\left[-12,12\right]$ in velocity space are adopted.
The solution in the collisionless limit at time $t_e = 0.1$ is presented in Fig.\ref{fig:hit_solution}, where as expected ray effect is observed in the numerical solution with discrete velocity space.
When we employ a finer uniform mesh with $241$ discrete points in the velocity space, the ray effect is much mitigated as shown in Fig.~\ref{fig:hit_refine}.
In the previous Rayleigh case, by adjusting the distribution of the discrete velocities near the zero velocity point, a better solution can be obtained with reduced ray effect.
However, for this case it takes no obvious effect on the results in Fig.~\ref{fig:hit_stretch}, which non-uniform $61$ velocity points are used.
As shown in Fig.~\ref{fig:hit_vs}, the distribution function difference in the velocity space is plotted for the current case,
and the required mesh refinement regions are around the velocity points $u_k = \pm 2$ in this case instead of $u_k = 0$ in the previous one.
Therefore, the method through adjusting discrete velocity space to mitigate the ray effect is problem dependent, and the velocity mesh refinement around the zero velocity point doesn't always give better solutions.
While, for the discontinuities induced by a stationary boundary, the region with large difference in distribution function usually locates near the zero velocity point, so the discrete velocity space refinement in the low speed region seems reasonable for such case.

\subsection{Cavity flow}

\begin{figure}[H]
	\centering
	\subfigure[\label{fig:cavity_coarse}]
	{\includegraphics[width=0.48\textwidth]{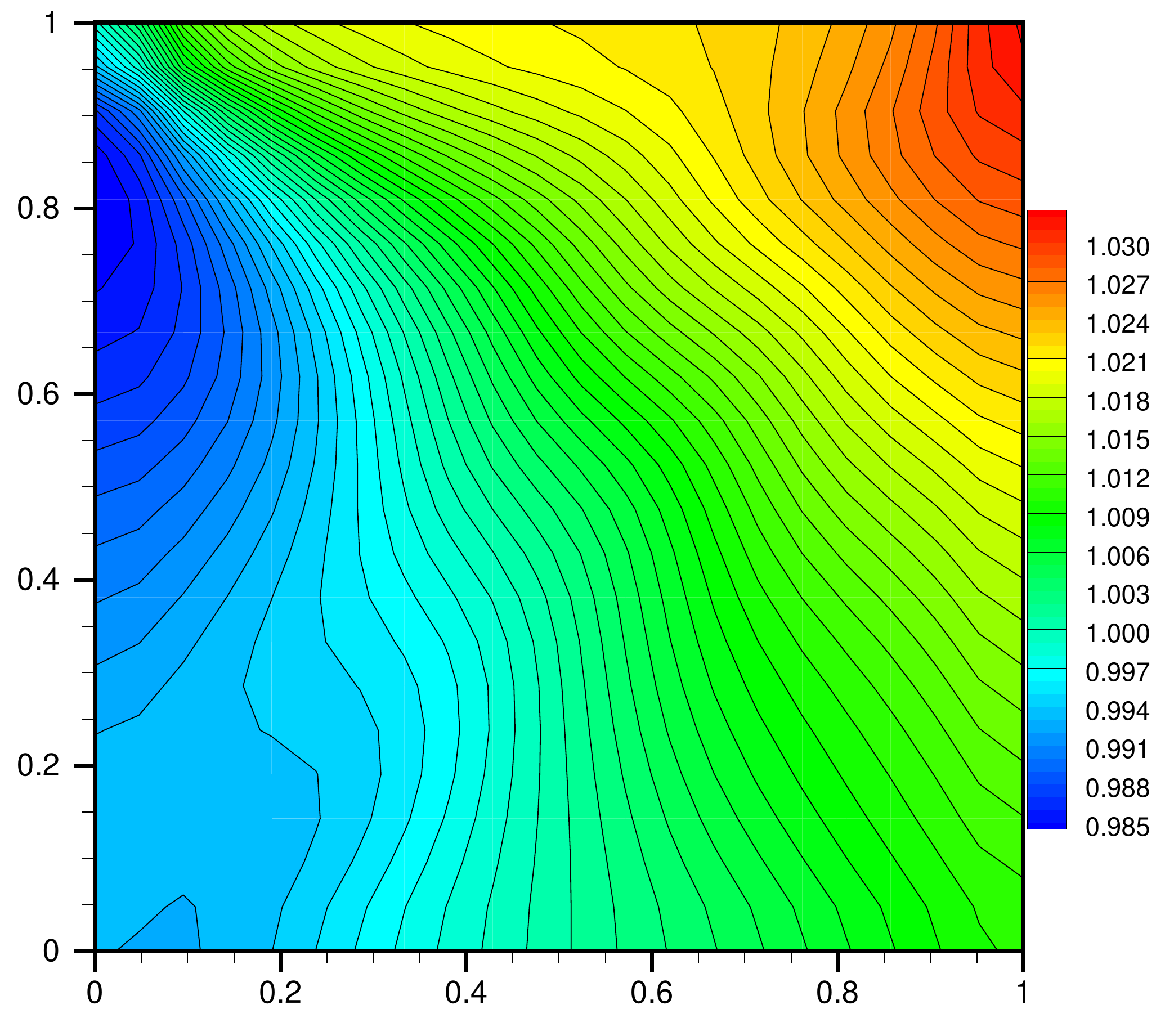}}
	\subfigure[\label{fig:cavity_refine}]
	{\includegraphics[width=0.48\textwidth]{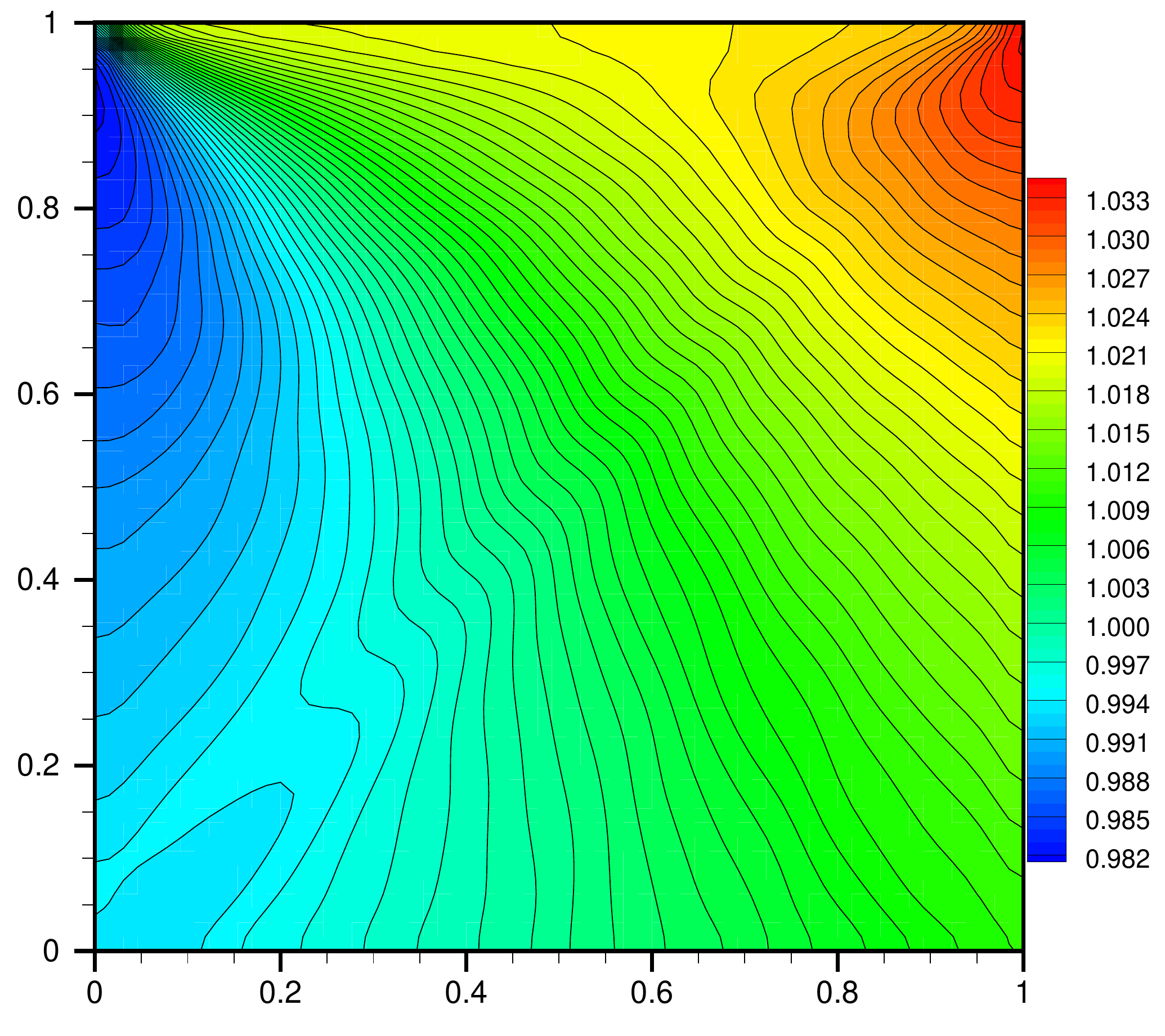}}\\	
	\subfigure[\label{fig:cavity_particle}]
	{\includegraphics[width=0.48\textwidth]{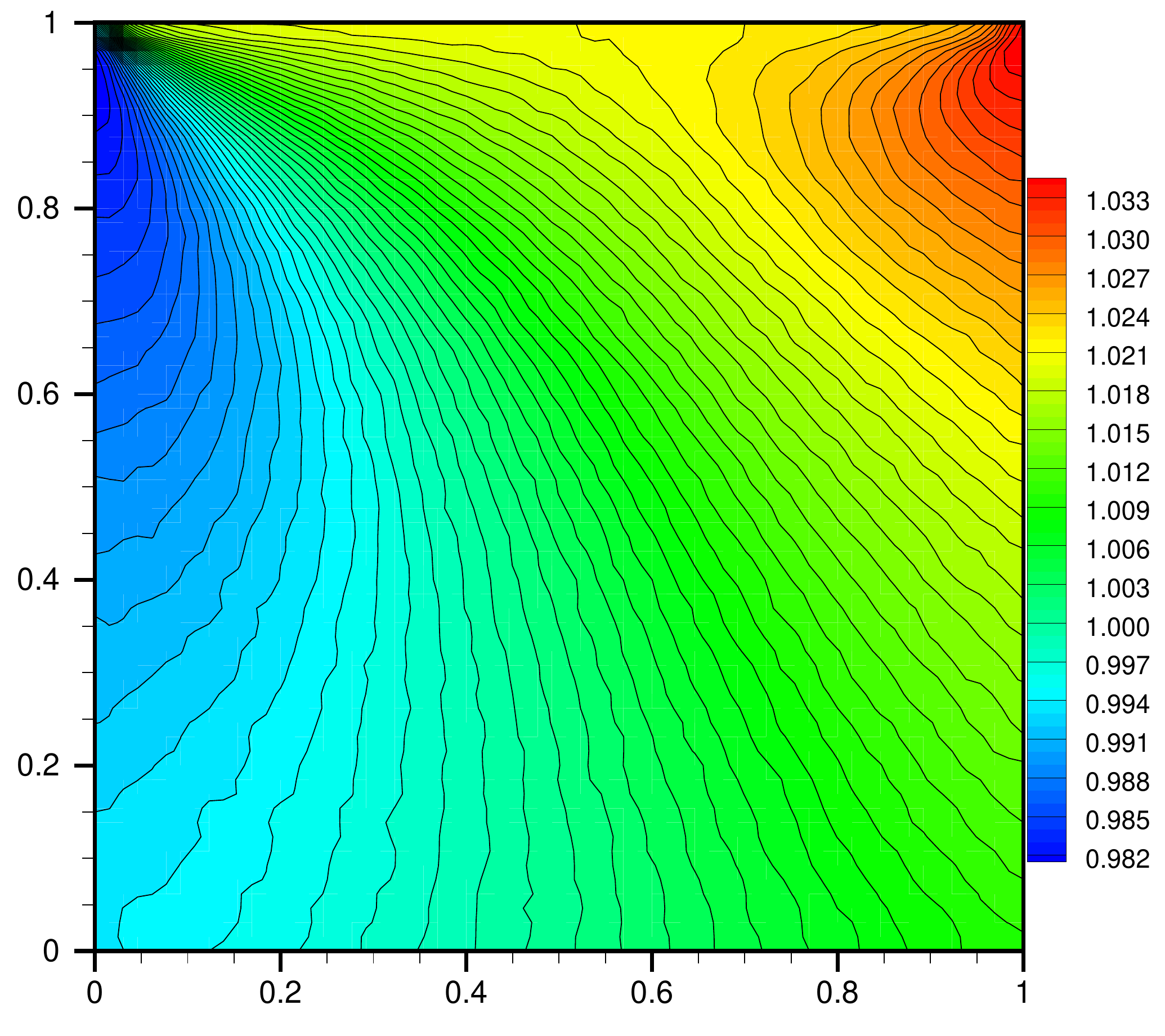}}
	\subfigure[\label{fig:cavity_special}]
	{\includegraphics[width=0.48\textwidth]{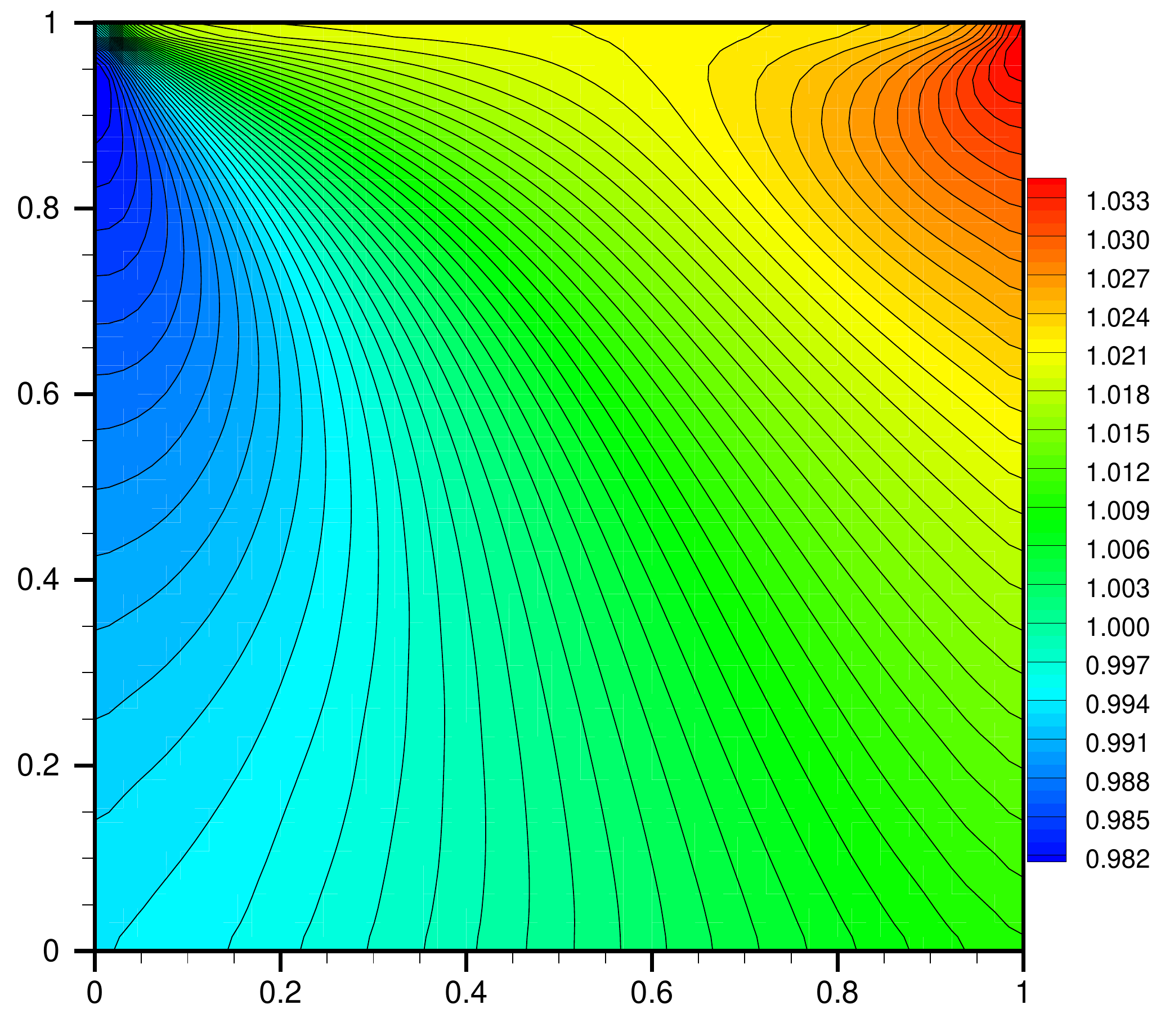}}
	\caption{\label{fig:cavity_mitigation}Cavity flow recomputed by (a) using coarse mesh in physical space; (b) adopting nonuniform velocity points in the velocity space; (c) employing stochastic particle method; (d) using a special designed discretization of velocity space.}
\end{figure}

\begin{figure}[H]
	\centering
	\subfigure[\label{fig:cavity_nonuniform_vs}]
	{\includegraphics[width=0.32\textwidth]{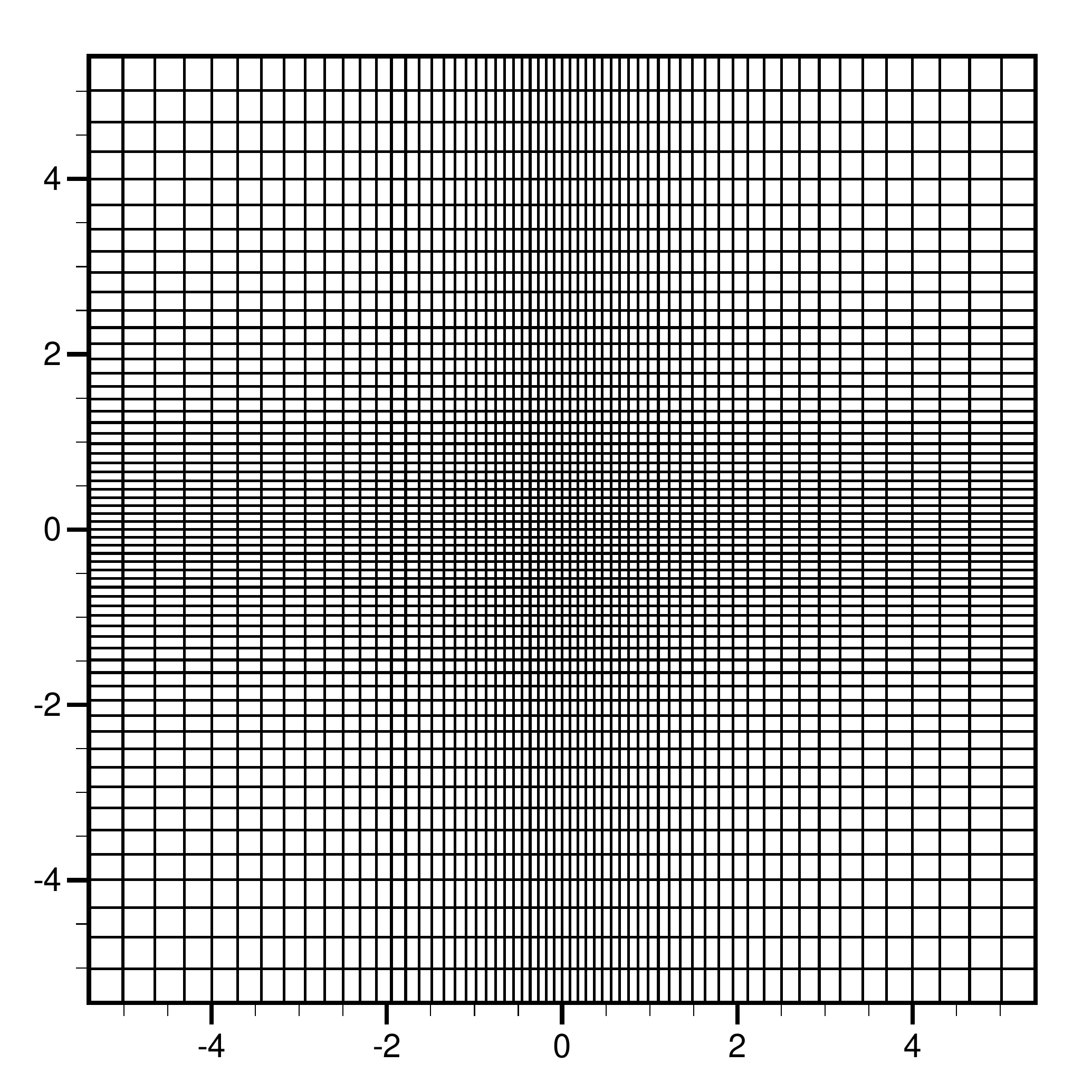}}
	\subfigure[\label{fig:cavity_special_vs}]
	{\includegraphics[width=0.32\textwidth]{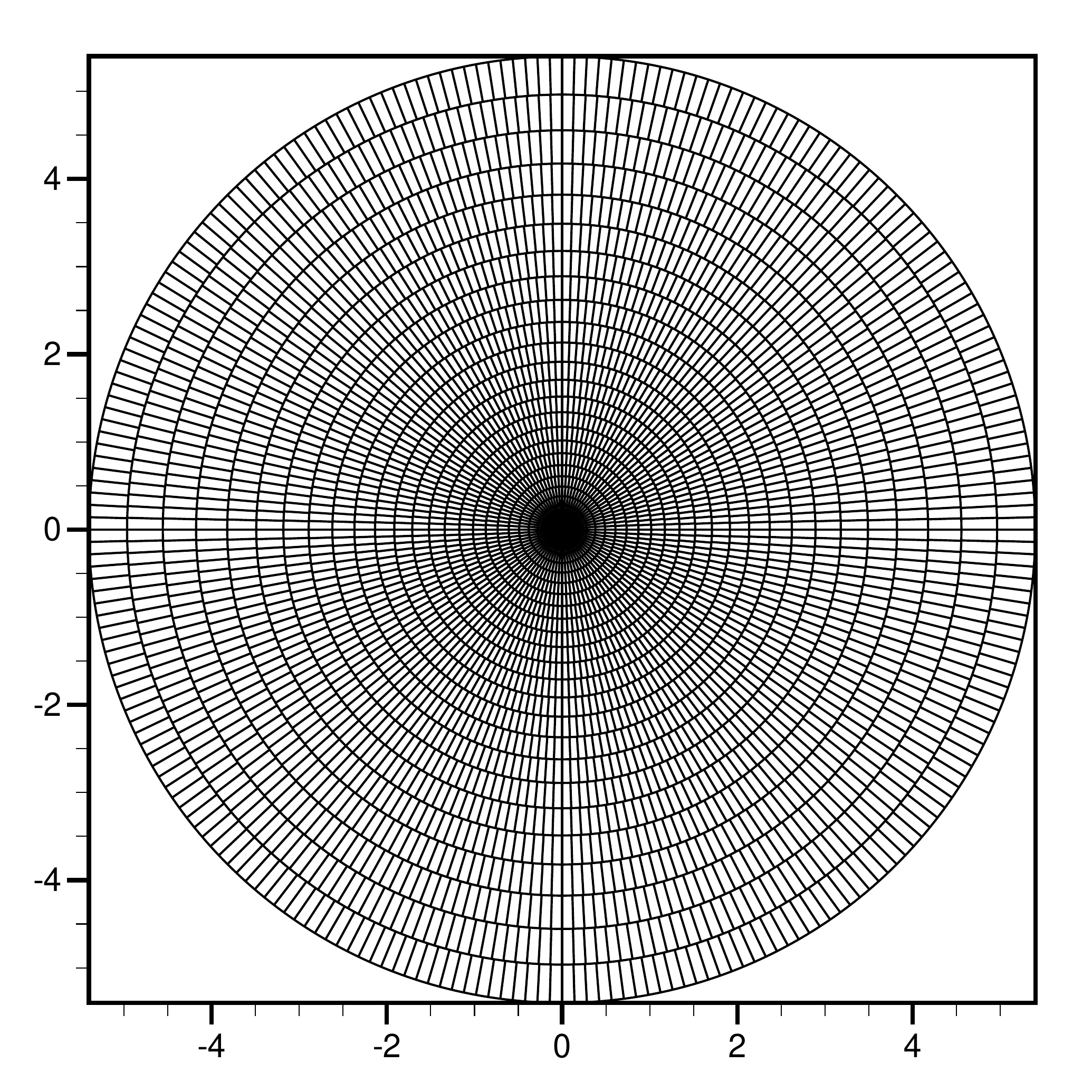}}
	\subfigure[\label{fig:cavity_special_vs_enlargement}]
	{\includegraphics[width=0.32\textwidth]{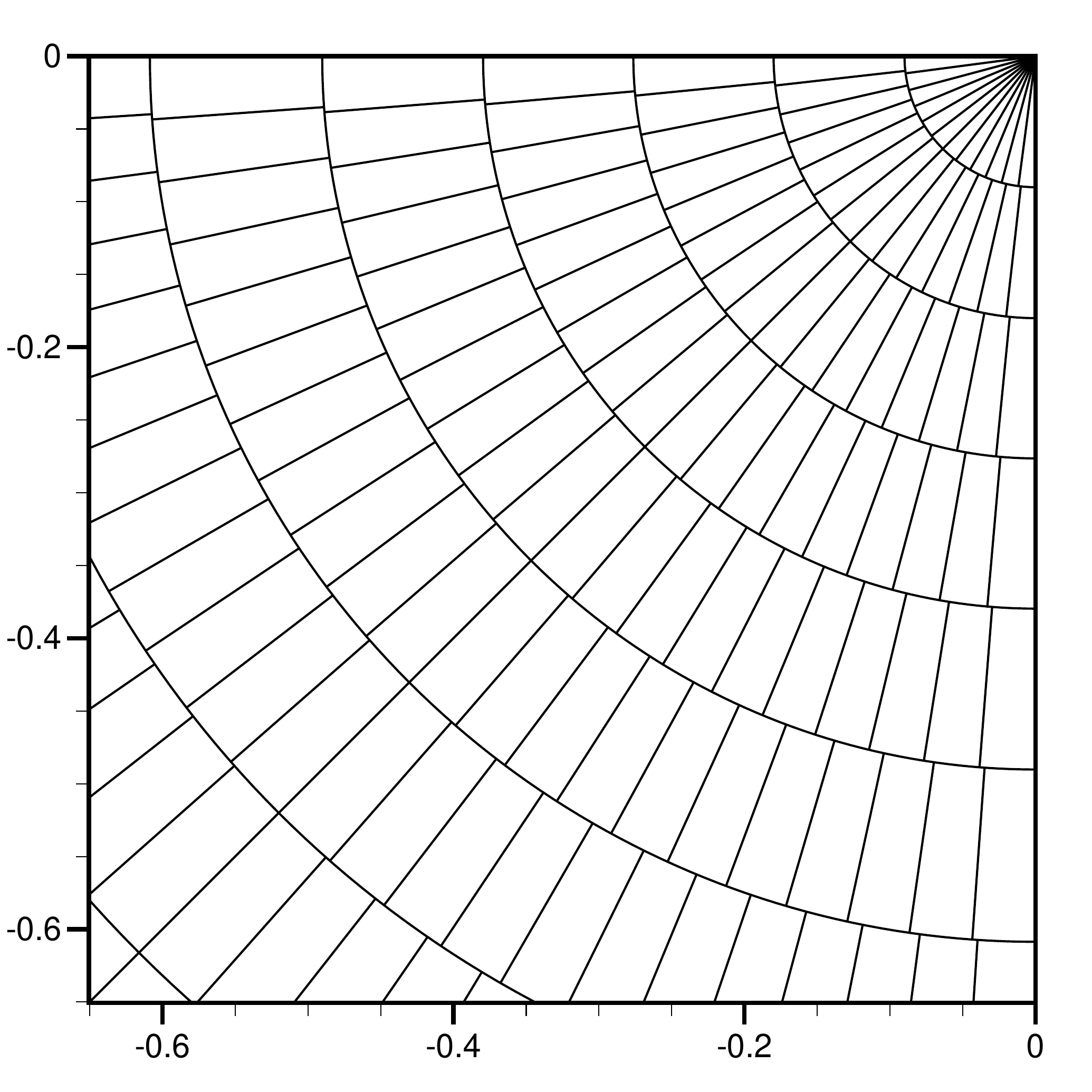}}
	\caption{\label{fig:cavity_vs_mesh}Discretization in the velocity space for mitigation of the ray effect. (a) Non-uniform velocity mesh; (b) special designed discretization of velocity space; (c) local enlargement.}
\end{figure}

The cavity flows are recomputed by employing the mitigation techniques, including using coarse mesh in the physical space, adopting non-uniform mesh in the velocity space, employing the stochastic particle method, and using a special designed velocity space discretization.
The results of temperature distribution based on different remedy are shown in Fig.~\ref{fig:cavity_mitigation}.
For the result in Fig.~\ref{fig:cavity_coarse}, the uniform mesh with $21\times21$ cells is used for the spatial discretization, which is about $3$ times coarser than those in the other computations.
It can be found that the wiggle in the flow structure is much reduced, but the solution is less accurate due to the lower spatial resolution.
For the solution in Fig.~\ref{fig:cavity_refine}, the original $60\times60$ uniform discrete velocity points are shrunk near the zero velocity point, which use smaller integration weights for the low speed region, see  Fig.~\ref{fig:cavity_nonuniform_vs}.
The ray effect is effectively mitigated, however, since too many discrete velocity points are lying on the $45\degree$ discontinuity line, the enhanced wiggles along the diagonal direction can be still observed.
This refinement in velocity space doesn't consider the principles B and C, where only jumps of ray effect are reduced, but the enhanced step along the diagonal direction are not well removed.
For stochastic particle method, we employ $3600$ particles in each cell and take $5000$ steps in averaging for the steady state solution.
In Fig.~\ref{fig:cavity_particle}, we can hardly see the ray effect.
Now, we give a well designed discretization of the velocity space as shown in Fig.~\ref{fig:cavity_special_vs}.
This discretization satisfies all the principles for adjusting the discrete velocity points.
For third quadrant of the discrete velocity space shown in Fig.~\ref{fig:cavity_special_vs_enlargement}, we discretize the radial direction into $25$ rings  where $\Delta r$ increases from the center to outside so that the integration weights in the low speed region are smaller.
The number of the discrete angular directions increases from $12$ to $60$ from the center to outside, which gives a finer resolution in the angular direction than that with regular Cartesian mesh.
Different numbers of discrete angular directions at different rings could avoid multiple discrete velocities lying on the same discontinuity line.
Therefore, it gives the best numerical solution without ray effect in Fig.~\ref{fig:cavity_special}.
However, since the optimal discretization of the velocity space is quite problem dependent, it requires special design case by case.
For steady flows, due to the adaptive property and irregular discretization in the velocity space, the particle method satisfies all the principles for velocity discretization as well and could provide the same good results without ray effect.
For numerical simulations with complex configurations, special discretization of the velocity space becomes difficult, so the particle methods provide a better and easy choice to avoid the ray effect in rarefied regime.

\section{Conclusion}

In this paper, a detailed theoretical analysis and numerical computations of the ray effect have been conducted in the collisionless limit
for the discrete velocity method with a continuous physical space.
The ray effect is coming from the use of a finite number of discrete velocity points to approximate a continuous velocity space, where the limited resolution cannot revolve the possible discontinuity in the velocity space.
As a result, the flow field is propagating in a finite number of directions with different speeds, which is accompanied with the resolution reduction in the physical space.
The theoretical analysis shows that the jumps in the flow variables in the physical space due to the ray effect are proportional to the resolution in the velocity space, which are equal to the moments of the differences of the distribution functions at the velocity points lying around the discontinuity line.
For unsteady flow computations, the limited resolution $\Delta u$ in the velocity discretization results in unresolved flow structure in the physical space on the scale of $t\Delta u$.
In practical computation, the compatible discretization of the physical space and velocity space, i.e., $\Delta x \ge t \Delta u$ or $\Delta u \le \Delta x / t$, is required for the simulation of highly rarefied flow.
With the inclusion of particle collision, the jumps of ray effect decay exponentially with time evolution by $e^{-t/\tau}$.
The theoretical analysis have been quantitatively validated in the numerical computations.

Several mitigation techniques have been investigated in reducing the ray effect, such as using coarse computational mesh, employing lower order flux solver to reduce the spatial resolutions, and adjusting the discrete velocity points to improve the resolution in the velocity space.
Three principles are used as guidelines to adjust the velocity space discretization in order to mitigate the ray effect.
Specifically, the discrete velocity points should be refined in the region where the moments of differences of the distribution functions are large; more discrete velocity points should be added in the direction perpendicular to the discontinuity line in the velocity space 
so as to increase the resolutions; and fewer discrete velocity points are allocated on the discontinuity line itself to reduce the number of moments on the discontinuous distribution functions there.
The mitigation techniques have been applied in the numerical tests and the effectiveness from different approaches have been compared.
Employing coarse computational mesh and low-order flux solver could reduce the ray effect, but provide lower accurate results.
Adjusting the velocity discretization is very effective, especially with the optimal designed discretization in the velocity space. 
However, the design is quite problem dependent, and the optimal choice is hard to be obtained in complex geometry. 

As a special velocity space discretization, stochastic particle method is also studied.
Due to the self-adaptive property and irregular discretization in the velocity space, 
all three guidelines for velocity adjustment are automatically satisfied by the particle method.
For unsteady flows, the error introduced by ray effect can be ignored because of large statistical noises.
For steady flow computation with averaged processes, the particle method can provide the same good results as the optimized discrete velocity method.
Moreover, the adaptive property in the discretization of velocity space from particle method makes it a better choice for general highly rarefied flow computations. 

\section*{Acknowledgement}

The work of Zhong is supported by the National Natural Science Foundation of China (Grant No. 11472219), the 111 Project of China (B17037) as well as the ATCFD Project (2015-F-016).
The research of Xu and Zhu as a visiting research assistant at HKUST is supported by Hong Kong research grant council (16206617) and National Natural Science Foundation of China (11772281, 91852114).

\section*{Reference}

\bibliography{ray_effect}

\clearpage



\begin{figure}[H]
	\centering
	\subfigure[Desity]
	{\includegraphics[width=0.32\textwidth]{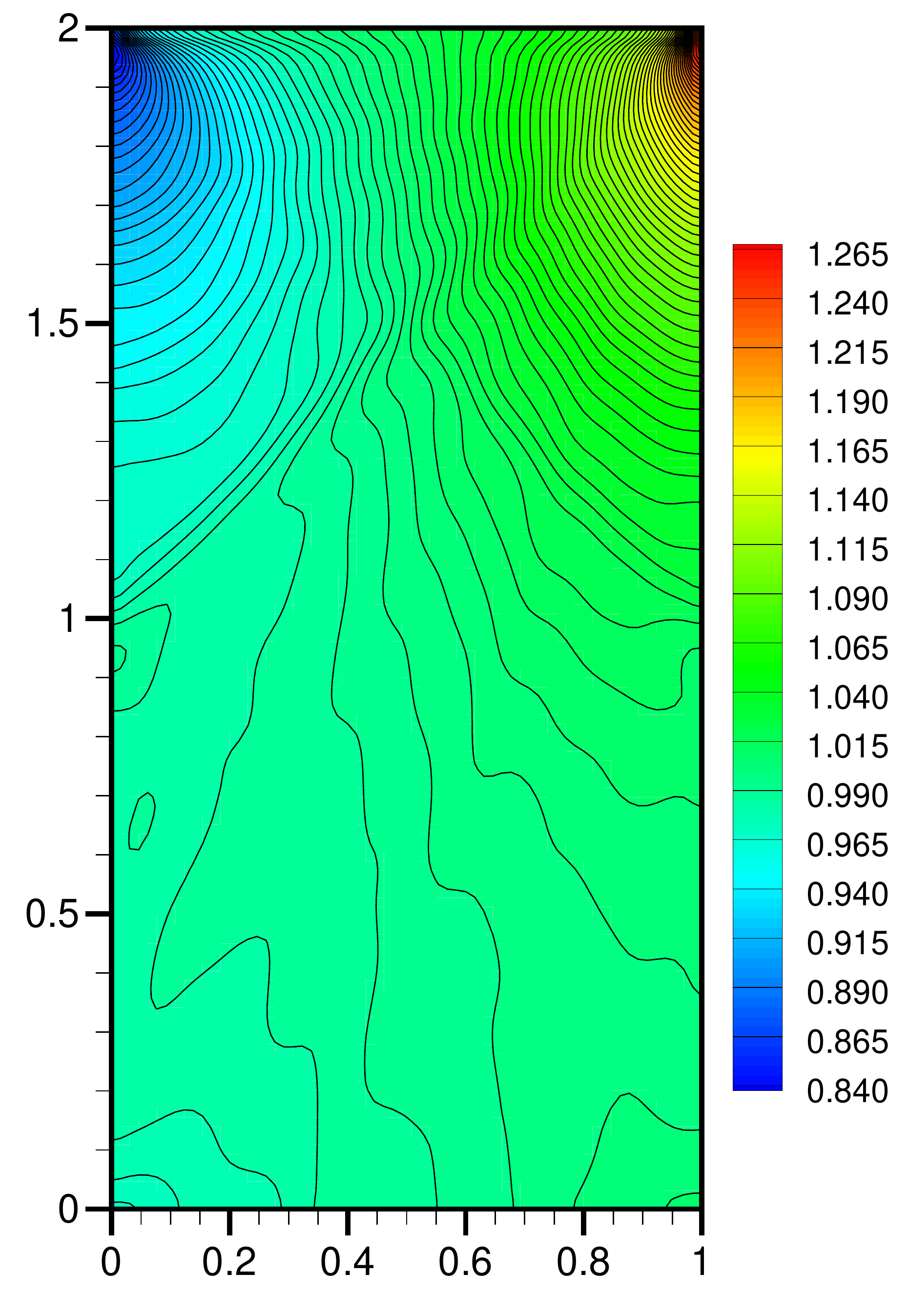}}
	\subfigure[Temperature]
	{\includegraphics[width=0.32\textwidth]{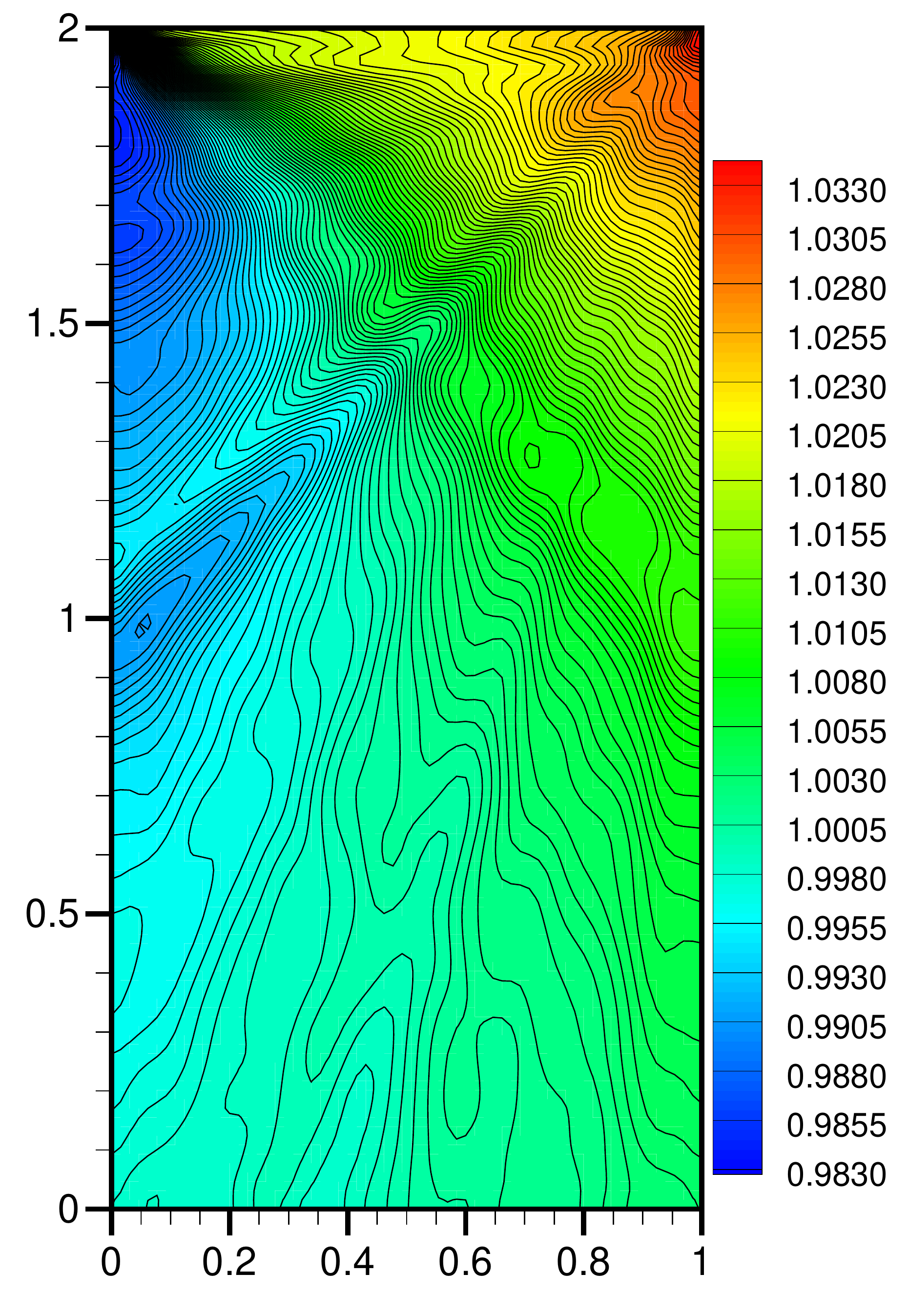}}
	\subfigure[Velocity]
	{\includegraphics[width=0.32\textwidth]{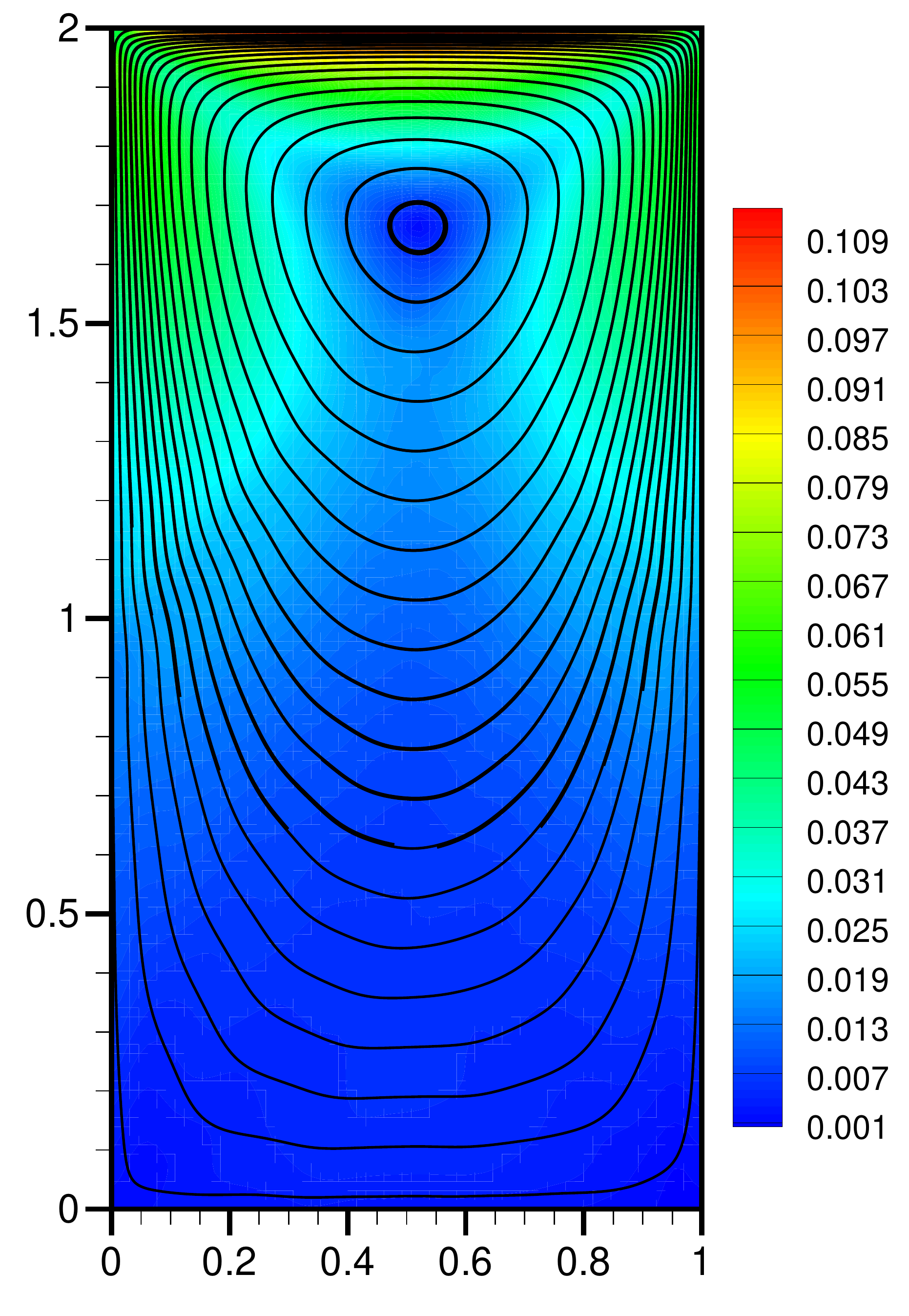}}
	\caption{\label{fig:cavity_high} Flow in a deep cavity using uniform mesh in velocity space.}	
\end{figure}

\begin{figure}[H]
	\centering
	\subfigure[Desity]
	{\includegraphics[width=0.32\textwidth]{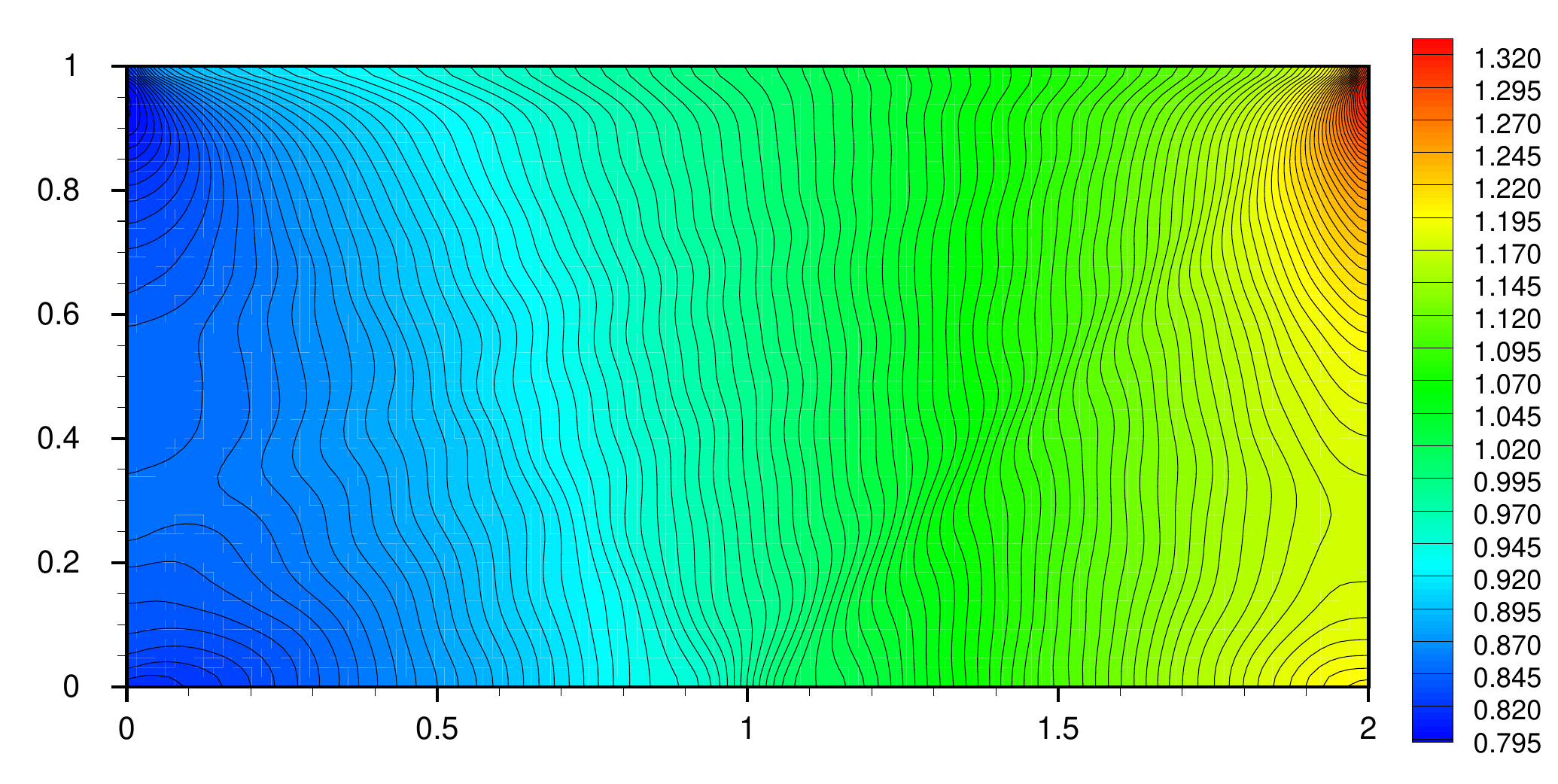}}
	\subfigure[Temperature]
	{\includegraphics[width=0.32\textwidth]{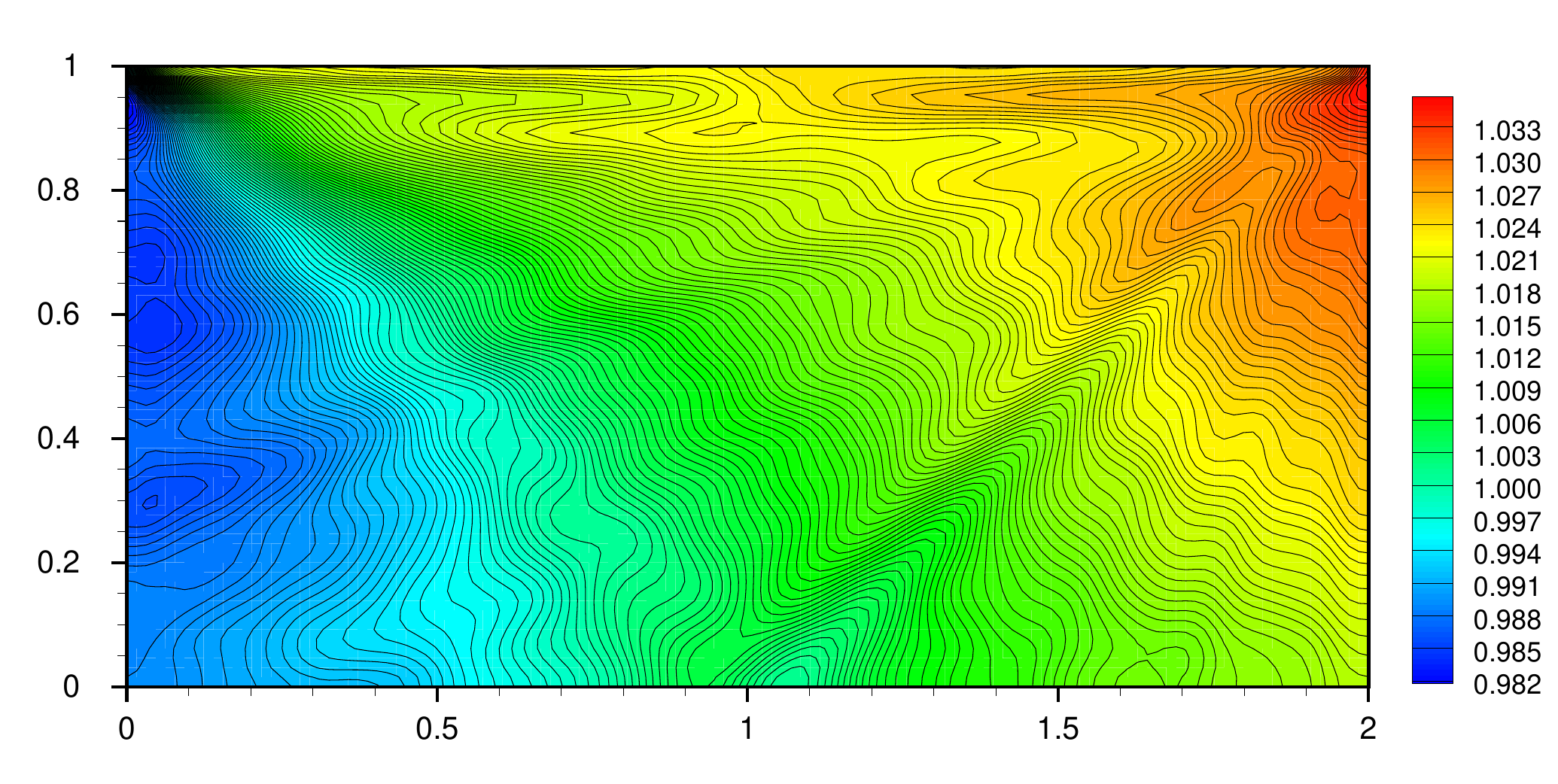}}
	\subfigure[Velocity]
	{\includegraphics[width=0.32\textwidth]{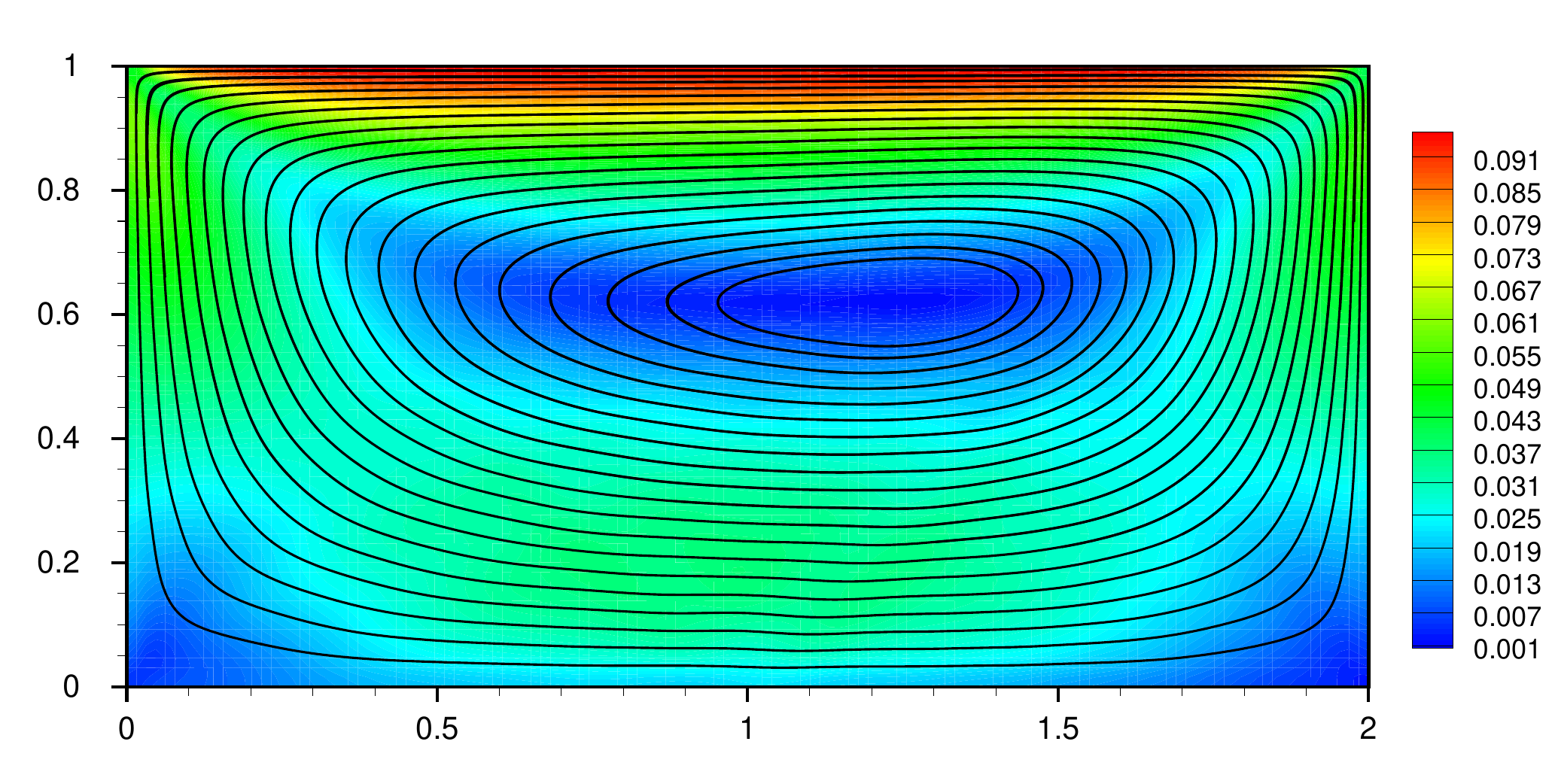}}
	\caption{\label{fig:cavity_long} Flow in a wide cavity using uniform mesh in velocity space.}	
\end{figure}


\clearpage

\end{document}